\pdfoutput=1
\documentclass[12pt,nofootinbib]{iopart}
\usepackage{iopams} 
\RequirePackage{doi}

\usepackage[sorting=none,style=numeric-comp,maxbibnames=99,doi=false,url=false,isbn=false]{biblatex}
\AtEveryBibitem{\clearfield{month}}
\AtEveryBibitem{\clearfield{day}}
\DeclareFieldFormat{pages}{#1}

\DeclareFieldFormat{doilink}{%
\iffieldundef{doi}{#1}%
{\href{http://dx.doi.org/\thefield{doi}}{#1}}}

\DeclareBibliographyDriver{article}{%
  \usebibmacro{bibindex}%
  \usebibmacro{begentry}%
  \usebibmacro{author/translator+others}%
  \setunit{\labelnamepunct}\newblock
  \usebibmacro{title}%
  \newunit
  \newunit\newblock
  \usebibmacro{byauthor}%
  \newunit\newblock
  \newunit\newblock
  \printfield{version}%
  \newunit\newblock
  \printtext[doilink]{%
  \usebibmacro{journal}%
  \newunit
  \iffieldundef{maintitle}
  { \textbf{\printfield{volume}} \printfield{pages} }.
  \newunit
  \printfield{year}}%
  \newunit\newblock
  \iftoggle{bbx:isbn}
    {\printfield{issn}}
    {}%
  \newunit\newblock
  \usebibmacro{doi+eprint+url}%
  \newunit\newblock
  \usebibmacro{addendum+pubstate}%
  \setunit{\bibpagerefpunct}\newblock
  \usebibmacro{pageref}%
  \usebibmacro{finentry}}

\AtBeginBibliography{\small}

\usepackage{hyperref}
\hypersetup{
    colorlinks,
    linkcolor={red!50!black},
    citecolor={blue!50!black},
    urlcolor={blue!80!black}
}

\usepackage{xcolor,amsthm,amsxtra,amsfonts,dsfont,graphicx,bm,tikz}

\usepackage{titlesec}
\usepackage{comment}

\usepackage[toc,page]{appendix}

\newcommand{\bea}{\begin{eqnarray}}
\newcommand{\eea}{\end{eqnarray}}
\newcommand{\bse}{\begin{subequations}}
\newcommand{\ese}{\end{subequations}}

\newtheorem{thm}{Theorem}[section]

\newtheorem{lem}[thm]{Lemma}

\newtheorem{result}[thm]{Result}

\newcommand{\ket}[1]{\vert#1\rangle}

\newcommand{\1}{\mathds{1}}

\makeatletter

\newcommand\mathcircled[1]{%
  \mathpalette\@mathcircled{#1}%
}
\newcommand\@mathcircled[2]{%
  \tikz[baseline=(math.base)] \node[draw,circle,inner sep=1pt] (math) {$\m@th#1#2$};%
}
\makeatother

\usepackage{tcolorbox}
\tcbuselibrary{skins,xparse}

\begin{document}
\title{Exact solution of long-range stabilizer Rényi entropy in the dual-unitary XXZ model
 }
\author{Jordi Arnau Montañà López\textsuperscript{1,2} and Pavel Kos\textsuperscript{1}}
\address{\textsuperscript{1}Max-Planck-Institut f{\"{u}}r Quantenoptik,
Hans-Kopfermann-Str.\ 1, 85748 Garching, Germany\\
\textsuperscript{2}Electrical and Computer Engineering, University of Washington, Seattle, Washington 98195, USA}
\ead{jordiaml@uw.edu and pavel.kos@mpq.mpg.de}
\vspace{10pt}
\begin{indented}
\item[]May 2024
\end{indented}

\begin{abstract}
Quantum systems can not be efficiently simulated classically due to the presence of entanglement and nonstabilizerness, also known as quantum magic. Here we study the generation of magic under evolution by a quantum circuit.
To be able to provide exact solutions, we focus on the dual-unitary XXZ model and a measure of magic called stabilizer Rényi entropy (SRE). Moreover, we focus also on long-range SRE, which cannot be removed by short-depth quantum circuits.
To obtain exact solutions we use a ZX-calculus representation and graphical rules for the evaluation of the required expressions. We obtain exact results for SRE after short-time evolution in the thermodynamic limit and for long-range SRE for all times and all Rényi parameters for a particular partition of the state. 
Since the numerical evaluation of these quantities is exponentially costly in the Rényi parameter, we verify this numerically for low Rényi parameters and accessible system sizes and provide numerical results for the long-range SRE in other bipartitions. 
\end{abstract}
\textit{This paper is dedicated to the memory of Marko Medenjak.}
\section{Introduction}
\label{sec:introduction}

Describing relevant many-body quantum states is a crucial task that is being intensively investigated. 
Remarkable headway has been achieved through the application of tensor networks, which can efficiently describe low entangled states, such as ground states~\cite{cirac2021matrix}. However, dynamics typically generate entanglement, making this approach inefficient as time progresses.   

Interestingly, the realm of efficiently describable quantum states extends beyond those with low entanglement. Another compelling example is states that are close to being stabilizer states, i.e. states prepared with Clifford operations from trivial state $\ket{00\dots}$.
These states might have large entanglement. The hardness of this particular expressivity is called nonstabilizerness, also called magic~\cite{m1,beverland2020lower,m2,m3,liu2022many}. 
Note that both entanglement and nonstabilizerness are necessary resources for universal quantum computation and for obtaining a quantum computational advantage~\cite{gottesman-knill,Bravyi_magic}. 
Typically, we expect that dynamics generate not only entanglement but also nonstabilizerness. If and how this happens is the motivating question for present work.


Significant effort has been put into developing tools to quantify nonstabilizerness by magic monotones~\cite{beverland2020lower,m3,liu2022many,haug_scalable_2023}. 
Most of these measures are, however, difficult to compute, especially for many-body systems.
One of the most promising ways to quantify nonstabilizness turned out to be the stabilizer Rényi entropy (SRE)~\cite{SRE_2022}, with its own resource theory~\cite{haug_stabilizer_2023,leone_nonstabilizerness_2023}. 
It is an entropy of the distribution of state coefficients in the Pauli basis.
SRE can be relatively cheaply computed for matrix product states (MPS)~\cite{quantifyingmps,tarabunga2024nonstabilizerness}. 
Moreover, it can be measured in experimental setups~\cite{Oliviero_2022, haug2023efficient}.

Magic contrasts with entanglement, since one can generate close to maximal density of magic by a single layer of T gates. 
This leads to the natural question of long-range magic, i.e. magic that cannot be removed by short-depth quantum circuits~\cite{tarabunga_many-body_2023,sewell2022mana,frau2024non}. 

The amount of nonstabilizerness has important physical consequences; in particular, low nonstabilizerness signals low complexity of state. 
This led to investigations of magic and particularly SRE in ground states
~\cite{haug_stabilizer_2023,tarabunga2024nonstabilizerness,lami_quantum_2023,oliviero_magic-state_2022,odavic2023complexity,tarabunga2023magic,tarabunga2023critical,passarelli2024nonstabilizerness}, especially in connection with quantum phase transitions, where complexity is expected to increase.  
Another way to observe the increase of complexity is after a quantum quench,  where we start evolving some particular initial state with low magic
~\cite{tarabunga2024nonstabilizerness,lami_quantum_2023,rattacaso_stabilizer_2023,goto2022probing,ahmadi2024quantifying,fux2023entanglementmagic}. 
All of the results cited above are limited to numerics for small system sizes or short times. Therefore, the problem of generation of magic asks for exact solutions, especially in the context of interacting many-body systems.
Some possible candidates, which proved fruitful in related contexts, are random unitary circuits~\cite{Fisher_2023}, dual-unitary circuits~\cite{bertini_exact_2018,Bertini_2019}, and Bethe ansatz integrable models~\cite{Baxter1982zz}. Here, we will focus on the intersection of the last two examples, the dual-unitary XXZ model.

In order to obtain exact solutions we will  use ZX calculus, which is a formalism for writing and transforming tensor diagrams~\cite{coecke2008interacting,coecke2011interacting}. The graphical transformation rules are useful for simplifying complicated tensor contractions. ZX-calculus has been used, for example,  in quantum circuit optimisation~\cite{Kissinger_2020}, measurement-based quantum computation~\cite{duncan2012graphical} and quantum error correction~\cite{khesin2023graphical}.

In this work, we investigate the evolution of long-range nonstabilizerness after a quantum quench for dynamics given by the dual-unitary XXZ model. Leveraging the ZX-calculus, we streamline the intricate tensor network contractions yielding insights into the system's behavior post-quench. Thus we obtain SRE after one layer of time evolution, and long-range magic at arbitrary times and Rényi indexes. These findings not only provide new physical insights into the generation of magic but also open up a new avenue for studying nonstabilizerness using the ZX-calculus.

The rest of this manuscript is organized as follows. In Sec.~\ref{sec:sre} we provide the definitions of SRE and long-range SRE. In Sec.~\ref{sec:model} we introduce the dynamics and states of interest, followed by Sec.~\ref{sec:methods}, where we introduce the ZX-calculus and its expressions for our quantities of interest. Sec.~\ref{sec:results} contains our results, with supporting details relegated to the Appendices. Finally, we provide conclusions and outlook in Sec.~\ref{sec:conclusions}.


\section{Stabilizer Rényi entropy}
\label{sec:sre}

We consider a one dimensional chain of $N$ qubits with a Hilbert space $\mathcal{H}= (\mathbb{C}^{2})^{\otimes N}$. Given a pure $N$-qubit state $|\psi\rangle \in\mathbb{C}^{2N}$, its coefficients in the Pauli basis are the \textit{Pauli spectrum}, $\text{spec}(|\psi\rangle)=\{\langle\psi|P|\psi\rangle, P\in\mathcal{P}_N\}$, where $\mathcal{P}_N = \{\sigma_{\alpha_1}\otimes...\otimes \sigma_{\alpha_N}|\sigma_{\alpha_i} \in \{\sigma_0,\sigma_x,\sigma_y,\sigma_z\}\}$ are all possible Pauli strings of length $N$. The Pauli spectrum yields two probability distributions,
one over the Pauli strings $P\in\mathcal{P}_N$, $\Xi_P$ referred to as the \textit{characteristic function} \cite{zhu2016clifford},
and one over expectation values $x_P=\langle\psi|P|\psi\rangle\in[-1,1]$, $\Pi(x)$ called the Pauli spectrum \cite{turkeshi2023pauli} as well:
\begin{align}
    &\Xi_P:=\frac{1}{2^N}\langle\psi|P|\psi\rangle^2, \ \ \ \Pi(x):=\frac{1}{4^N}\underset{x_P\in\text{spec}(|\psi\rangle)}{\sum}\delta(x-x_P)
\end{align}
The Rényi entropy of the characteristic funciton $\Xi_P$\footnote{Up to an offset of $-N\log(2)$.} gives a measure of nonstabilizerness, or quantum magic, called \textit{stabilizer Rényi entropy} (SRE)~\cite{SRE_2022}: 
\begin{align}
    &M_n(|\psi\rangle) =\frac{1}{1-n}\log(\zeta_n(|\psi\rangle)), \text{ where }\zeta_n(|\psi\rangle) := \frac{1}{2^N} \underset{P\in \mathcal{P}_N}{\sum}\langle\psi|P|\psi\rangle^{2n},
    \label{eq:Mn}
\end{align}
and $n$ is the Rényi parameter, which we take to be an integer $n \geq 1$. Here $\zeta_n(|\psi\rangle)$ is called the \textit{stabilizer purity} of state $|\psi\rangle$\cite{leone2024stabilizer} and corresponds to the moments of the Pauli spectrum: $\zeta_n = 2^N\int dx \Pi(x)x^{2n}$ \cite{turkeshi2023pauli}. 
In the thermodynamic limit, we will consider the SRE density~\cite{SRE_2022}
\begin{align}
    &m_{n}(|\psi\rangle) = \underset{N\to\infty}{\lim} \frac{1}{N}M_n(|\psi\rangle).
\end{align}
SRE has been extended to mixed states $\rho$ for $n=2$~\cite{SRE_2022} as
\begin{align}
    &\tilde{M}_2(\rho) = -\log\bigg(\frac{\sum_{P\in\mathcal{P}_N}\tr(P\rho)^{4}}{\sum_{P\in\mathcal{P}_N}\tr(P\rho)^{2}}\bigg)=-\log\bigg(\frac{\zeta_2(\rho)}{\zeta_1(\rho)}\bigg), \ \ \zeta_n(\rho):=\frac{1}{2^N}\underset{P\in\mathcal{P}_N}{\sum}\tr(P\rho)^{2n}. \label{eq:Mtil}
\end{align}
The term $\sum_{P\in \mathcal{P}_N}\tr(P\rho)^2 = \tr(\rho^2)$ is the suitable normalization by the purity. When $\rho$ is a reduced density matrix of a subsystem, $\tilde{M}_2(\rho)$ measures the SRE in that subsystem.

%

Next, let us define \emph{long-range SRE} between two regions $A,B$. It is the difference between $n=2$ SRE in the joint state $\rho_{AB}$ and the product state $\rho_A\otimes \rho_B$~\cite{tarabunga_many-body_2023}:
\begin{align}
    &L(\rho_{AB}) = \tilde{M}_2(\rho_{AB})-\tilde{M}_2(\rho_{A})-\tilde{M}_2(\rho_{B}). \label{eq:L(rhoAB)}
\end{align}
Note that $L(\rho_{AB})$ quantifies the amount of magic that can not be removed by short-depth quantum circuits~\cite{tarabunga_many-body_2023,white_conformal_2021}.

Let us now survey the computational complexities involved in quantifying magic.
First of all, SRE can be computed more efficiently than other measures of nonstabilizerness~\cite{SRE_2022}, as it does not require costly minimization procedures. Several algorithms have been developed to compute it exactly and approximately. When $|\psi\rangle$ is an $N$-qubit Matrix Product State (MPS) of bond dimension $\chi$, its $n$-Rényi SRE can be computed exactly with cost $O(N\chi^{6n})$~\cite{quantifyingmps}.  It can also be approximated by sampling algorithms with a cost $O(N\chi^3)$~\cite{haug_stabilizer_2023,lami_quantum_2023} or by MPSs in the Pauli basis with a cost $O(\chi^4)$~\cite{tarabunga2024nonstabilizerness}. 
Long-range SRE can be approximated using a sampling algorithm and Tree-Tensor Networks with a cost $O(\log(N)\chi^4)$~\cite{tarabunga_many-body_2023}.

\section{Dual-Unitary XXZ model}
\label{sec:model}
In this work, we will be interested in the nonstabilizerness of states produced by the \emph{dynamics} of the many-body systems. Typically, we will consider the evolution starting from a simple two-site product states. In particular, we will focus on the product of Bell pairs $\ket{\phi^+}=\frac{1}{\sqrt{2}}\sum_{i=0}^1 \ket{ii}$:
\begin{align}
    \ket{\psi(0)}=|\phi^+\rangle^{\otimes N/2}.
\end{align}
They are an example of so-called solvable states in the context of dual-unitary dynamics~\cite{piroli_exact_2020}.
The dynamic is given by a brick-wall quantum circuit, i.e. a periodically driven Floquet time evolution resulting from applying layers of two-qubit unitaries. Even layers are given by $\mathcal{U}_e = U_e^{\otimes N/2}$, where $U_e\in \mathbf{U}(4)$ is a two-qubit local unitary. Odd layers are given by $\mathcal{U}_o = \Pi U_o^{\otimes N/2} \Pi^\dagger$, where $U_o\in \mathbf{U}(4)$ and $\Pi$ is an $N$-periodic shift by one qubit: $\Pi|i_1...i_N\rangle = |i_2...i_Ni_1\rangle$. One Floquet timestep consists of an even and an odd layer: $\mathcal{U}_e\mathcal{U}_o$. Time evolution for positive integer Floquet time $t$ is then given by the following propagator:
\begin{align}
    &\mathcal{U}(t) = (\mathcal{U}_e\mathcal{U}_o)^t.
\end{align}

\begin{figure}
    \centering
    \includegraphics[width=0.5\linewidth]{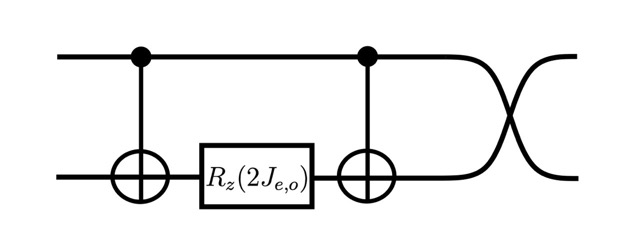}
    \caption{Dual-unitary XXZ gate in Eq.~\eqref{eq:V(J)DUXXZ} corresponding to the even and odd layers of the Floquet-dual-unitary circuit, up to a global phase. Here we wrote $e^{-iJ_{e,o}\sigma_z\otimes\sigma_z} = e^{-iJ_{e,o}}\text{CNOT}(\1 \otimes R_z(2J_{e,o}))\text{CNOT}$, where $R_z(2J_{e,o})=\begin{pmatrix}
    e^{-iJ_{e,o}} & 0\\ 0 & e^{iJ_{e,o}}    
    \end{pmatrix}$.}
    \label{fig:Uoe}
\end{figure}

The dynamics, in general, prove to be intricate. To achieve an exact solution for nonstabilizerness we must narrow down the scope of the local gates $U_e$ and $U_o$. Restricting to general dual-unitary models was not sufficient to compute SRE. This led us to reduce the models even more and focus on \emph{dual-unitary XXZ gates}~\cite{OpEE, dowling2023scrambling, rampp2023haydenpreskill,holdendye2023fundamental}:
\begin{align}
    U_{e,o}= \exp(-iJ_{e,o}\sigma_z\otimes \sigma_z)\cdot \text{SWAP} \label{eq:V(J)DUXXZ},
\end{align}
where $\text{SWAP}$ is the swap gate $\text{SWAP}\ket{ab} =\ket{ba}$. We show the decomposition of this gate in terms of standard gates in Fig. \ref{fig:Uoe}. 

The above-mentioned dynamic is not only dual-unitary but also Bethe ansatz integrable. It corresponds to a particular Trotterization of the spin-$\frac{1}{2}$ Heisenberg XXZ chain~\cite{ljubotina2019ballistic}. Since it is integrable, it does not exhibit quantum chaos, but it does exhibit operator scrambling~\cite{dowling2023scrambling}. Its time evolution has
only two parameters: the Ising interaction strengths $J_e,J_o$ for the odd and even layers. Our states of interest are:
\begin{align}
    &\ket{\psi(t)}=|\psi(J_o,J_e,t)\rangle := (\mathcal{U}_e(J_e)\mathcal{U}_o(J_o))^t|\phi^+\rangle^{\otimes N/2}. 
    \label{eq:psiJ1J2t}
\end{align}

\section{Methods: ZX-calculus expressions for SRE}
\label{sec:methods}
Let us now express the (long-range) SRE of the states from Eq.~\eqref{eq:psiJ1J2t} in a way that allows us to obtain exact expressions. We will follow the method for computing SRE in MPS states from~\cite{quantifyingmps} and combine it with the tools of ZX-calculus.

\begin{figure}
    \centering
    \includegraphics[width=0.7\linewidth]{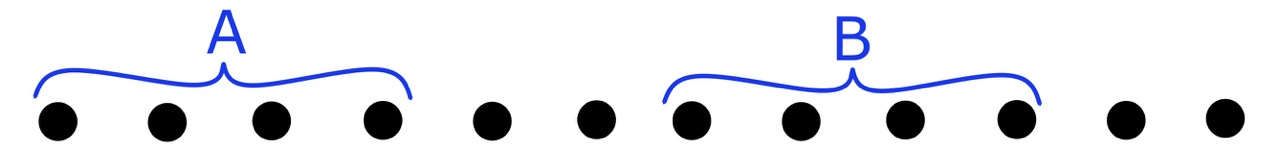}
    \caption{Bipartition of $N=12$ qubits into regions $A,B$ of $4$ qubits each, separated by two qubits.
    }
    \label{fig:regionsAB}
\end{figure}

The $N$-qubit state $|\psi(t)\rangle$ is partitioned into three regions: $A$, $B$, and the remaining portion, as illustrated in Figure \ref{fig:regionsAB}.
This leads us to define the corresponding reduced density matrices $\rho_{AB} = \tr_{N\setminus A,B}|\psi(t)\rangle\langle \psi(t)|,\rho_{A},\rho_{B}$. Our goal will be to compute the moments:
\begin{align}
    &\zeta_{n}(\rho_{AB})=\underset{P\in\mathcal{P}_{AB}}{\sum}\frac{\tr(P\rho_{AB})^{2n}}{2^N},\zeta_{n}(\rho_A)= \underset{P\in\mathcal{P}_{A}}{\sum}\frac{\tr(P\rho_{A})^{2n}}{2^N},\zeta_{n}(\rho_B)= \underset{P\in\mathcal{P}_{B}}{\sum}\frac{\tr(P\rho_{B})^{2n}}{2^N}.\label{eq:averagePaulis}
\end{align}
In order to find the long-range SRE from Eq.~\eqref{eq:L(rhoAB)} we only need to compute these expressions for Rényi parameters $n=1,2$. 

Naively evaluating the expressions in Eq.~\eqref{eq:averagePaulis}  is computationally expensive, as there are exponentially many Pauli strings to sum over. For a partition of the state, we will find an exact analytic expression for $\zeta_n(\rho_{AB}),\zeta_n(\rho_{A}),\zeta_n(\rho_{B})$ for all $n$ based on the exact numerical method in \cite{quantifyingmps} and verify it for small system sizes using the sampling algorithm for mixed states, as suggested in ~\cite{lami_quantum_2023}. See Appendix \ref{app:samplingmixed} for details. 

The exact numerical method in ~\cite{quantifyingmps} will be the starting point of our derivations.
There, the authors recast the sum of expectation values of Pauli strings in $M_n(|\psi\rangle)$ as the expectation value of a tensor $(\Lambda^{(n)})^{\otimes N}$ for a replica state $(|\psi\rangle|\psi^*\rangle)^{\otimes n}$:
\begin{align}
    &\underset{P\in\mathcal{P}_N}{\sum}\frac{\langle \psi|P|\psi\rangle ^{2n}}{2^N} =(\langle \psi|\langle \psi^*|)^{\otimes n}(\Lambda^{(n)})^{\otimes N}(|\psi\rangle|\psi^*\rangle)^{\otimes n},\label{eq:LambdaExp}
\end{align}
where $\Lambda^{(n)} = \frac{1}{2}\sum_{\alpha=0}^3 (\sigma_\alpha\otimes \sigma_\alpha^*)^{\otimes n}$. Eq.~\eqref{eq:LambdaExp} can be extended to mixed states $\rho$ as follows:
\begin{align}
    &\zeta_n(\rho)=\underset{P\in\mathcal{P}_N}{\sum}\frac{\tr(P\rho) ^{2n}}{2^N} =\tr((\rho\otimes \rho^*)^{\otimes n}(\Lambda^{(n)})^{\otimes N}). \label{eq:mixedLambda}
\end{align}

Our approach to derive an analytical expression for $L(\rho_{AB})$ involves recasting Eqs. ~\eqref{eq:averagePaulis}, ~\eqref{eq:mixedLambda} as ZX-calculus diagrams and then applying the ZX-calculus rules to streamline them. 
Let us now introduce parts of ZX calculus that we need, with more details aviable in  Appendix \ref{app:ZX} and~\cite{Montana2023}.

\begin{figure}[t]
    \centering
    \includegraphics[width=0.9\linewidth]{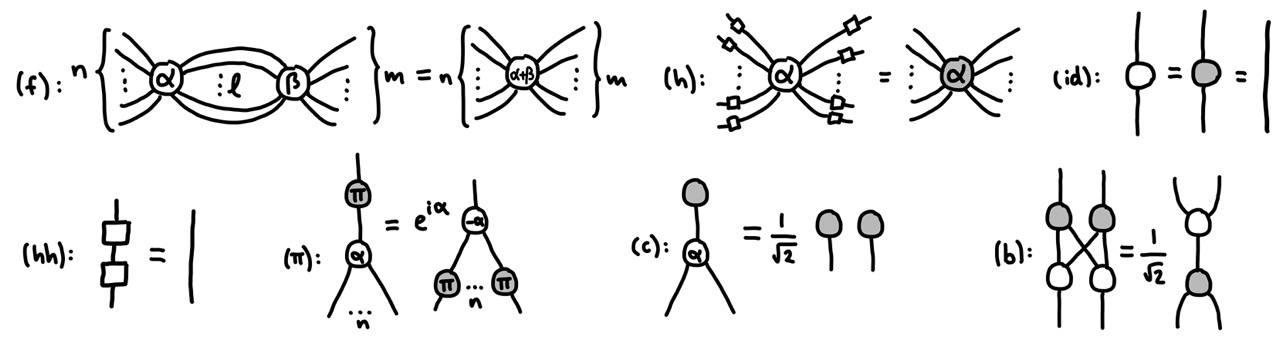}
    \caption{Graphical transformation rules of the ZX-calculus (adapted from~\cite{vandewetering2020zxcalculus}). Fusion rule (f): same-colored spiders joined by a leg add their phases. Hadamard rule (h): applying Hadamards on all gates of a spider changes its color. Identity removal (id): phaseless spiders with an incoming and an outgoing leg are identity. Hadamard removal (hh): Hadamard is its own identity. $\pi$-commute rule ($\pi$): a $\pi$-spider applied on a spider of a different color negates its phase and is applied to the remaining legs, with a global phase. $\pi$-copy rule (c): applying a phaseless state spider on an opposite-colored spider copies the state on the outgoing legs. Bialgebra rule (b).
    }
    \label{fig:ZX_rules}
\end{figure}

We first summarize the building blocks and graphical rules of ZX-calculus, following closely~\cite{vandewetering2020zxcalculus}. The building blocks are $Z$-spiders and $X$-spiders, written as  white and grey tensors, with a phase $\alpha$:
\begin{align}
    \includegraphics[width=0.9\linewidth]{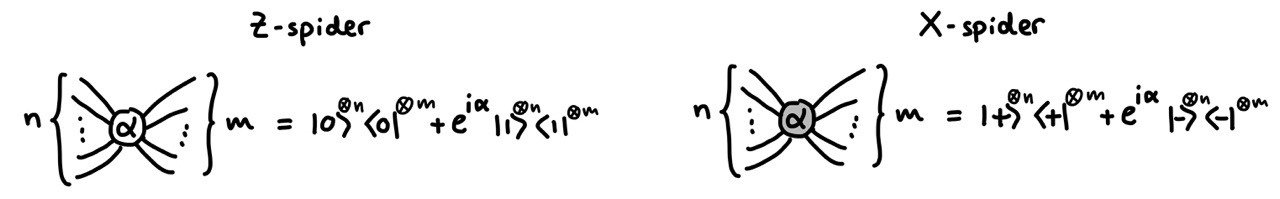}.
\end{align}
If the phase is zero, we omit it in the diagrams. 
The spiders comprise many common parts of a quantum circuit. For example, an $X$-spider with only one leg is $\sqrt{2}|0\rangle$, which is a common initial state. Pauli matrices can be written in terms of $Z$- and $X$-spiders with two legs and a phase $\pi$. 
In principle, the Hadamard gate can be written in terms of spiders through its Euler decomposition. 
But it turns out to be convenient to define it as its own symbol, a box: 
\begin{align}
    \includegraphics[width=0.4\linewidth]{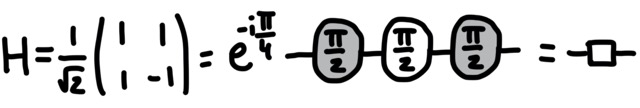}.
\end{align}
Another common gate is the CNOT:
\begin{align}
    \includegraphics[width=0.27\linewidth]{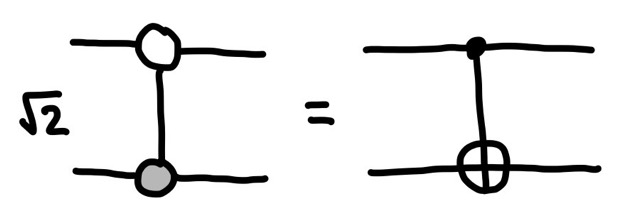}.
\end{align}

Once a tensor contraction is written using the spiders (and Hadamards) we use the graphical rules from Figure \ref{fig:ZX_rules} to simplify them. For example, using these rules the following useful identities can be derived:
\begin{align}
    \includegraphics[width=0.5\linewidth]{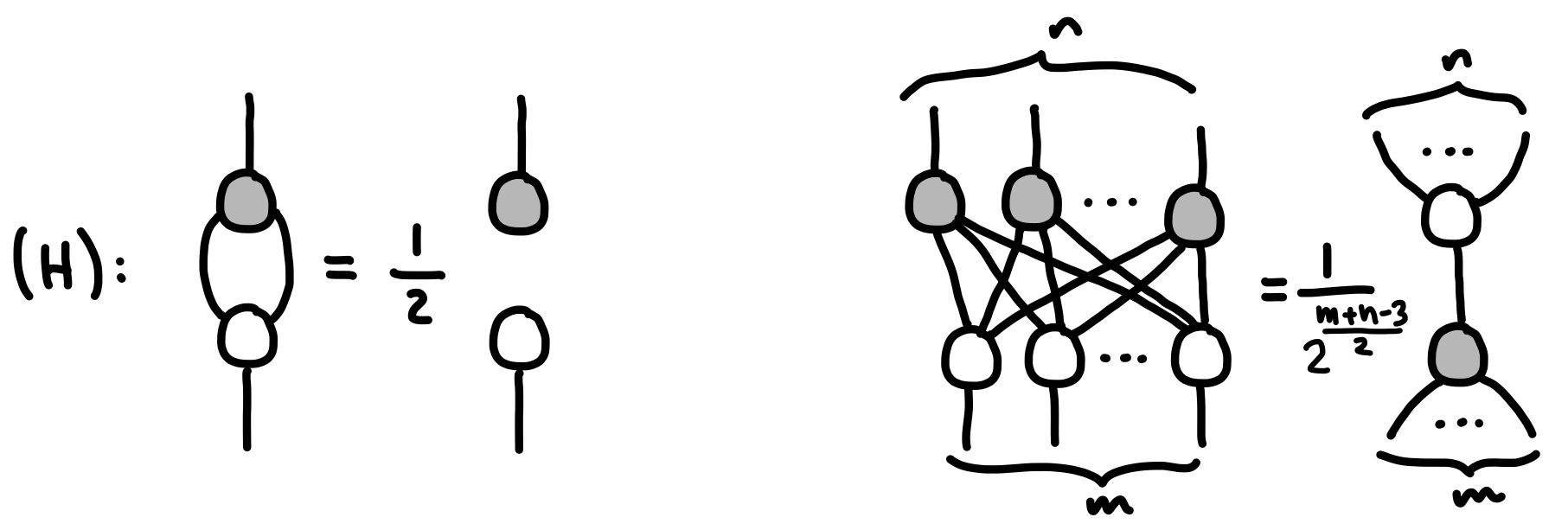}. \label{eq:hopf_bi}
\end{align}

Let us now express the local gates of time evolution using ZX calculus. Writing $\alpha := 2J_o, \beta:=2J_e$ for brevity, we have:
\begin{align}
    &U_o=\exp\Big(-i\frac{\alpha}{2}\sigma_z\otimes \sigma_z\Big)\cdot \text{SWAP} = \raisebox{-2.0\baselineskip}{%
        \includegraphics[
        width=0.45\linewidth,
        keepaspectratio,
        ]{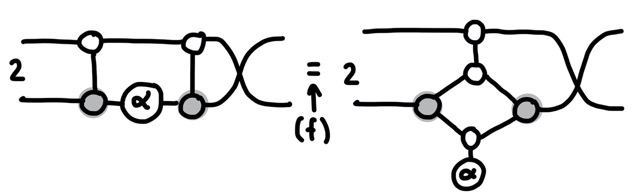}%
    }%
    \\
    &\raisebox{-1.0\baselineskip}{%
        \includegraphics[
        width=0.7\linewidth,
        keepaspectratio,
        ]{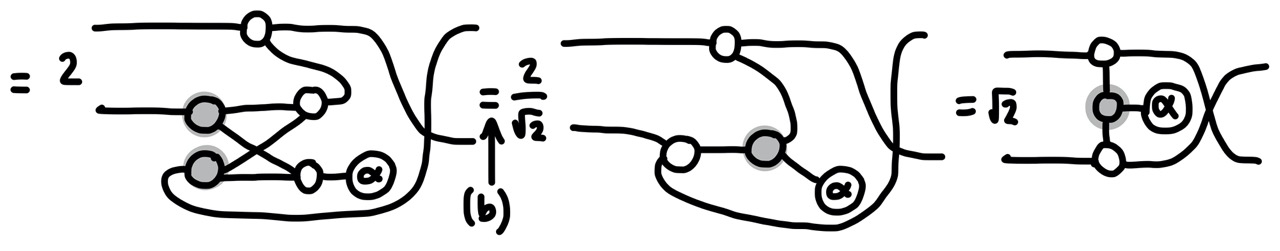}%
    }. \label{eq:Uo}
\end{align}
In the first line we use the fusion rule $(f)$ from Figure \ref{fig:ZX_rules}. In the second line we move around the tensors so it is clear we can apply the bialgebra rule $(b)$ and present the diagram in a clearer way. The last equality says that we can write $U_o$ as a product of a \textit{phase gadget} and a SWAP gate. 
It is instructive to see how the unitary simplification $U_o U_o^\dagger = \mathds{1}$ happens in the ZX-calculus: 
\begin{align}
    \includegraphics[width=0.9\linewidth]{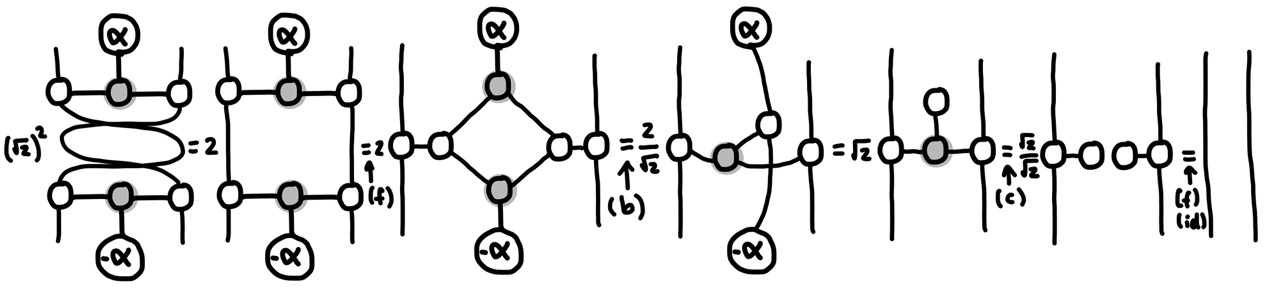}. \label{eq:gadget_unitary}
\end{align}
We first reshape the diagram using the fusion rule $(f)$ such that we can apply the bialgebra rule $(b)$. Then we fuse the $\alpha$ and $-\alpha$ spiders (leaving a $0$-phase spider) and apply the copy rule $(c)$. Finally we apply the fusion $(f)$ and identity removal $(id)$ rules.

\section{Results}
\label{sec:results}

We first present the simpler results on SRE with shallow circuits, to illustrate the techniques we use. Then we proceed to our main result, which is the exact expression of long-range SRE for all times and Rényi indices for a specific non-trivial bipartition.

\subsection{Exact result for half time step SRE}\label{sec:results_halfstep}

\begin{figure}
    \centering
    \includegraphics[width=0.8\linewidth]{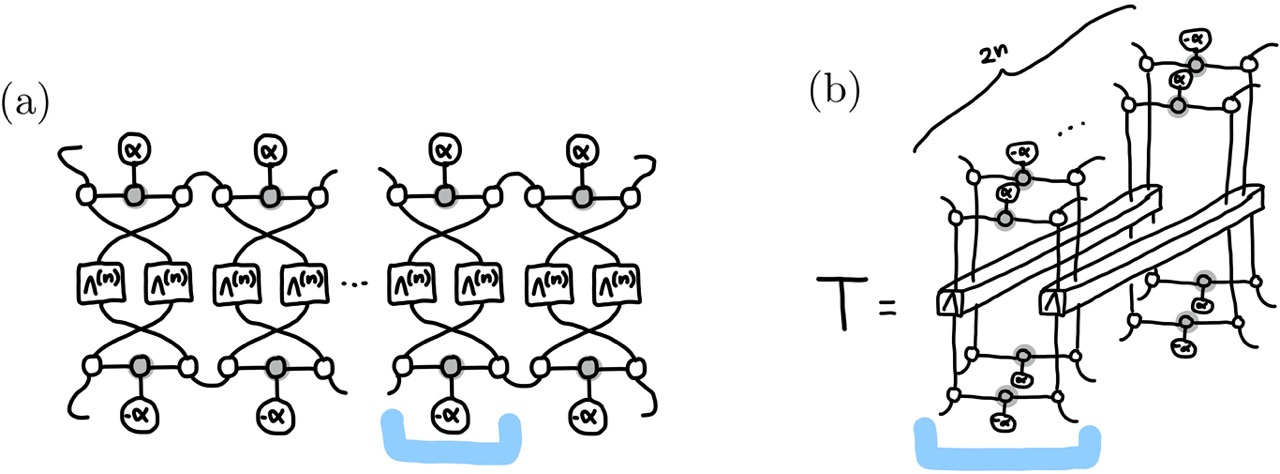}
    \caption{(a) shows Eq. ~\eqref{eq:averagePaulis} for the state $|\psi(J_o,0,\frac{1}{2})\rangle = \mathcal{U}_o(J_o)|\phi^+\rangle ^{\otimes N/2}$, with Ising interaction $J_o = \frac{\alpha}{2}$ and Rényi parameter $n$. Only the first (top) Rényi copy is shown. 
    (b) shows the transfer tensor $T$ showing all $2n$ Rényi copies, also indicated in blue on the left.}
    \label{fig:halfstep}
\end{figure}

Let us first illustrate the usefulness of ZX expression for computing SRE after one layer of gates and any Rényi index.
For translationally invariant systems, the SRE contraction in Eq.~\eqref{eq:averagePaulis} can be obtained by repeatedly applying a transfer tensor $T$~\cite{quantifyingmps}, e.g. see Fig. \ref{fig:halfstep}. In the thermodynamic limit, if the transfer tensor has a unique leading eigenvector $|\lambda_0\rangle$, the contraction will converge to the leading eigenvalue $\lambda_0$:
\begin{align}
    &\zeta_n(|\psi\rangle) 
    = \tr(T^{N/2})\approx \tr(\lambda_0^{N/2}|\lambda_0\rangle\langle \lambda_0|).
\end{align}
Note that $T,\lambda_0$ depend on $n$, but we omit it for clarity. Thus we obtain the SRE density:
\begin{align}
    & m_n(|\psi\rangle) = \underset{N\to \infty}{\lim}\frac{1}{N}\frac{1}{1-n}\log\big(
    \zeta_n(|\psi_n\rangle)
    \big) = \frac{1}{2(1-n)}\log(\lambda_0).
\end{align}

In the following result, we show that for the half time-step example from Figure \ref{fig:halfstep}, with $|\psi(J_o,0,\frac{1}{2})\rangle = \mathcal{U}_o(J_o)|\phi^+\rangle^{\otimes N/2}$, we can obtain the leading eigenvector and eigenvalue exactly. We could not prove its uniqueness, but we checked numerically for $n=2,3$ that it is indeed unique. 
\begin{result}
\label{thm:halfstep}
    For a half time-step of the dual-unitary XXZ evolution, the density of magic in the thermodynamic limit  is:

    $$m_{n}(|\psi(J,0,\frac{1}{2})\rangle) =\frac{1}{2(1-n)}\log\bigg(\frac{1+\cos^{2n}(2J) + \sin^{2n}(2J)}{2}\bigg).$$
\end{result}

Here we show the main ideas of the proof, more details can be found in Appendix \ref{app:halfstep}. The strategy is to check that the candidate eigenvector, vectorization of $\Lambda^{(n)}$, satisfies the eigenvector equation. 

We start by expressing Eq.~\eqref{eq:averagePaulis} for the half-time time-step with state $|\psi(J_o,0,\frac{1}{2})\rangle= \mathcal{U}_o|\phi^+\rangle^{\otimes N/2}$ using ZX-diagrams and $\Lambda^{(n)}$, which we show in Figure \ref{fig:halfstep}. 
Next we express the tensor $\Lambda^{(n)}$  with spiders. Note that $\Lambda^{(n)}= 2\Lambda_x^{(n)}\Lambda_z^{(n)}$, where $\Lambda_x^{(n)}:=\frac{1}{2}(\sigma_0^{\otimes 2n}+\sigma_x^{\otimes 2n}),\Lambda_z^{(n)}:=\frac{1}{2}(\sigma_0^{\otimes 2n}+\sigma_z^{\otimes 2n})$ are commuting projectors. Their ZX-calculus diagrams are:
\begin{align}
    \includegraphics[width=0.8\linewidth]{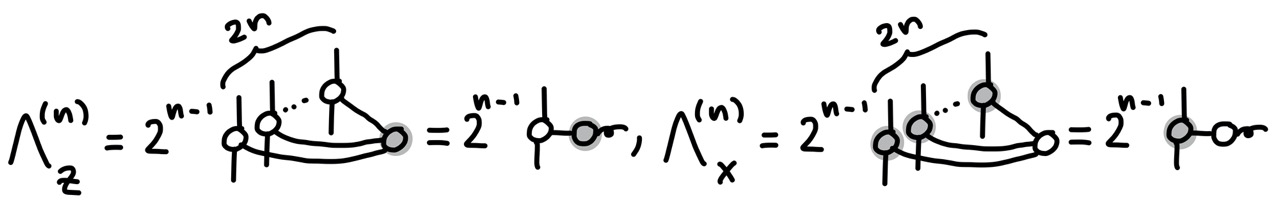},\label{eq:Lsquiggle}
\end{align}
which we derive in Appendix \ref{app:ZX}. In the last equality we introduced the notation \includegraphics[width=0.1\linewidth]{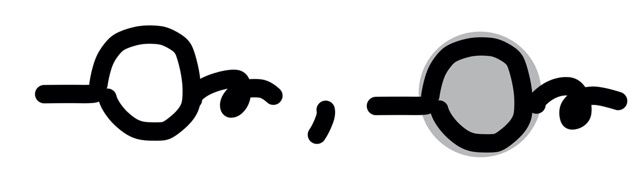}. The tensor contraction required to compute SRE, Eq.~\eqref{eq:LambdaExp}, requires contracting tensors $\Lambda^{(n)}$ with $2n$ replicas of the state. Drawing all Rényi copies in each diagram would be cumbersome so
with this notation we only write the first Rényi copy, while remembering that the rest of the copies that are not shown are still connected to \includegraphics[width=0.1\linewidth]{figures_compressed/squiggle.jpeg} and that the even copies are complex conjugated. We will show all Rényi copies explicitly in the following computation, but will use the compact notation from Eq.~\eqref{eq:Lsquiggle} in Sec.\ref{sec:LSRE}.\\

We start by applying the candidate eigenvector, vectorized $\Lambda^{(n)}$, to the transfer tensor $T$ from the right:
\begin{align}
    \includegraphics[width=0.9\linewidth]{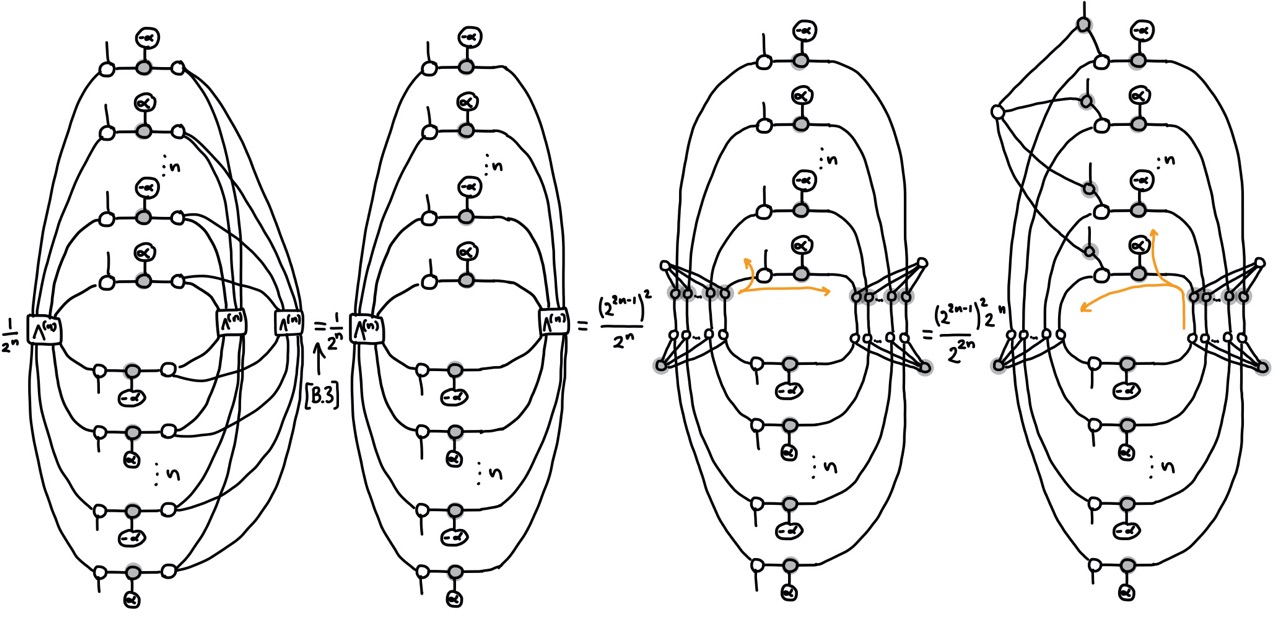}.
\end{align}
 Since applying two $\Lambda^{(n)}$ tensors to the same $Z$-spider is the same as applying only one (Lemma \ref{lem:Lambda2}), in the first equation we remove one of the two $\Lambda^{(n)}$ on the right. Next we write the diagram completely in ZX-calculus notation and indicate with orange arrows that we would like to ``pass" the $\Lambda^{(n)}_x$ through the $Z$-spiders. Using properties of the $\Lambda^{(n)}_x$ tensor, we arrive at the diagram on the right, where we managed to move $\Lambda^{(n)}_x$ to the top output leg. Since we claim $\Lambda^{(n)}$ is the eigenvector, we also need to ``push" a $\Lambda^{(n)}_z$ component to this output leg. First, we indicate with orange arrows that we try to push the $\Lambda^{(n)}_z$ on the right through.
\begin{align}
    \includegraphics[width=0.9\linewidth]{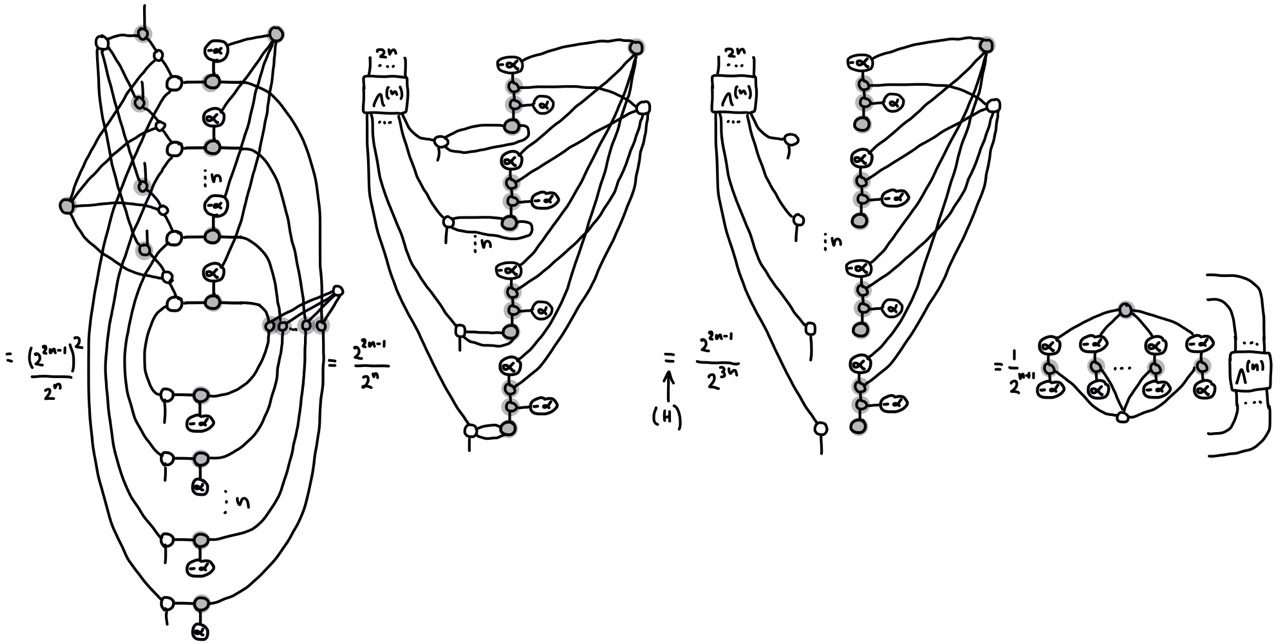},
\end{align}
Using properties of $\Lambda^{(n)}_z$, we arrive at the first diagram, where the rightmost $\Lambda^{(n)}_z$ ended up attached to the phase spiders and the leftmost one was moved to the top output leg. The first equality repackages the $\Lambda^{(n)}_x,\Lambda^{(n)}_z$ on the top legs into $\Lambda^{(n)}$ with the appropriate scalar. It also presents the diagram in a way that makes it clear that the candidate eigenvector $\Lambda^{(n)}$ is joined to the rest of the diagram only by $Z$-spiders on the left side connected by two edges to $X$-spiders on the right. Using the Hopf rule $(H)$ from Figure \ref{fig:ZX_rules}, all these double connections split, so we are left with a vector $\Lambda^{(n)}$ and a subdiagram that has no inputs or outputs, i.e. a scalar. In the last equality, we use the identity removal rule $(id)$ on $\Lambda^{(n)}$ to remove the $Z$-spiders underneath $\Lambda^{(n)}$ and vectorize it. We also present the scalar subdiagram in a clearer way. Thus we obtain $\Lambda^{(n)}$ multiplied by a scalar, the eigenvalue in its ZX-calculus representation. We find the numerical representation of the eigenvalue in Appendix \ref{app:halfstep}.

\subsection{Long-range SRE}
\label{sec:LSRE}

We proceeded to compute long-range SRE $L(\rho_{AB})$ and moments $\zeta_n(\rho_{AB})$, exactly for all times and all Rényi parameters $n$ for a specific partition $B_0$ of the state $|\psi(t)\rangle$. 
Firstly, we derive an expression, which can be evaluated efficiently numerically for fixed Rényi index $n$, but is exponentially costly in $n$. Secondly, we also evaluate this expression analytically for any $n$.
Thirdly, we check that both results agree with the numerical sampling algorithm for the accessible times.
Lastly, we discuss other partitions.
\\

In this subsection it is convenient to denote by $T=2t$ the number of layers of evolution.
Let us now define our first (time depended) partition of interest. We will always partition $N$-qubit state $|\psi(t)\rangle$  into regions $A,B$ of equal number of qubits $N_A=N_B$ and the rest with $N-N_A-N_B$ sites. We consider periodic boundary conditions, and regions $A$ and $B$ are separated by $d=(N-N_A-N_B)/2$ sites on both sides.
We then trace out the rest of the system and obtain $\rho_{AB}=\tr_{AB^c}|\psi(t)\rangle\langle \psi(t)|$.

If $d\geq 2T$, the lightcones of regions A and B will not intersect, so $\rho_{AB}=\rho_A\otimes \rho_B$ and thus $L =0$. 
This leads us to define a partition $B_0$ with separation $d=2(T-1)$. 
 We further specify the partition by
\begin{align}
    B_0: \quad N_A=N_B=2T,\quad d=2(T-1), \quad N=8T-4.
    \label{eq:B0}
\end{align}
In this case, $\rho_A = \frac{1}{2^{N_A}}\1$, so $\tilde{M}(\rho_A)=\tilde{M}(\rho_B)=0$ and the only contribution to long range magic is from the joint state, $\tilde{M}(\rho_{AB})$:
\begin{align}
    &L_{B_0}(\alpha,\beta,t) = \tilde{M}(\rho_{AB}(\alpha,\beta,t)).
\end{align}
For $B_0$, using ZX-calculus we can simplify the contraction of $\zeta_n(\rho_{AB})$ in Eq.~\eqref{eq:mixedLambda} to
\begin{result}\label{result52}
    \begin{align}
        \!\!L_{B_0}(\alpha,\beta,T)\!=\!-\log\!\Bigg(\!\!\ 2^{8T+6}\raisebox{-2.85\baselineskip}{
        \includegraphics[
        width=0.60\linewidth,
        keepaspectratio,
        ]{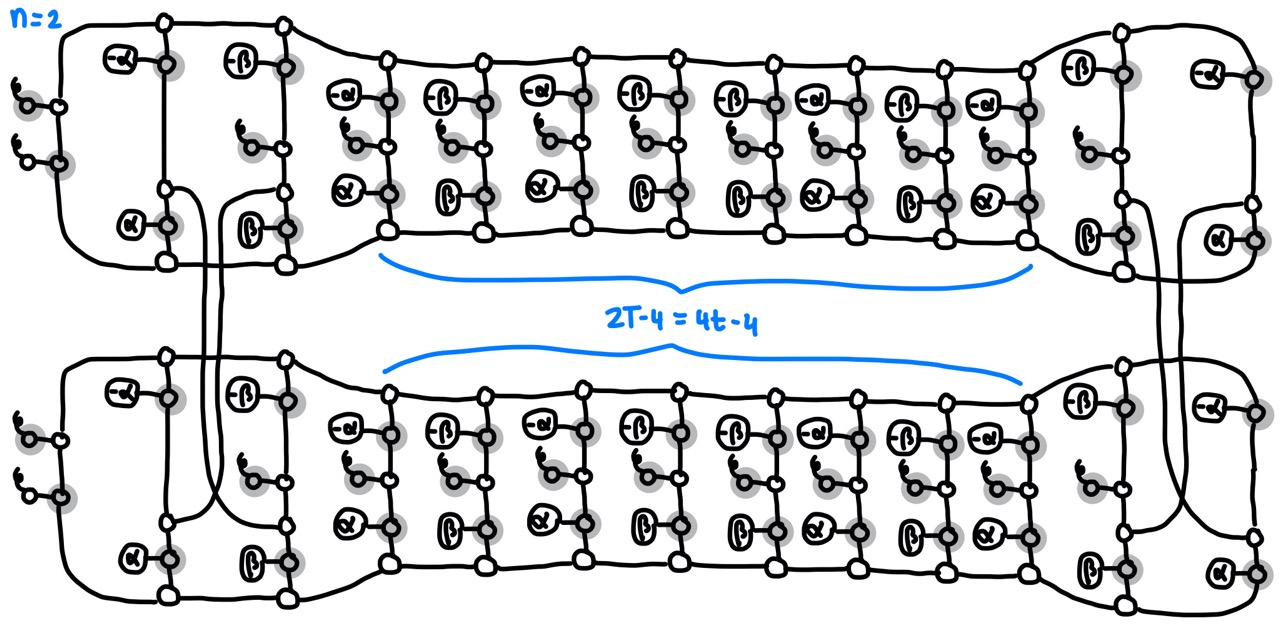}} \Bigg).
    \end{align}
\end{result}

Here we provide the main steps of the derivation, shown for $T=6$. The result is derived by first using unitarity and then using ZX-calculus rules to simplify further. These often involve keeping track of global scalars. We omit them here for clarity, but can be found in Appendix \ref{sec:app_longSRE}. Thus, the following diagrams are equivalent up to a global scalar.

We start by writing out the expression for $\zeta_n(\rho_{AB})$ in Eq.~\eqref{eq:mixedLambda} as a tensor diagram:

\begin{align}
    &\raisebox{-4\baselineskip}{\includegraphics[width=0.99\linewidth]{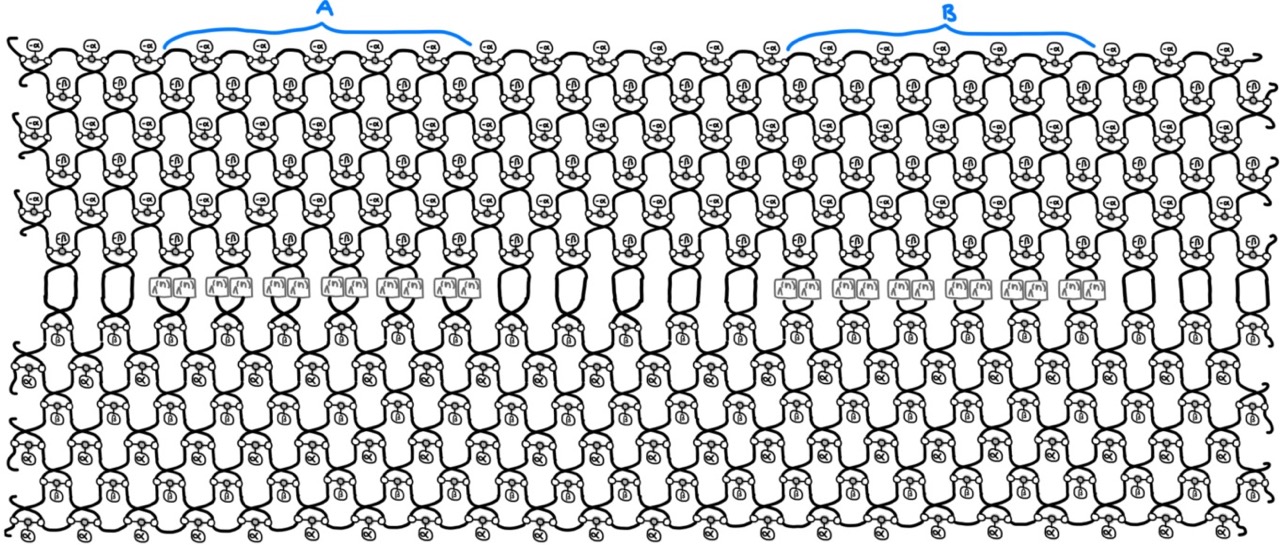}}
\end{align}
Using unitarity in the regions between A and B, we simplify two backwards lightcones:
\begin{align}
    &\raisebox{-4\baselineskip}{\includegraphics[width=0.9\linewidth]{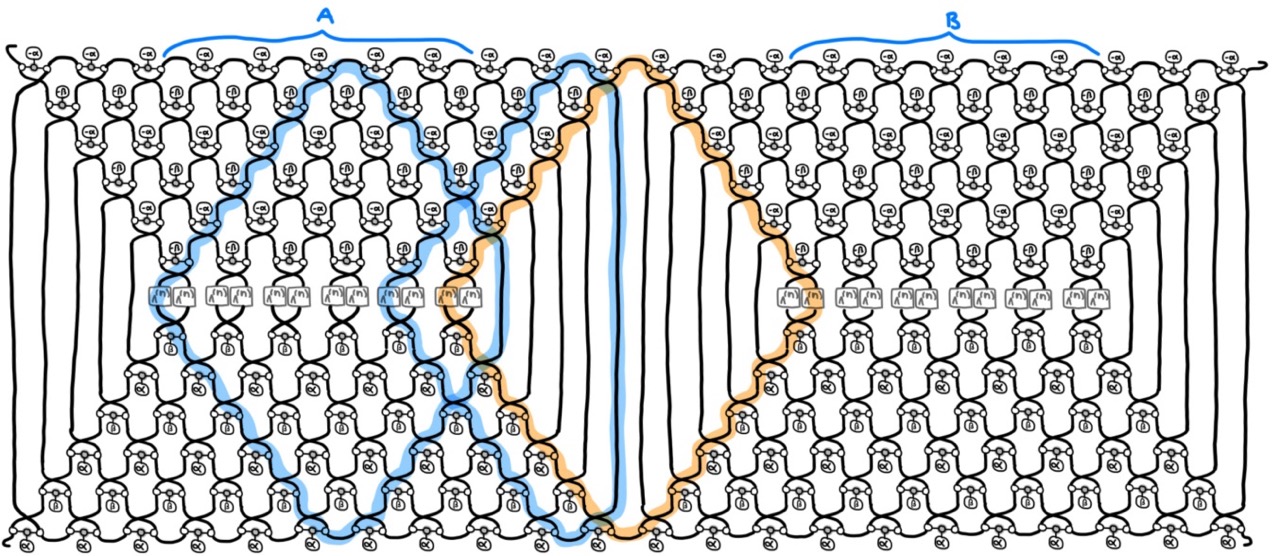}}
\end{align}
We highlight in blue some $\Lambda^{(n)}$ that will be simplified to $\Lambda^{(n)}_z$. Along each blue loop there is only one $\Lambda^{(n)}$ and by Lemma \ref{lem:LzI}, the $\Lambda^{(n)}_x$ component is projected out and only $\Lambda^{(n)}_z$ remains. Not all $\Lambda^{(n)}$ are simplified like this, in orange we show a loop where we can not apply Lemma \ref{lem:LzI}. Using the ZX-calculus representation for $\Lambda^{(n)},\Lambda^{(n)}_z$ gives:
\begin{align}
    &\raisebox{-2.2\baselineskip}{\includegraphics[width=0.9\linewidth]{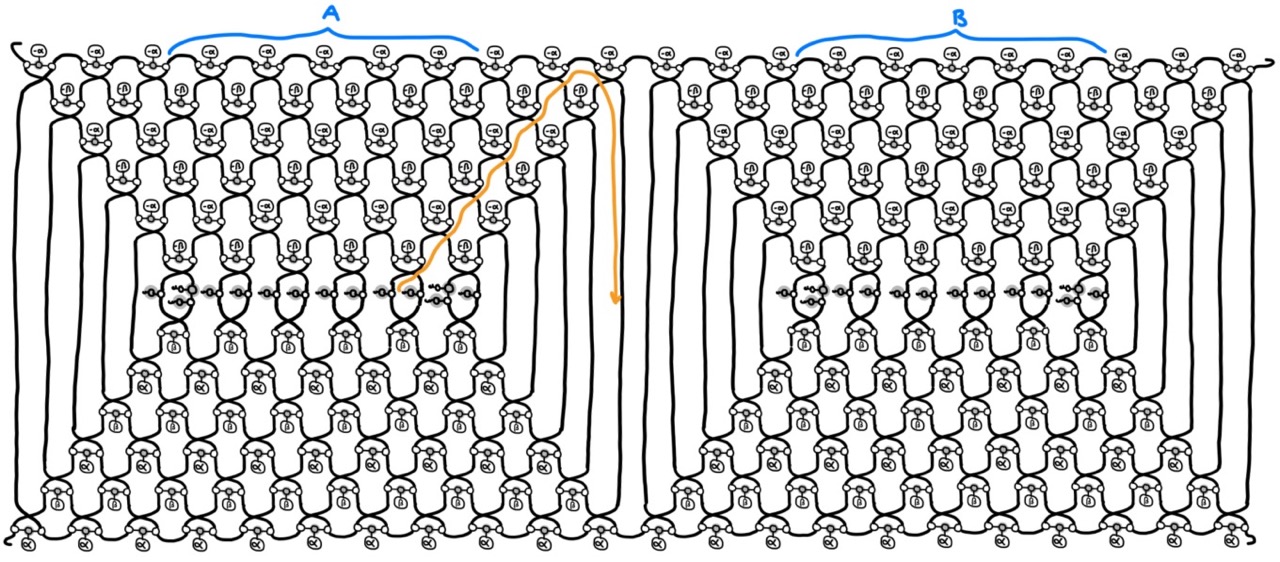}}
\end{align}
We indicate with an orange arrow that we slide a $\Lambda_z^{(n)}$ outside of region $A$. This makes use of one of the key properties of the dual-unitary XXZ gate: it is a product of a term that is diagonal in the computational basis, the phase gadget or Ising interaction, and a $\text{SWAP}$. Thus we slide $\Lambda_z^{(n)}$ along the $\text{SWAP}$s, commmute it with the phase gadgets and slide it outside of region $A$. Doing this for all $\Lambda_z^{(n)}$ that are not multiplying a $\Lambda_x^{(n)}$ gives:
\begin{align}
    &\raisebox{-4\baselineskip}{\includegraphics[width=\linewidth]{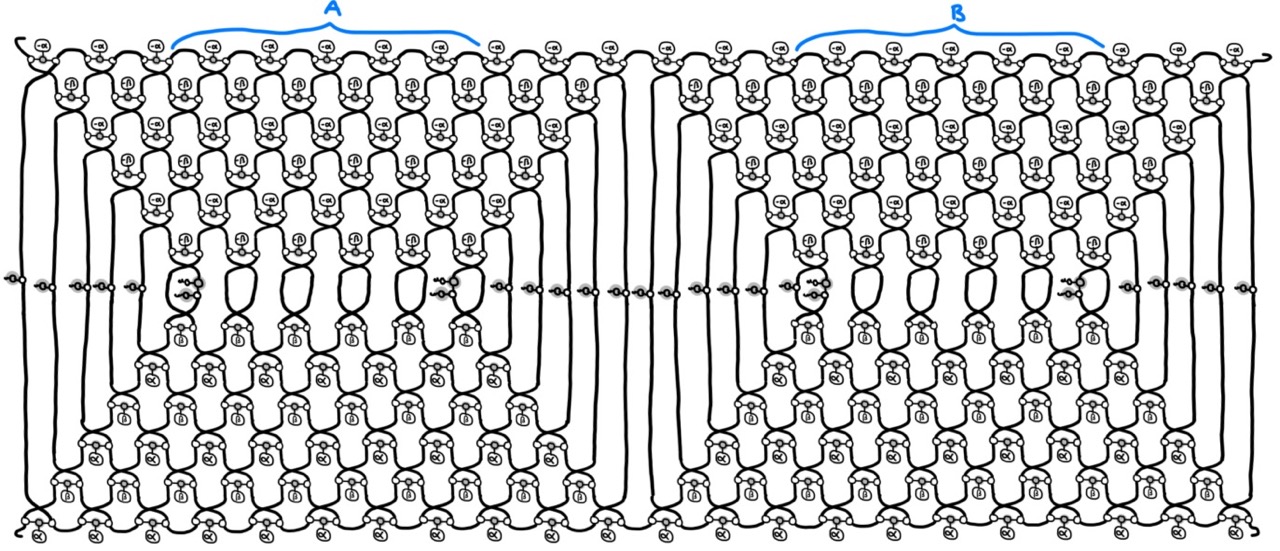}}
\end{align}
The bulk qubits in regions $A,B$ are now connected to their conjugates through identities. Therefore, we can use unitarity as in Eq.~\eqref{eq:gadget_unitary} 
in the regions $A,B$ to simplify a backwards lightcone. 
\begin{align}
    &\raisebox{-4\baselineskip}{\includegraphics[width=\linewidth]{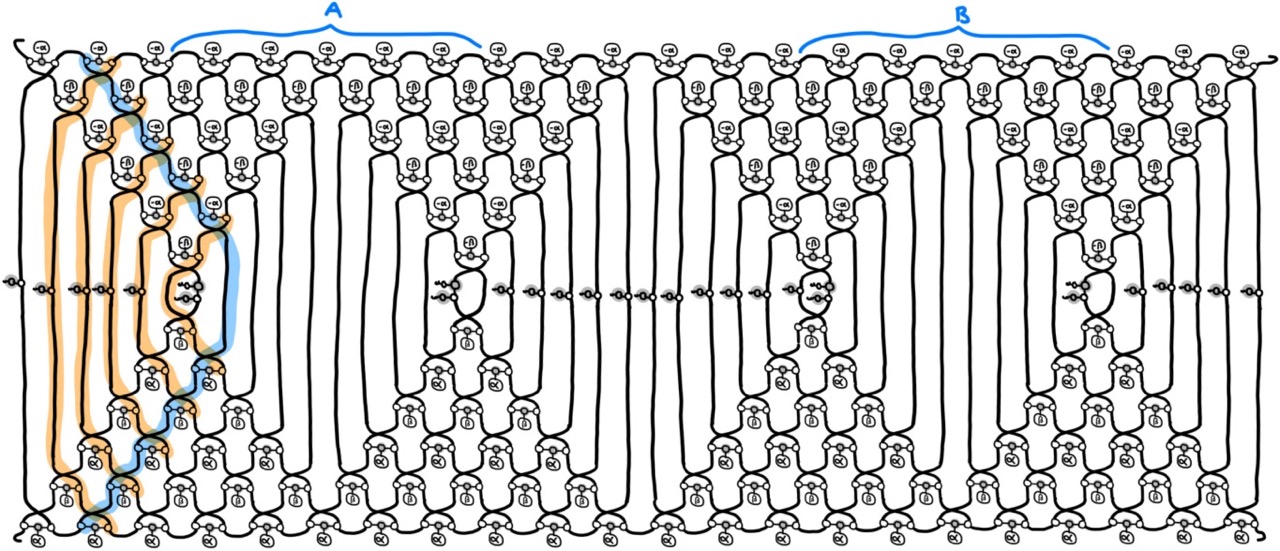}}
\end{align}

The blue and orange highlights indicate the legs of a diagonal of phase gadgets. Along the highlighted legs there are only $Z$-spiders, so we can use unitarity to simplify each phase gadget with its conjugate. See Appendix \ref{sec:app_longSRE} for details on this step. This allows us to simplify all phase gadgets except for those connected to their conjugates via $\Lambda_x^{(n)}$.

\begin{align}
    &\raisebox{-2.2\baselineskip}{\includegraphics[width=0.92\linewidth]{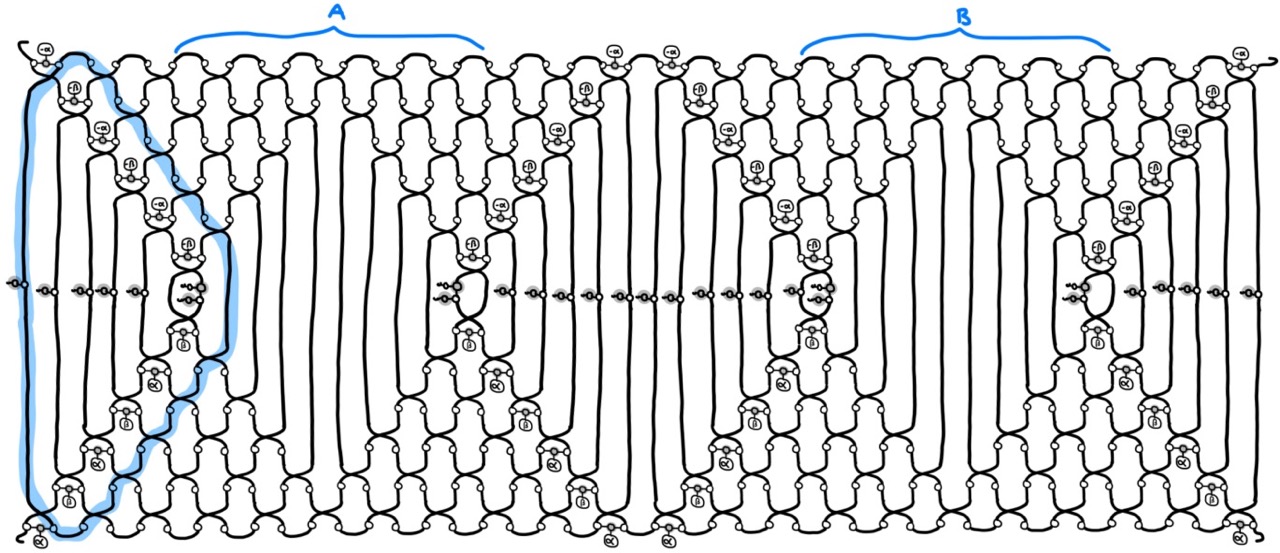}}
\end{align}
We highlight in blue a loop that only contains $Z$-spiders, so the right half of the loop can be contracted to a phaseless $Z$-spider. See Appendix \ref{sec:app_longSRE} for details. Doing this for all loops that don't contain a $\Lambda_x^{(n)}$ we get:
\begin{align}
    &\raisebox{-2.2\baselineskip}{\includegraphics[width=0.92\linewidth]{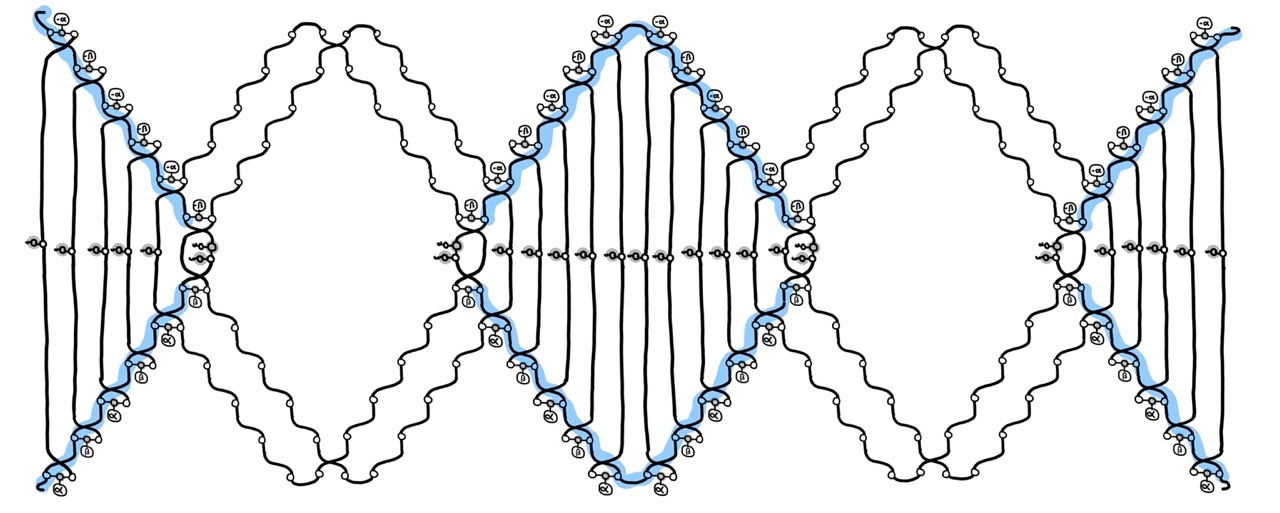}}
\end{align}

Along the blue lines we can remove two pairs of $\Lambda_x^{(n)},\Lambda_z^{(n)}$ using Lemmas \ref{lem:LxSquared} and \ref{lem:Lacross}. Then, removing the identities with $(id)$ and reshaping the diagram for clarity we obtain:
\begin{align}
    &\raisebox{-2.2\baselineskip}{\includegraphics[width=0.7\linewidth]{figures_compressed/LB0n2.jpeg}}
\end{align}
The middle sections of the diagram are dependent on time: more time-steps correspond to making the middle section longer. It can be efficiently  contracted numerically as a quantum channel with complexity linear in $t$~\footnote{For fixed $t$ it can actually be reduced to $\log(t)$ if necessary via exponentiation by squaring.}. The left and right endcaps are independent of $t$, so the complexity to contract the endcaps with the middle sections is independent of $t$. Since the contraction for a given Rényi parameter $n$ requires contracting $2n$-Rényi copies that are coupled through $\Lambda^{(n)},\Lambda^{(n)}_z$, the tensor that needs to be stored to perform the contraction has a size that grows exponentially in $n$. This limits numerical results to low $n=2,3$.  

Nevertheless, we can find an analytical expression for the diagram for all $n$, as detailed in Appendix \ref{sec:app_longSRE}, to obtain our main result:
\begin{result}\label{res:Lresult}
For partition $B_0: N_A=N_B=2T,d=2(T-1),T=2t$, the long-range SRE is: 
\begin{align}
    &L_{B_0}(J_o,J_e,T)= \tilde{M}_2(\rho_{AB}) -2\tilde{M}_2(\rho_A)= -\log\bigg(\frac{\zeta_2(\rho_{AB})}{\zeta_1(\rho_{AB})}\bigg), \label{eq:LB0}
\end{align}
where we exactly evaluate all of the moments $\zeta_n(\rho_{AB})$ as: 
\begin{align}
    &\zeta_n(\rho_{AB})=\frac{1}{2^{4T-2}}\big(1+2(f_n(J_o)f_n(J_e))^T+g_n(J_o,J_e)(f_n(J_o)f_n(J_e))^{2T-4}\big), \label{eq:B0n>1}
\end{align}
and we have defined
\begin{alignat}{2}
    &f_n(J) := \cos^{2n}(2J)+\sin^{2n}(2J),\notag\\
    &g_n(J_o,J_e) := \frac{1}{2^{2n}}\underset{k=0}{\overset{4}{\sum}}{4\choose k}\underset{m=0}{\overset{1}{\sum}}&&\big(\cos(2J_o+2J_e)^{4-k}\sin(2J_o+2J_e)^k+\notag\\
    & &&+(-1)^{m}\cos(2J_o-2J_e)^{4-k}\sin(2J_o-2J_e)^k\big)^{2n}.
\end{alignat}
\end{result}

\begin{figure}
    \centering
    \includegraphics[width=0.9\linewidth]{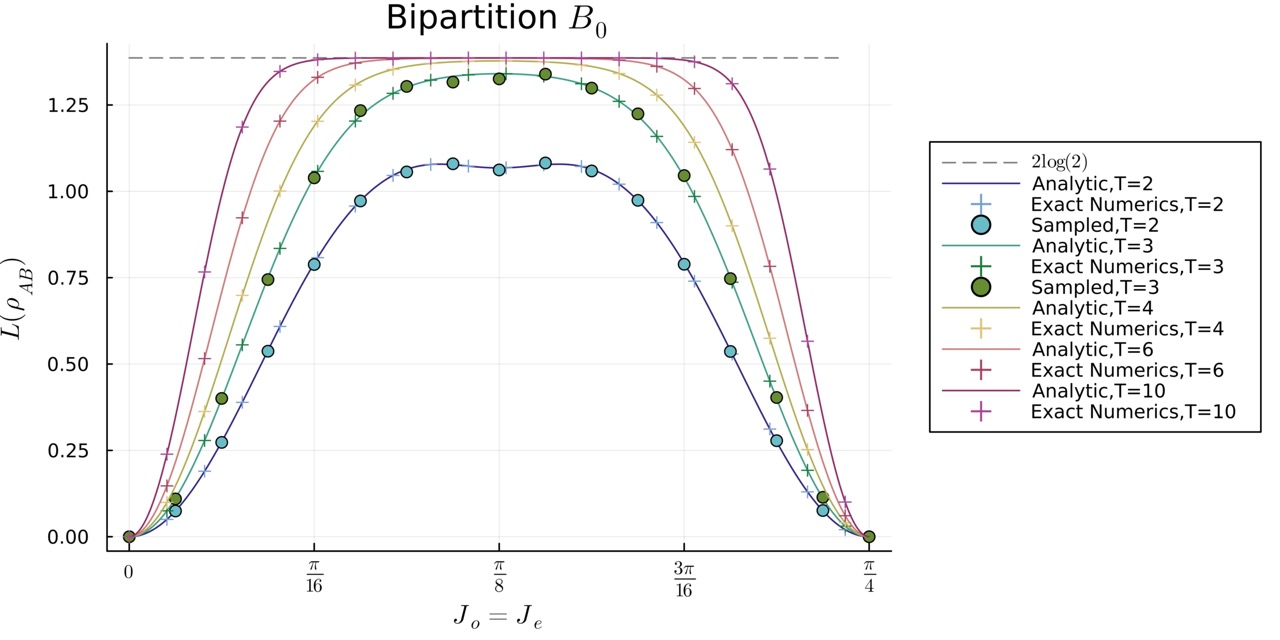}
    \caption{Long range magic between regions $A,B$ of the dual-unitary XXZ time evolution of the state $|\phi^+\rangle^{\otimes N/2}$ with periodic boundary conditions for partition $B_0$ defined in Eq.~\eqref{eq:B0}. 
    The analytical result (line), Result \ref{res:Lresult}, and the efficient numerical contraction of Result \ref{result52}  (crosses) agree up to numerical precision.
    For $T=2,3$ we also approximate it via sampling algorithm (dots) with $20000$ samples. 
    }
    \label{fig:longsreB0}
\end{figure}

\begin{figure}
    \centering
    \includegraphics[width=0.6\linewidth]{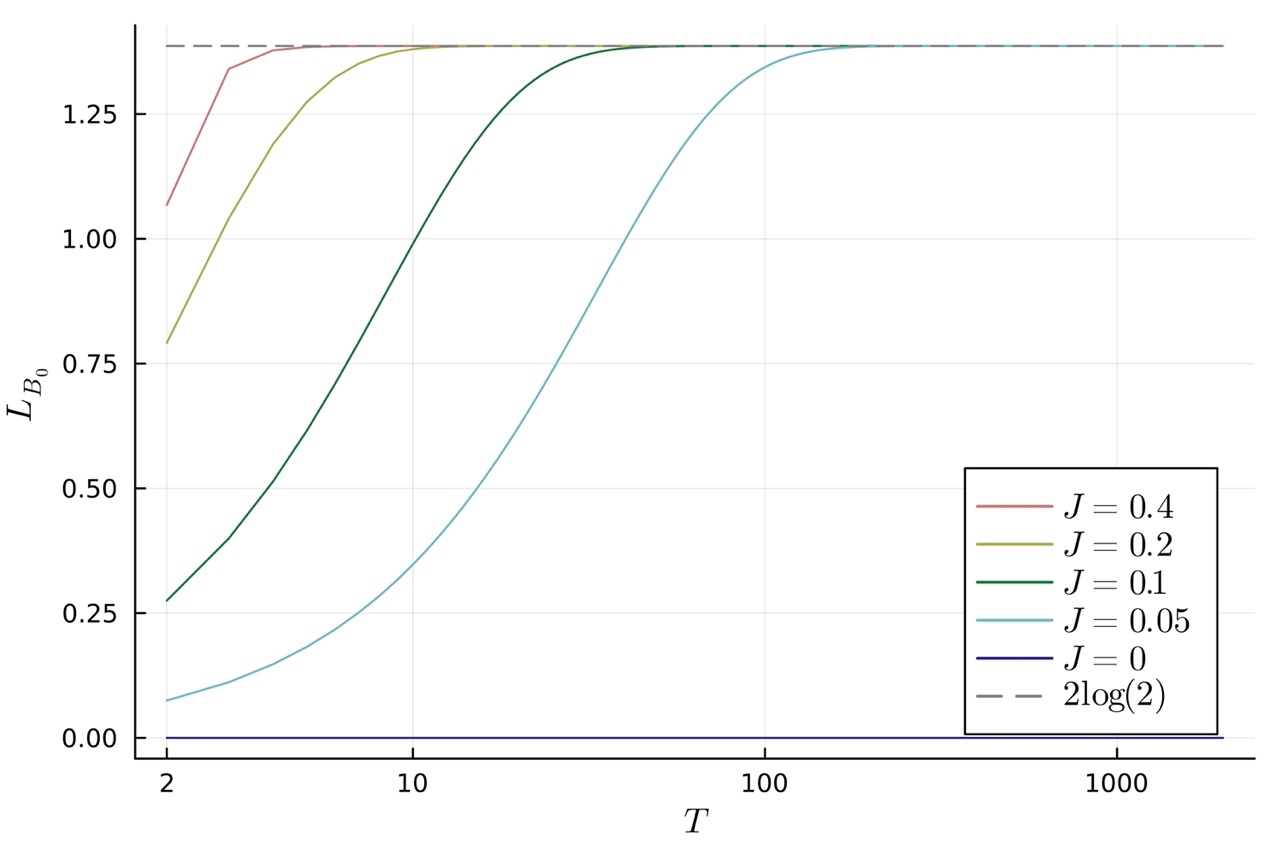}
      
    \caption{Equilibration  of the long-range SRE for partition $B_0$ at $J=J_o=J_e=0.4,0.2,0.1,0.05$ for $T=2,...,1000$, from Result \ref{res:Lresult}.}
    \label{fig:Lequil}
\end{figure}

\begin{figure}
    \centering
    \includegraphics[width=0.9\linewidth]{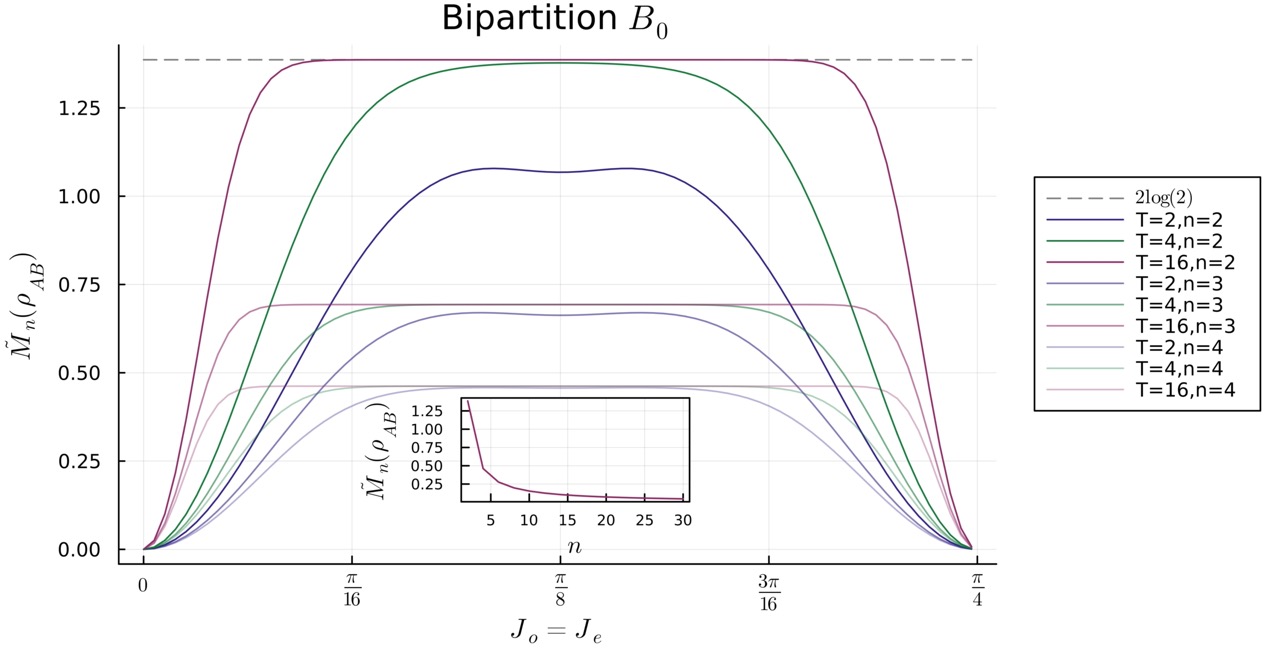}
    \caption{
    Magic in the joint state $\rho_{AB}$ for different $n=2,3,4$ given by the analytic expression in Result \ref{res:Lresult}. The state is a product of dual-unitary XXZ circuit applied to $|\phi^+\rangle^{\otimes N/2}$ with periodic boundary conditions. We take partition $B_0$ and number of layers $T=2,4,16$. The case $n=2$ corresponds to $L$ in Figure \ref{fig:longsreB0}. The inset shows the dependence of SRE on $n$ for $J_o=J_e=\frac{\pi}{8}$ and $T=16$.} 
    \label{fig:MnABB0}
\end{figure}

We plotted this result and checked its correctness with numerical methods in the accessible regimes in Figures \ref{fig:longsreB0}, \ref{fig:Lequil} and \ref{fig:MnABB0}. 
In Figure \ref{fig:longsreB0} we show long-range SRE versus parameters of the gate $J=J_o=J_e$ ($J=\alpha/2=\beta/2$).
Both the analytical expression from  Result \ref{res:Lresult} and the exact efficient numerical contraction of Result \ref{result52} agree with the results obtained via the sampling algorithm for the accessible times and sizes ($T=2,3$, i.e. $N = 12,20$). 
At long times, the long-range SRE for partition $B_0$ saturates to a constant $2\log(2)$ for all $J=J_o=J_e$ except $J=0$ (corresponding to the $XX$ model) and $J=\frac{\pi}{4}$ (corresponding to $\text{SWAP}$ gates), which are both Clifford circuits and thus generate no magic. 
From the analytical expression in Result \ref{res:Lresult} we see that for $J\neq 0,\frac{\pi}{4}$, the long-range SRE approaches equilibrium in the form $-\log(2^2(1+a\cdot b^T +c\cdot b^{2T-4}))$, where $a,b,c >0$ and $b<1$, as can be seen in Figure~\ref{fig:Lequil}. Note also that small nonzero values of $J$ take longer to equilibrate. Recall that in partition $B_0$, as $T$ grows so does the system size $N$, so this plot does not show the equilibration of $L$ for a particular state after long time.

In Figure \ref{fig:MnABB0} we look at the SRE of $\rho_{AB}$:
\begin{align}
    &\tilde{M}_n(\rho_{AB}) = \frac{1}{1-n}\log\bigg(  \frac{\zeta_{n}(\rho_{AB})}{\zeta_{1}(\rho_{AB})}  
    \bigg), \label{eq:MnABB0}
\end{align}
which is defined for all Rényi parameter $n$.
Plotting the analytical expression in Result \ref{res:Lresult}, we show equilibration for Rényi parameters $n=2,3,4$. Moreover, we show the decrease of the equilibration value for higher $n$ in the inset.

%

\begin{figure}
    \centering
    \includegraphics[width=0.8\linewidth]{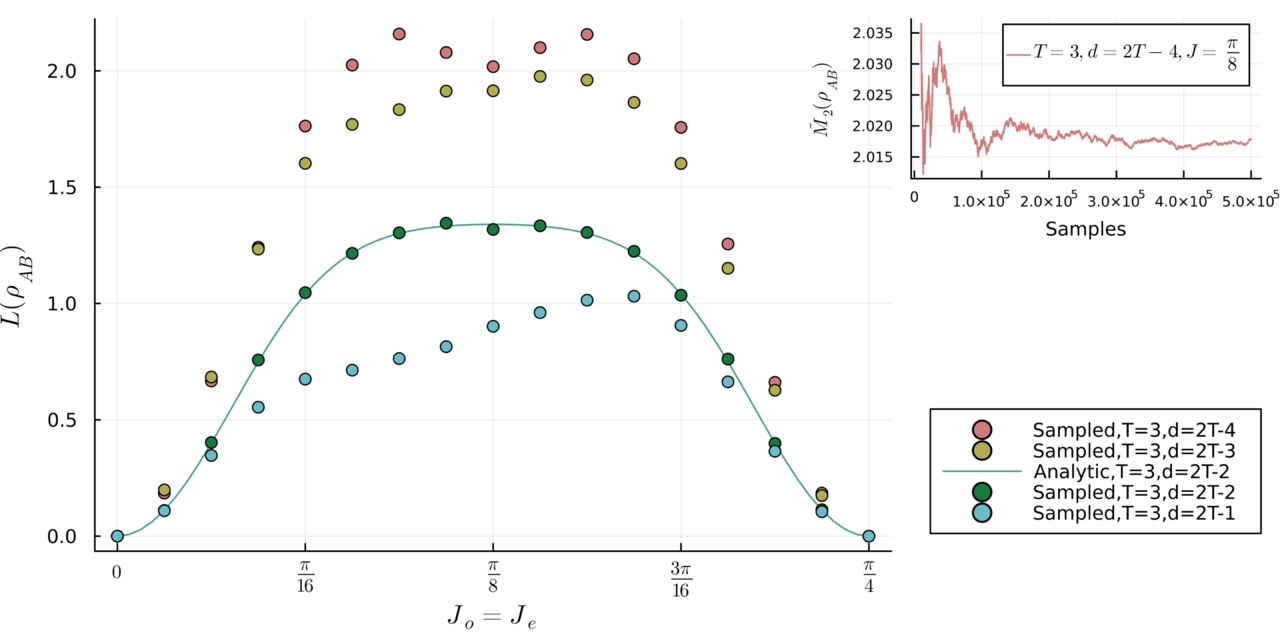}
    \caption{
    Long range magic between regions $A,B$ of the dual-unitary XXZ time evolution of the state $|\phi^+\rangle^{\otimes N/2}$ with periodic boundary conditions for $T=3$ and partitions with $N_A=N_B=2T$ and separations $d=2T-1,2T-2,2T-3,2T-4$. We provide sampling results for $20000$ samples and analytical results for $d=2(T-1)$ as well. In the inset we show convergence results for $T=3,d=2T-4,J=\frac{\pi}{8}$ up to $5\cdot 10^5$ samples.}
    \label{fig:LT3D1234}
\end{figure}

Let us now discuss what happens for other partitions, more concretely for smaller separations $d$ between regions. As $d$ gets smaller, the overlap between light cones becomes bigger and $L$ can attain larger values. The diagrams can still be simplified, but the end result is exponentially costly in $2T-d$ to numerically evaluate. In Figure \ref{fig:LT3D1234} we keep $N_A=N_B=2T$ for $T=3$ and decrease the distance between regions $d=4,3,2,1$, seeing an increase in $L$.


\begin{figure}
    \centering
    \includegraphics[width=\linewidth]{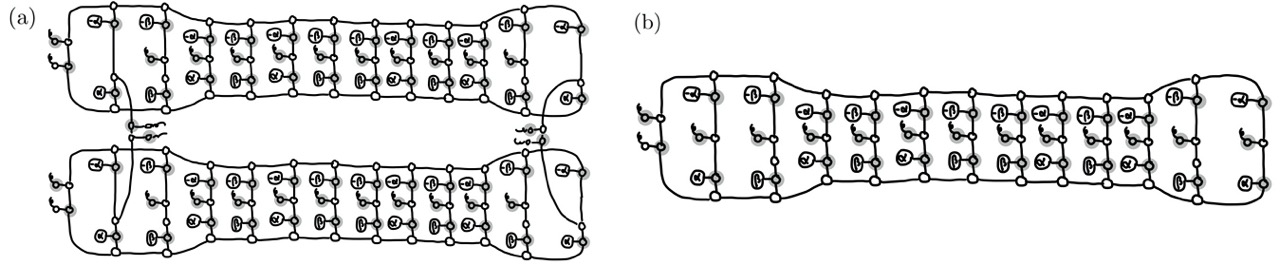}
    \caption{Tensor contractions required to compute Eq.~\eqref{eq:mixedLambda} for partition $B_1$ and $T=6$, for the a) joint state of regions $A,B$, $\rho_{AB}$, and b) reduced state of region A, $\rho_{A}$. 
    The simplifications follow similar steps as the one for the partition $B_0$ presented in the main text. The final expressions can be numerically efficiently contracted for any $T$ and small $n$.
    }
    \label{fig:B1channel}
\end{figure}

\begin{figure}
    \centering
    \includegraphics[width=0.9\linewidth]{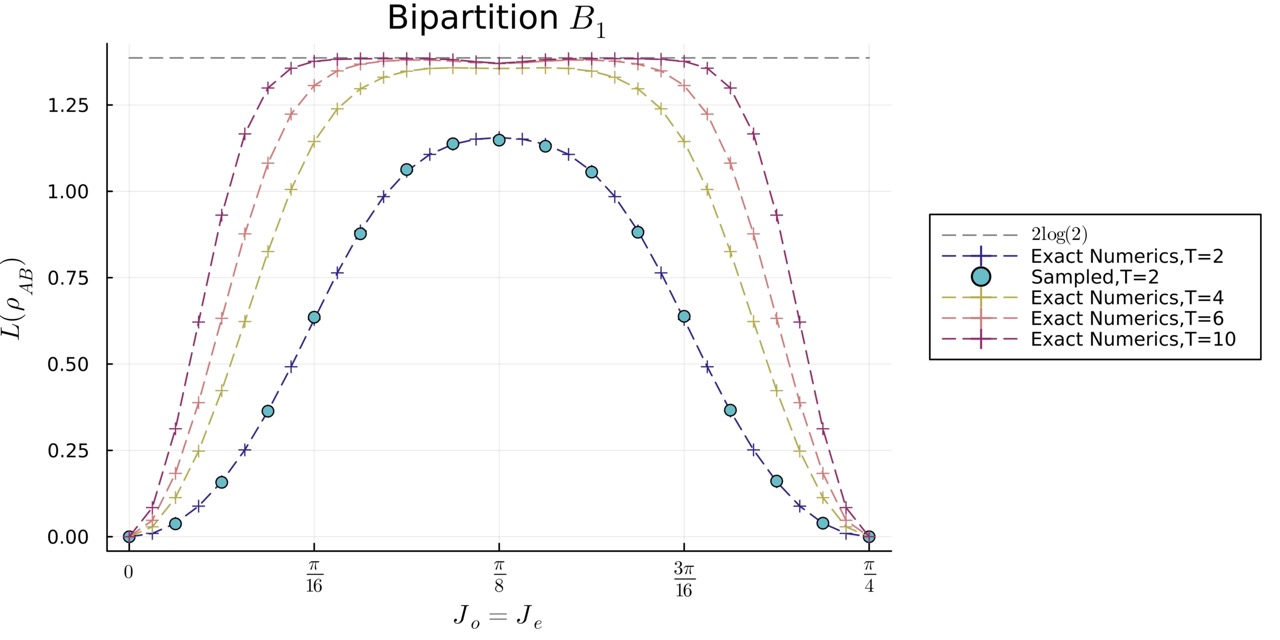}
    \caption{
    Long range magic between regions $A,B$ of the dual-unitary XXZ time evolution of the state $|\phi^+\rangle^{\otimes N/2}$ with periodic boundary conditions for partition $B_1$ and times $T=2,4,6,10$. The numerical tensor contraction (cross) agrees with the sampling result (dot) with $10^5$ samples for $T=2$. We consider the number of qubits of regions $A,B$ to be $N_A=N_B = 2(T+1)$ and the distance between regions $d=2(T-1)$.}
    \label{fig:LSRE1}
\end{figure}

Another partition that is accessible to exact numerics is
\begin{align}
    B_1:  \quad N_A=N_B=2(T+1),\quad d=2(T-1),\quad N=8T.
    \label{eq:B1}
\end{align}
The distance between regions $A,B$ is the same as in partition $B_0$, so the backwards lightcones of $N_A,N_B$ still only intersect at two Bell pairs, but unlike $B_0$ in this case $\rho_{A}$ is nontrivial. The contraction of Eq.~\eqref{eq:mixedLambda} can again be simplified to efficiently contractible 1D tensor network, see Figure \ref{fig:B1channel}. We plot the results in Fig.~\ref{fig:LSRE1}, checking these computations with the sampling algorithm for the accessible time, $T=2$.


\clearpage
\section{Conclusions and outlook}
\label{sec:conclusions}

In this work we provided exact solutions for the generation of (long-range) magic by many-body dynamics. Moverover, we were able to compute the moments of the Pauli spectrum $\zeta_n(\rho_{AB}(\alpha,\beta,T))$ for all $\alpha,\beta,T$ and  all Rényi parameters $n$. In principle, they allow us to deduce the Pauli spectrum \cite{turkeshi2023pauli}, Rényi-$2$ mutual information~\cite{tarabunga_many-body_2023} and their dynamics in time for partition $B_0$.
This was possible due to three reasons: firstly, by restricting to a dual-unitary and integrable Floquet XXZ model; secondly, focusing on a specific partition; and thirdly by adapting the techniques of ZX calculus to the novel setting presented here.


Nonetheless, we expect that our results can be straightforwardly generalized, to more general gates, partitions and observables. In particular, the derivations are expected to apply also for any gate which is a combination of a diagonal gate in the computational basis followed by a SWAP gate, see Appendix \ref{app:duxxz}.
In addition to SRE, we can compute correlations between any local observables with supports in the before-mentioned partitions.

Moreover, the graphical rules for $\Lambda^{(n)}$ open new avenues for the study of magic using the ZX-calculus. So far, ZX-calculus rules have been used to optimize the number of $T$-gates in a quantum circuit \cite{Kissinger_2020,debeaudrap2020fast,sutcliffe2024procedurally}. 
Apart from SRE, there are additional measures of magic that can be computed using $\Lambda^{(n)}$: the stabilizer linear entropy of a state, which is a strong magic monotone \cite{leone2024stabilizer}, and the nonstabilizing power of a unitary, which provides a lower bound on the number of $T$-gates \cite{SRE_2022}. We hope that the graphical rules for $\Lambda^{(n)}$ will simplify the contractions required to compute them.


A plethora of open questions remains.
So far, we did not succeed in computing long range SRE using only the dual-unitary property, which would provide exact solutions for all dual-unitary models, including chaotic examples. This raises an important question on whether the generation of magic is similar in chaotic models. To tackle this question, it would be worthwhile to try obtaining exact solutions in other models, such as
random unitary circuits~\cite{Fisher_2023}, $k$-doped random circuits~\cite{Leone_2021}, or  generalizations of dual unitarity~\cite{Yu_2024,Kos_2023}. 
Maybe an even more interesting question is the status of magic in localized models. How much magic is generated? What happens with long-range magic? One could try numerically studying many-body localized systems~\cite{abanin2019colloquium}. 
For analytical progress other models are more suitable, such as strongly localized models \cite{kos2021, bertini2022exact} and Floquet quantum east model~\cite{bertini2024localized}. 


\section{Acknowledgements}
\label{sec:acknowledgements}
We thank Tobias Haug, Lorenzo Piroli, Salvatore F.E. Oliviero and Ignacio Cirac for useful discussions. PK acknowledges financial support from the Alexander von
Humboldt Foundation. We acknowledge support from the German Federal Ministry of Education and Research through project THEQUCO within
the funding program quantum technologies - from basic research to market. This research is part of the Munich Quantum Valley (MQV), which is supported by the Bavarian state
government with funds from the Hightech Agenda Bayern
Plus.



\printbibliography

\begin{appendices}
  
\section{Sampling algorithm for mixed states}
\label{app:samplingmixed}
We implement the SRE sampling algorithm for mixed states from ~\cite{lami_quantum_2023}, which is a slight modification of their 
Algorithm 1 for pure states.

Given a mixed state $\rho$ we want to estimate the quantity $q_n=\sum_{P\in \mathcal{P}_N}\Xi_P(\rho)^n$
, where for mixed states the probability distribution $\Xi_P(\rho) = \frac{1}{\tr(\rho^2)}\frac{\tr(\rho P)^2}{2^N}$ is normalized by the purity, and use the following estimators $\tilde{q}_n$, as detailed in~\cite{lami_quantum_2023}:
\begin{align}
    &q_n = \begin{cases}
        \sum_{P\in\mathcal{P}_N}\Xi_P(\rho)\log\Xi_P(\rho), & n=1\\
        \sum_{P\in \mathcal{P}_N}\Xi_P(\rho)^n, & n > 1
    \end{cases}, &\tilde{q}_n = \begin{cases}
        \frac{1}{\mathcal{N}}\sum_{\mu=1}^\mathcal{N}\log\Xi_{P_\mu}(\rho),  & n=1\\
        \frac{1}{\mathcal{N}}\sum_{\mu=1}^\mathcal{N}\Xi_{P_\mu}(\rho)^{n-1}, & n > 1
    \end{cases}.
\end{align}

\begin{figure}[b]
    \centering
    \includegraphics[width=0.7\linewidth]{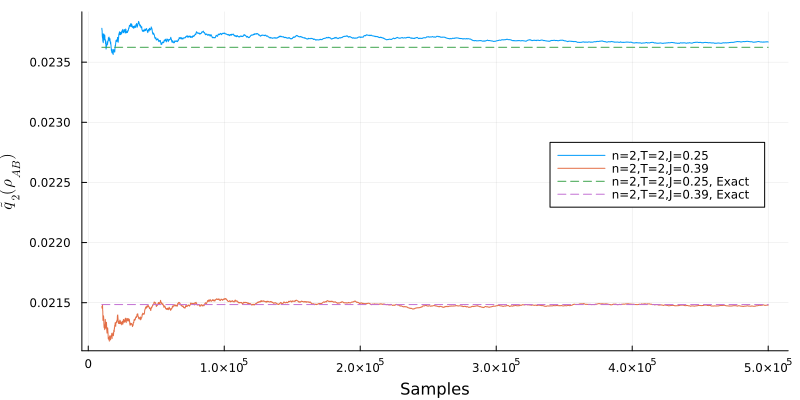}
    \caption{Convergence of the $\tilde{q}_2$ estimator for the partition $B_0$ at $T=2$ and $J=0.29,0.39$, for $500000$ samples. The exact result is obtained from Result \ref{res:Lresult}.}
    \label{fig:q2Conv}
\end{figure}

The convergence of $\tilde{q}_n$ to $q_n$ in a polynomial number of samples $\mathcal{N}=O(\frac{1}{\epsilon^2}\chi^3 N)$ is only guaranteed for $n=1$, which would correspond to computing the von Neumann stabilizer entropy; analogous convergence results for higher $n$ are not known \cite{haug_stabilizer_2023}. Nevertheless in practice we see convergence for any $n$. For example, in Figure \ref{fig:q2Conv} we can see that for $5\cdot 10^5$ samples the estimator $\tilde{q}_2$ converges to within 4 decimal digits of the exact value obtained from Result \ref{res:Lresult}:
\begin{align}
    &q_n=\frac{1}{\tr(\rho^2)^n2^{N(n-1)}}\zeta_n.
\end{align}

\section{ZX-calculus for computing SRE}\label{app:ZX}
In this section we introduce the representation of the tensor $\Lambda^{(n)}$, used to compute SRE in \cite{quantifyingmps}, in terms of the ZX-calculus. We show that it has a simple representation and prove several lemmas about the graphical rules it satisfies, following the thesis \cite{Montana2023}.

Let us start by proving the ZX-calculus diagrams for the $\Lambda^{(n)}_x,\Lambda^{(n)}_z$ tensors from the main text:
\begin{align}
    \includegraphics[width=0.8\linewidth]{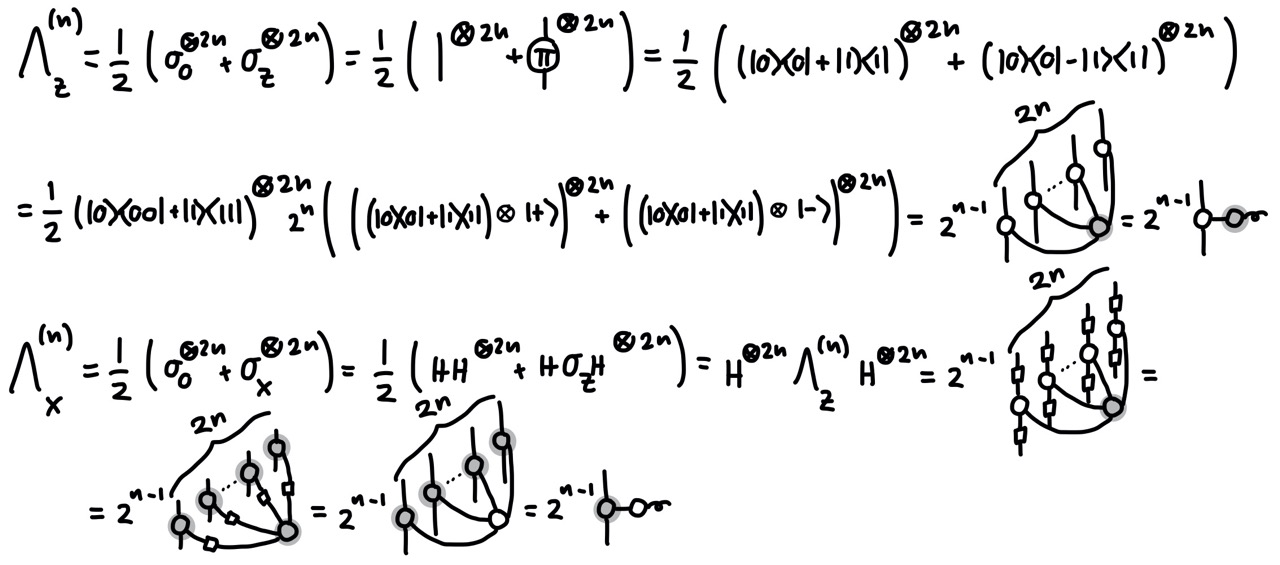}.\label{fig:LxLz_proof}
\end{align}

Recall that we want to evaluate Eq.~\eqref{eq:mixedLambda}, which involves contractions of the form $\tr((\rho\otimes \rho^*)^{\otimes n}(\Lambda^{(n)})^{\otimes N})$. Writing the $2n$ Rényi copies on each diagram would be cumbersome, so we use the notation \includegraphics[width=0.1\linewidth]{figures_compressed/squiggle.jpeg} for conciseness, which we introduced in the main text in Sec \ref{sec:results}. 
In this notation, $\Lambda^{(n)}$ can be written as:
\begin{align}
    \includegraphics[width=0.4\linewidth]{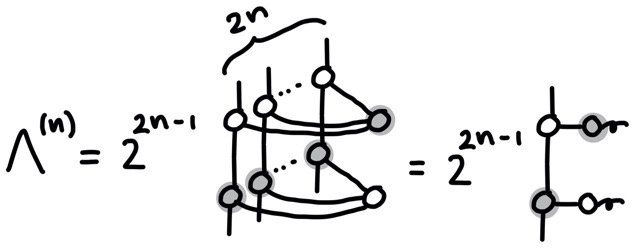}.\label{fig:L_ZX}
\end{align}

For example, if we wanted to compute the $n$-SRE of the state $|\psi\rangle = R_z(\alpha)\otimes \1 |\phi^+\rangle$, the rightmost expression below is clearer. The scalar factors account for $\Lambda^{(n)}$ and the Bell pairs:
\begin{align}
    &(\langle \psi|\langle \psi^*|)^{\otimes n}(\Lambda^{(n)})^{\otimes N}(|\psi\rangle|\psi^*\rangle)^{\otimes n} = \raisebox{-2.5\baselineskip}{%
        \includegraphics[
        width=0.5\linewidth,
        keepaspectratio,
        ]{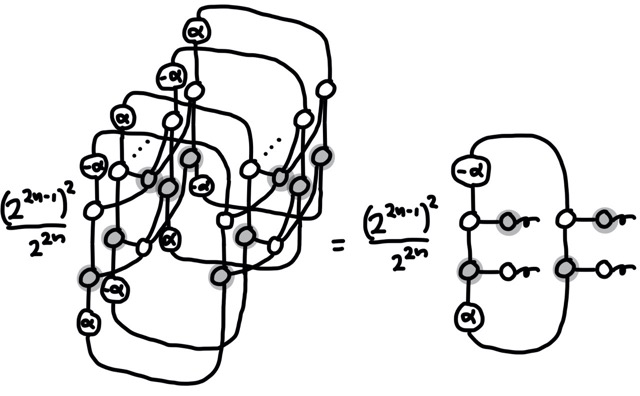}%
    }%
\end{align}

In the following lemmas we prove graphical transformation rules for $\Lambda^{(n)},\Lambda^{(n)}_x,\Lambda^{(n)}_z$. Any time we prove an identity for $\Lambda_z^{(n)}$, we can obtain an analogous result for $\Lambda_x^{(n)}$ by conjugating it with Hadamards.

The first identity, which we will use extensively, states that with the correct scalar $\Lambda_z^{(n)}$ is a projector. 
\begin{lem}
    \raisebox{-1\baselineskip}{%
        \includegraphics[
        width=0.37\linewidth,
        keepaspectratio,
        ]{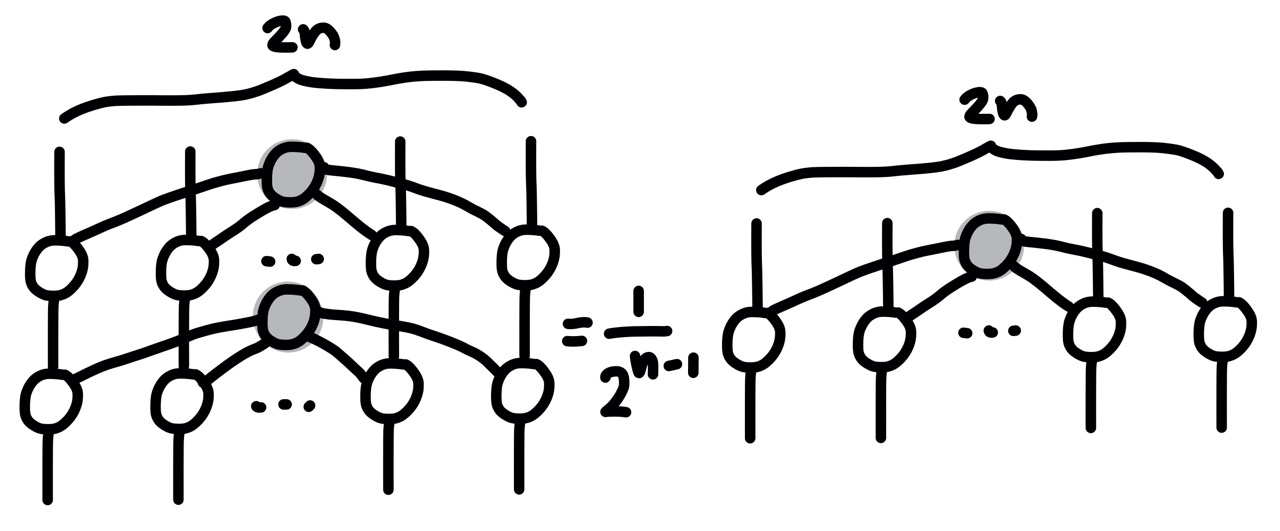}%
    }%
    .\label{lem:LxSquared}
\end{lem}
\begin{proof} \hspace{1mm}\\
\raisebox{-2\baselineskip}{%
    \includegraphics[
    width=0.8\linewidth,
    keepaspectratio,
    ]{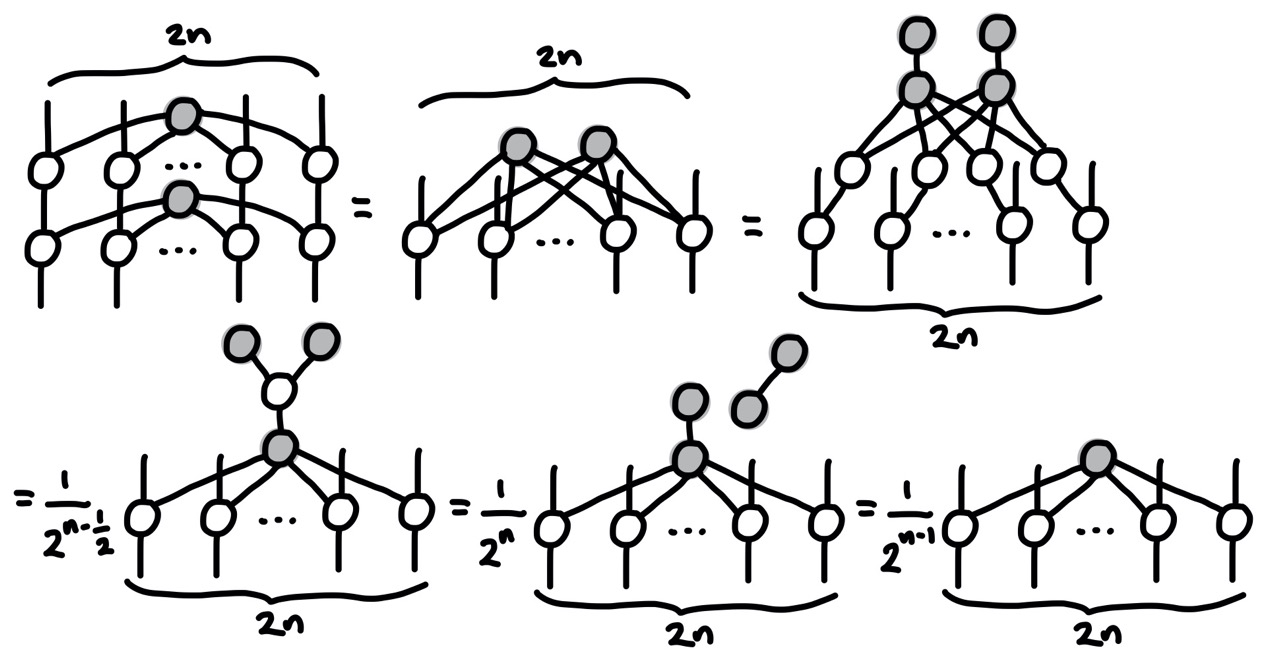}%
    
}.\\
In the first step we fuse the $Z$-spiders. In the second step we unfuse $Z$- and $X$-spiders so that we can apply the generalized bialgebra rule in Eq.~\eqref{eq:hopf_bi} in the next step. We then use the copy rule in the fourth step. Collecting the scalars and using the fusion rule yields the desired result. 
\end{proof}

\begin{lem}
    \raisebox{-1.5\baselineskip}{%
        \includegraphics[
        width=0.4\linewidth,
        keepaspectratio,
        ]{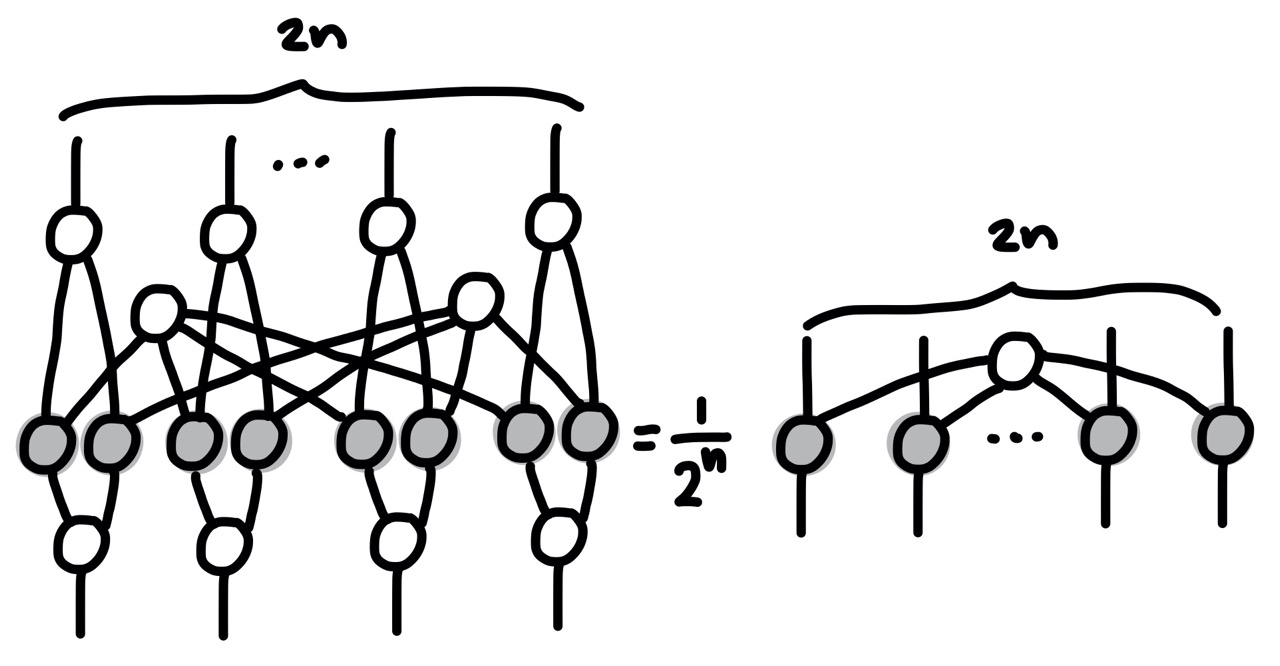}%
    }%
    .\label{lem:Lacross}
\end{lem}
\begin{proof} \hspace{2mm}\\
    \raisebox{-2\baselineskip}{%
        \includegraphics[
        width=0.95\linewidth,
        keepaspectratio,
        ]{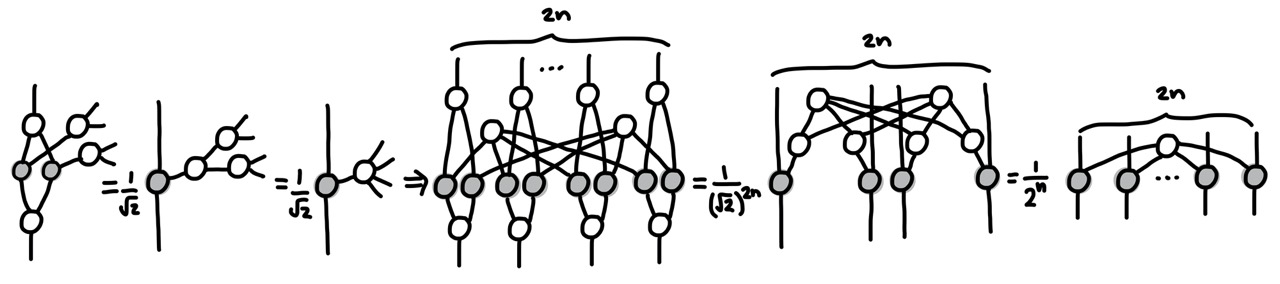}%
    }.\\
In the first step we use the bialgebra rule to remove the ``loop", then we fuse the $Z$-spiders. Applying this identity to each Rényi copy gives us the second to last equation. Fusing all the $Z$-spiders into one gives us the final result.
\end{proof}

\begin{lem}
    \raisebox{-3\baselineskip}{%
        \includegraphics[
        width=0.3\linewidth,
        keepaspectratio,
        ]{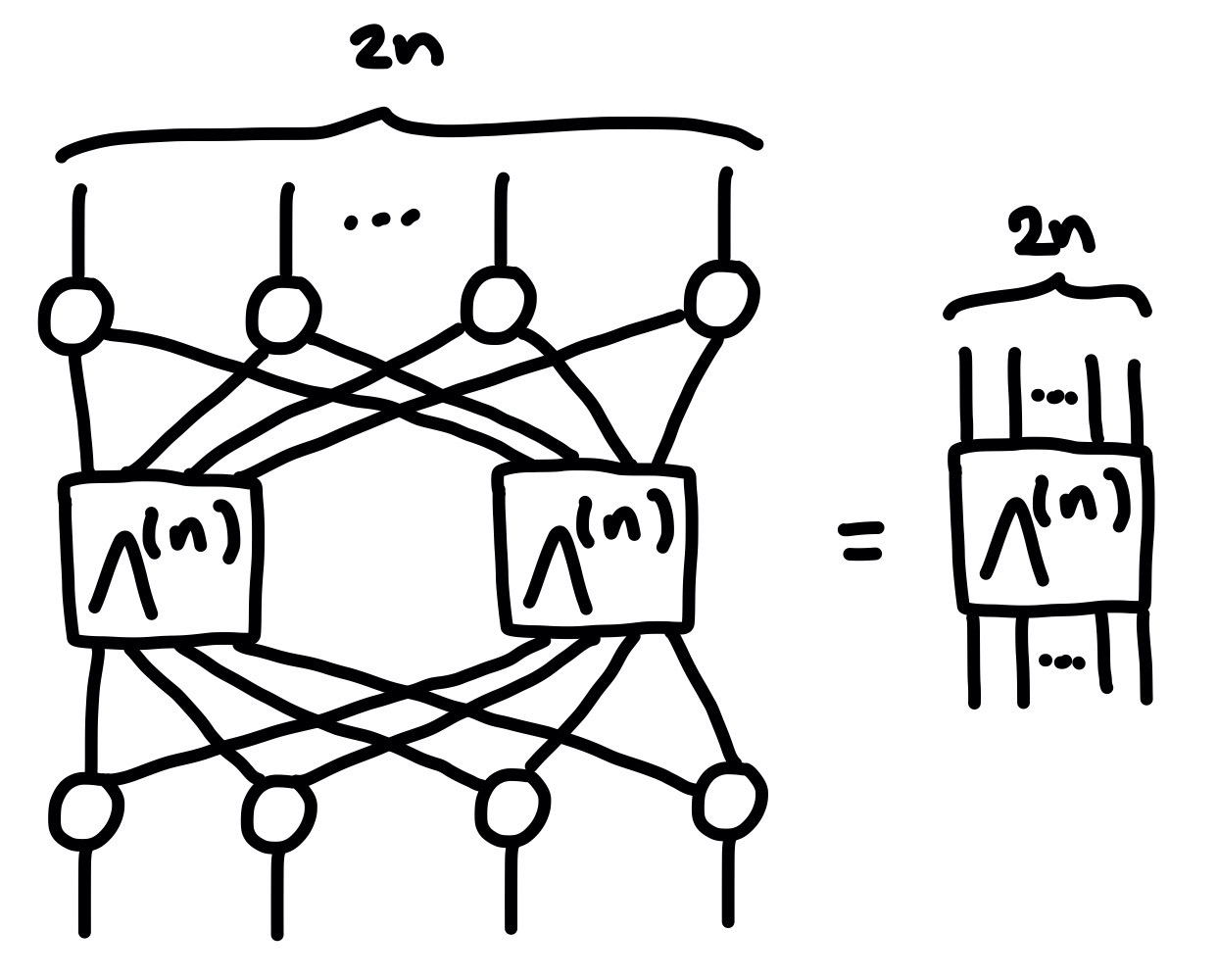}%
    }%
    .\label{lem:Lambda2}
\end{lem}
\begin{proof}\hspace{1mm}\\
\raisebox{-2\baselineskip}{%
        \includegraphics[
        width=0.99\linewidth,
        keepaspectratio,
        ]{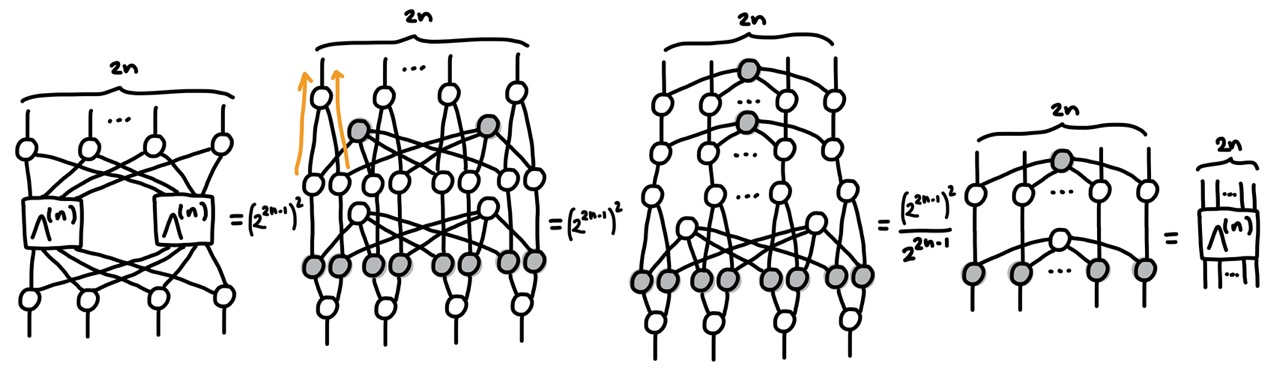}%
    }.\\
The first step is to rewrite the identity in ZX-calculus language with the correct scalars. We then slide the $\Lambda_{z}^{(n)}$ up through the $Z$-spider at the top, to remove it from the ``loop". The top part of the circuit can be simplified with Lemma \ref{lem:LxSquared} and the bottom part with Lemma \ref{lem:Lacross}, arriving at the final result.    
\end{proof}

\begin{lem}
    \raisebox{-2.5\baselineskip}{%
        \includegraphics[
        width=0.4\linewidth,
        keepaspectratio,
        ]{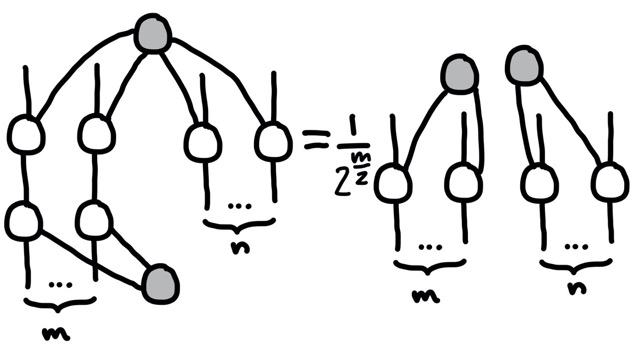}%
    }%
    .\label{lem:gadget_split}
\end{lem}
\begin{proof}\hspace{1mm}\\
    \raisebox{-2\baselineskip}{%
        \includegraphics[
        width=0.95\linewidth,
        keepaspectratio,
        ]{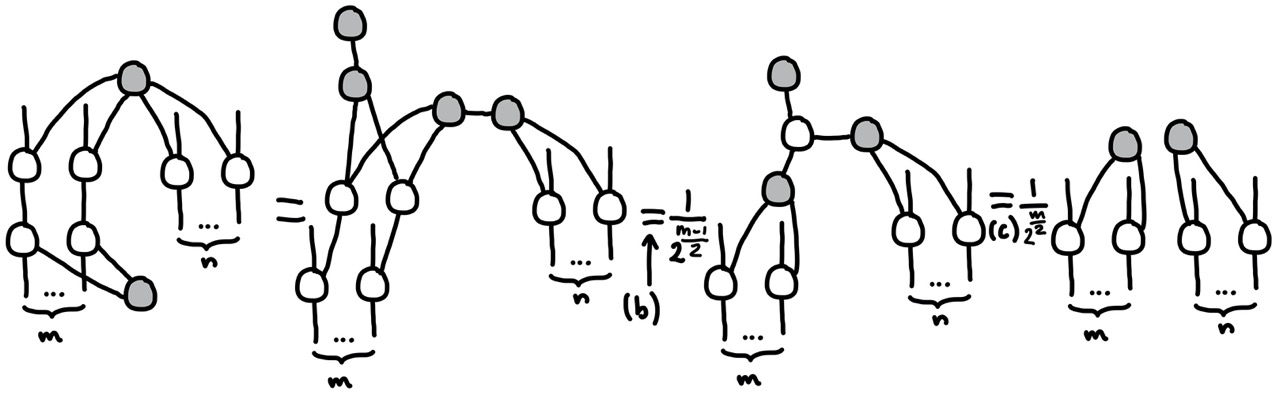}%
    }.\\
The first step is to unfuse the necessary spiders to shape the diagram in a way so that we can use the bialgebra rule in the next step. After that we apply the copy rule to split the diagrams and fuse the remaining grey spiders.
\end{proof}

\begin{lem}
    \raisebox{-2\baselineskip}{%
        \includegraphics[
        width=0.3\linewidth,
        keepaspectratio,
        ]{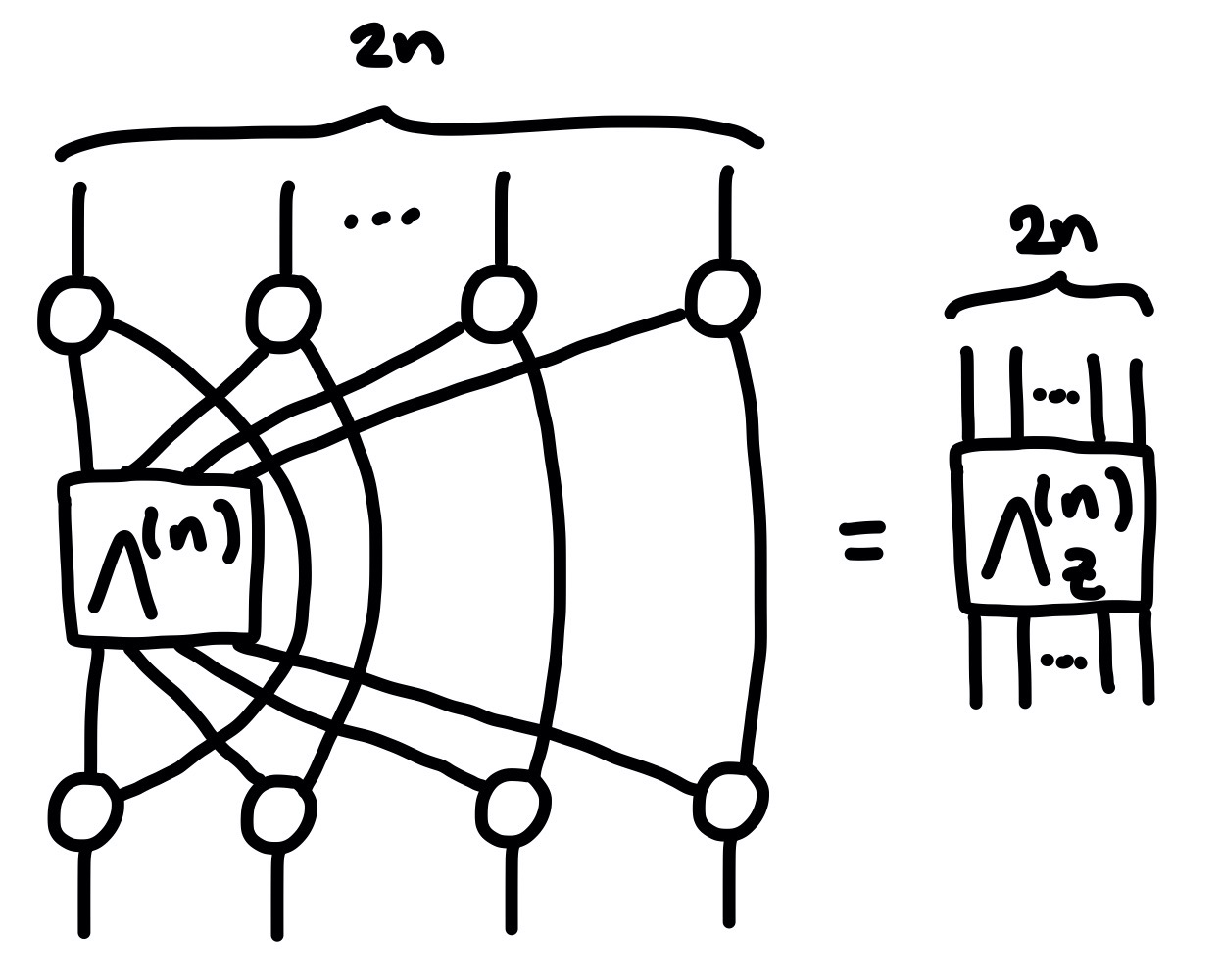}%
    }%
    .\label{lem:LzI}
\end{lem}
\begin{proof}\hspace{1mm}\\
    \raisebox{-2\baselineskip}{%
        \includegraphics[
        width=0.85\linewidth,
        keepaspectratio,
        ]{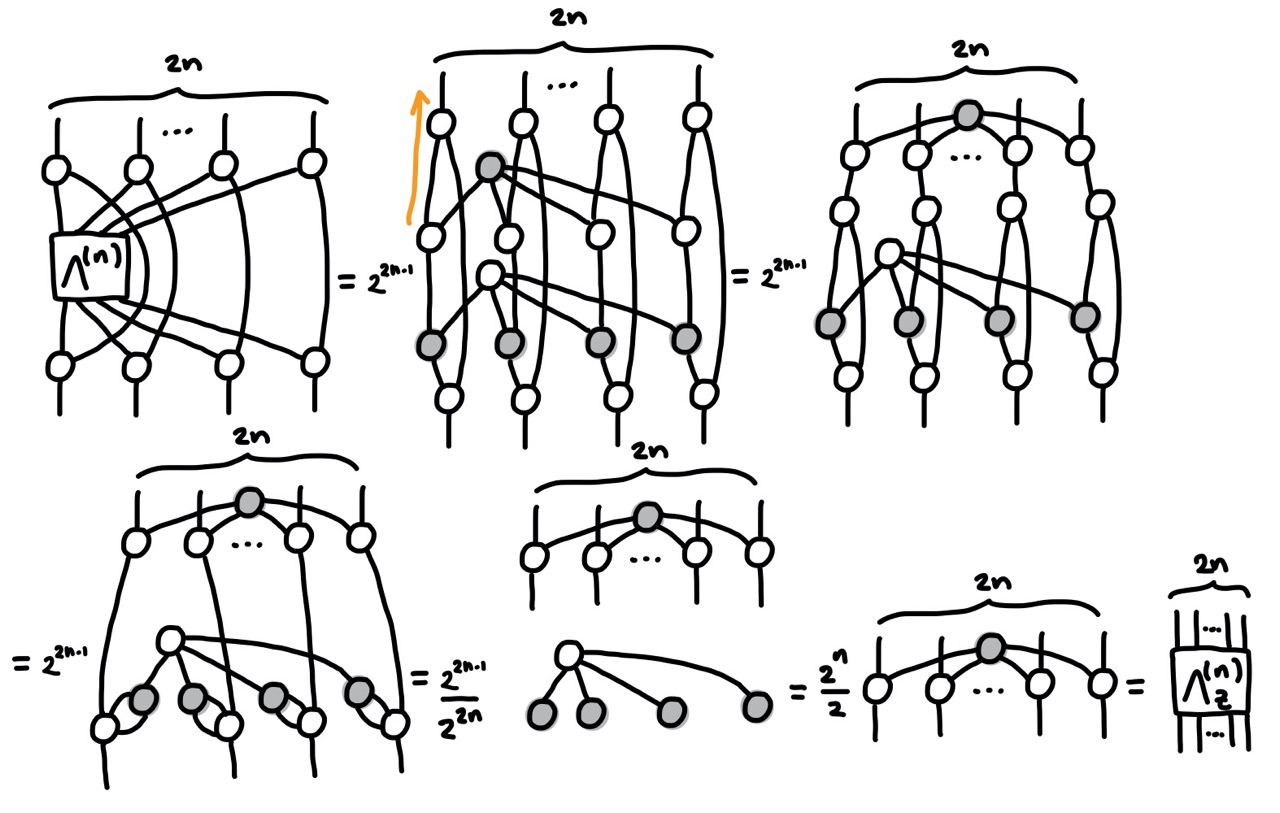}%
    }.\\
In the first step we write the $\Lambda^{(n)}$ in ZX-calculus and indicate in orange that we want to slide  $\Lambda_z^{(n)}$  up. In the next step we fuse the $Z$-spiders so it is clear that they are connected to the $X$-spiders at the bottom by two edges. We can thus use the Hopf rule to split them in the second equality of the second line. Using the copy rule we find the value of the scalar diagram on the bottom, obtaining the desired result.
\end{proof}

The following lemma says that if we have a $3$-legged $X$-spider with $\Lambda_z^{(n)}$ on two of its legs, we can slide one of them to the third leg.
\begin{lem}
    \raisebox{-2\baselineskip}{%
        \includegraphics[
        width=0.2\linewidth,
        keepaspectratio,
        ]{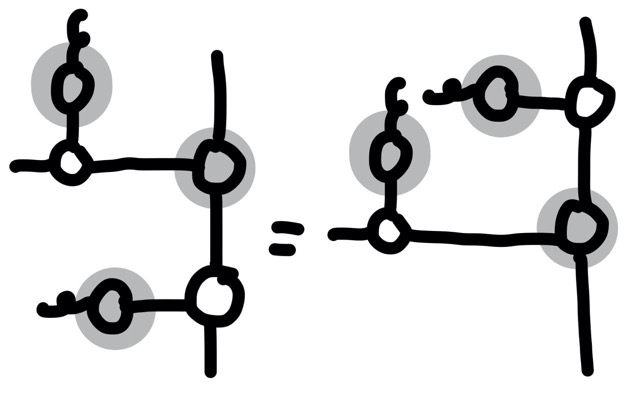}%
    }%
    .\label{lem:LzSlide}
\end{lem}
\begin{proof}\hspace{1mm}\\
    \raisebox{-2\baselineskip}{%
        \includegraphics[
        width=0.6\linewidth,
        keepaspectratio,
        ]{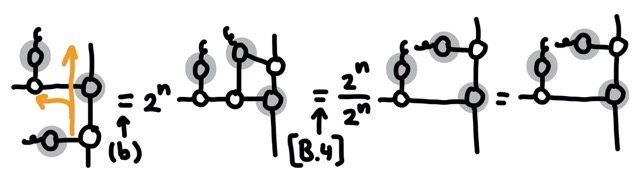}%
    }.\\
In the first equality we use the bialgebra rule, indicated with orange arrows. In the second equality we use Lemma~\ref{lem:gadget_split} to split the resulting $\Lambda^{(2n)}_z$.
\end{proof}

\section{SRE in the thermodynamic limit for short time}\label{app:halfstep}

Here we repeat the result from  Section \ref{sec:results_halfstep} and provide a complete proof of it.
Recall from Section \ref{sec:results_halfstep} that in order to find the SRE of the state time evolved with only one layer of gates, $|\psi(J,0,\frac{1}{2})\rangle$, it is sufficient to find the largest eigenvalue of the transfer matrix $T$~\cite{quantifyingmps}, from Figure \ref{fig:halfstep}. 
In Result \ref{thm:halfstep}, which we repeat here, we find it exactly.
\begin{result}
    For a half time-step of the dual-unitary XXZ evolution, the density of magic in the thermodynamic limit  is:

    $$m_{n}(|\psi(J,0,\frac{1}{2})\rangle) =\frac{1}{2(1-n)}\log\bigg(\frac{1+\cos^{2n}(2J) + \sin^{2n}(2J)}{2}\bigg).$$
\end{result}

\begin{proof}
We can obtain this result by noticing that vectorized $\Lambda^{(n)}$ also denoted by $|\Lambda^{(n)}\rangle$ is the leading eigenvector of the transfer matrix, and obtaining its eigenvalue.
\begin{align}
    \includegraphics[width=0.9\linewidth]{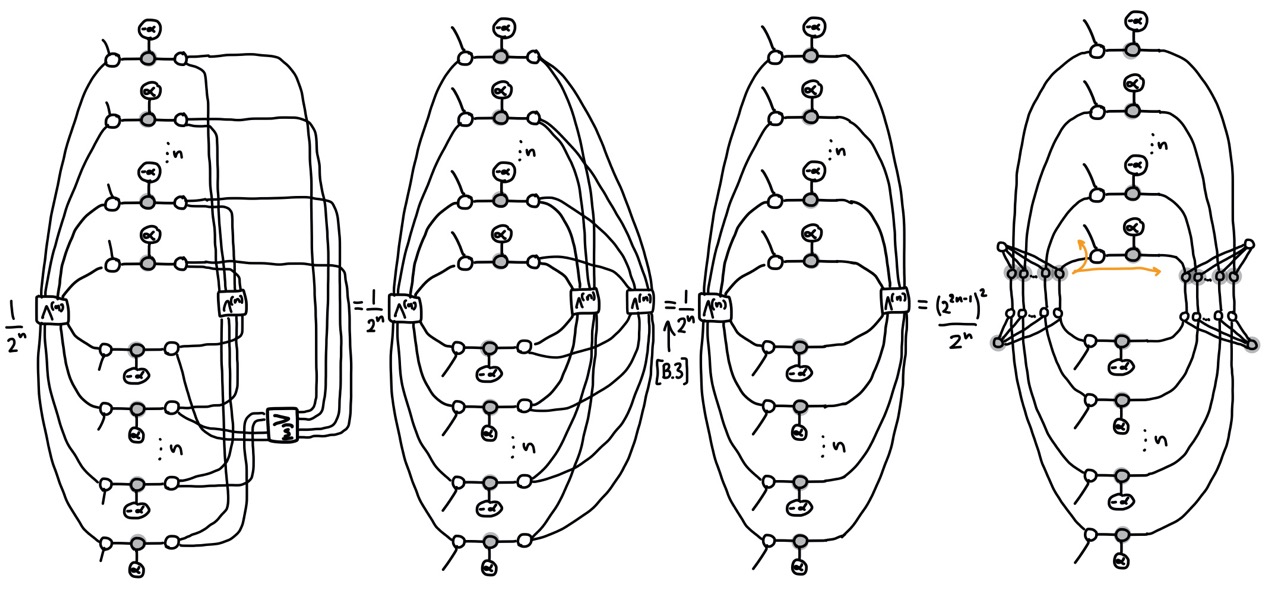},
\end{align}

The first diagram shows the transfer matrix for half a timestep being applied the vectorized $|\Lambda^{(n)}\rangle$ as an input on the right. Notice that the $\Lambda^{(n)}$ in the middle has the same ordering of the legs as the $\Lambda^{(n)}$ on the right. However, from the ZX-representation of $\Lambda^{(n)}$ we see that only connectivity matters, i.e. the corresponding top-bottom indices of $\Lambda^{(n)}$ need to match, but how we arrange them horizontally does not matter. Therefore, in the second step we rearrange the legs such that the connectivity is easier to understand. Since we have two $\Lambda^{(n)}$ applied to the same $Z$-spiders, in the third step we remove one of them using Lemma \ref{lem:Lambda2}. The last step is just writing the ZX-calculus representation of $\Lambda^{(n)}$. Moreover, we indicate in orange lines that we will try to ``pass" the left $\Lambda^{(n)}_x$ through the white $Z$-spider on top, using the bialgebra rule.
\begin{align}
    \includegraphics[width=0.9\linewidth]{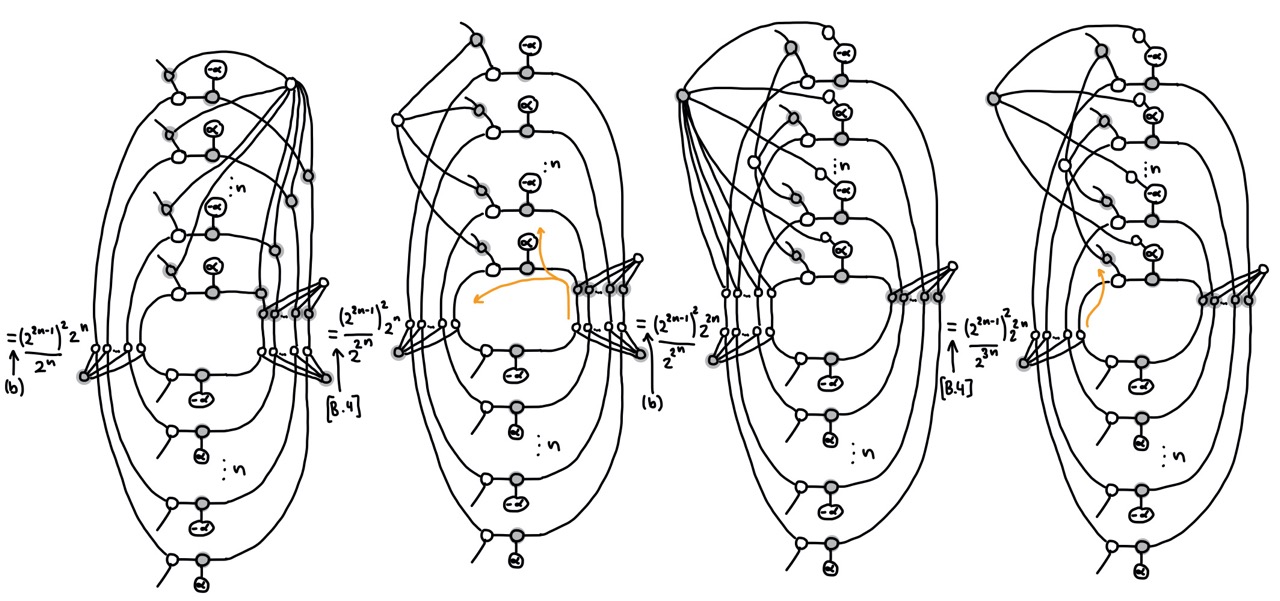},
\end{align}
In the first diagram we are left with a $\Lambda_x^{(2n)}$ that has half of its legs on the top-left output legs of the transfer tensor and half of them across the right hand side. Since there is already a $\Lambda^{(n)}$ on the right legs, in the second step we apply Lemma \ref{lem:gadget_split} to split the top $\Lambda_x^{(2n)}$, such that one $\Lambda_x^{(n)}$ remains in the top-left output legs. We will do the same thing with the $\Lambda_z^{(n)}$ on the right half: as signaled in orange, we pass it through the $X$-spider that has a phase at the end using the bialgebra rule. Again, it now has twice the original number of legs, but half of them are on the left, where there is already a $\Lambda_z^{(n)}$. Thus using Lemma \ref{lem:gadget_split} we can split it so that only the half connected to the $\alpha$ $Z$-spiders remains. In the last step, as indicated in orange we just move the left $\Lambda_z^{(n)}$ up through the $Z$-spider into the top-left output leg.

\begin{align}
    \includegraphics[width=0.99\linewidth]{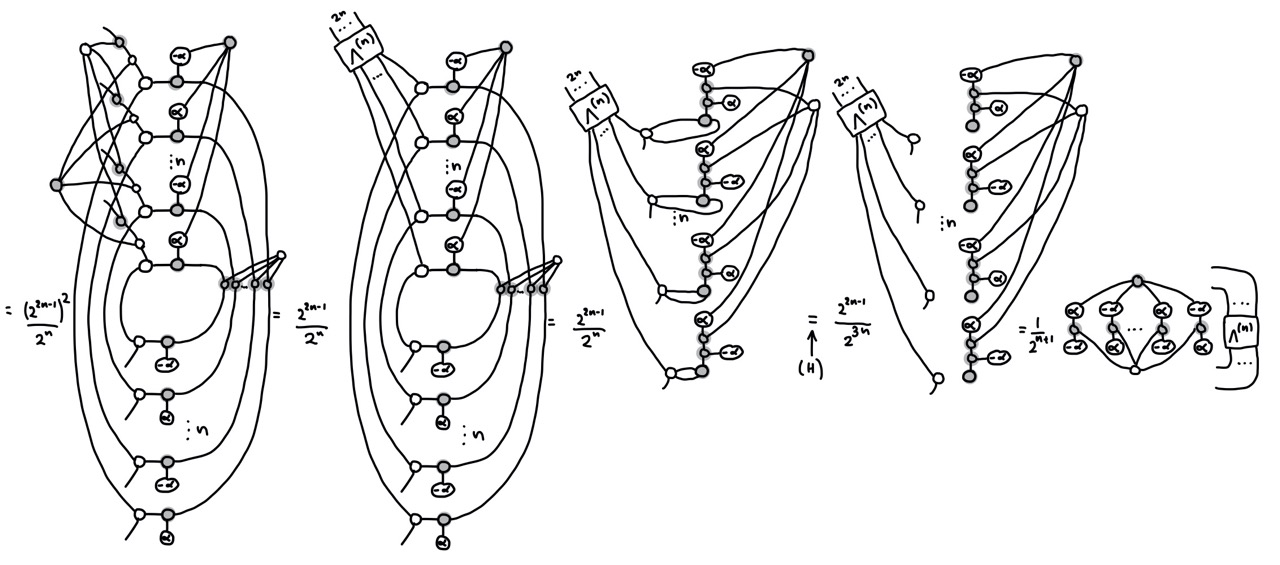},
\end{align}
First note that we have $\Lambda^{(n)}$ on the top-left output legs. It is still connected to the rest of the diagram, but if it is indeed an eigenvector, there must be some way to split it away, with the remaining diagram giving the scalar $\lambda_0^{(n)}$. Indeed, in the third step we prepare for the split by moving the $\Lambda_x^{(n)}$ on the right up through the $X$-spider on the top. We also move the bottom $X$-spider up through the top $X$-spider. This results in the left part of the circuit, which contains $\Lambda^{(n)}$, to be connected to the rest of the circuit only by $Z$-spiders, and the right part of the circuit to be connected to $X$-spiders. Since the $Z$-spiders and $X$-spiders are connected by pairs of edges, we can apply the Hopf rule to split away the $\Lambda^{(n)}$ from the rest of the circuit. The remaining diagram has no inputs or outputs, so it is a scalar, and together with the scalar prefactors we had accumulated it will give us the eigenvalue $\lambda_0^{(n)}$.
\begin{align}
    \includegraphics[width=0.9\linewidth]{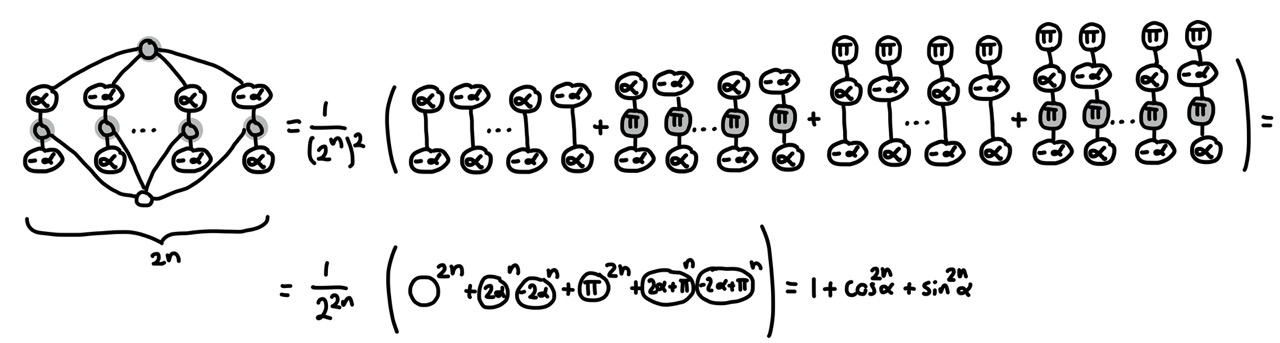},
\end{align}

The last step is to compute the scalar diagram to obtain the eigenvalue. We write $\Lambda_z^{(n)},\Lambda_x^{(n)}$ as sums of their terms and compute them individually. In each of the resulting terms we apply the phase fusion rule and the $\pi$-copy rule. Notice that since we have the same number of $\alpha$ and $-\alpha$, the global phases cancel. We end up with a sum of powers of $Z$-spiders, which give us each of the terms in the final equation, since they evaluate to $\mathcircled{+2\alpha}\mathcircled{-2\alpha} = 2^2\cos^2(\alpha),\mathcircled{0} = 2,\mathcircled{\pi} =0$.
\begin{align}
    \includegraphics[width=0.9\linewidth]{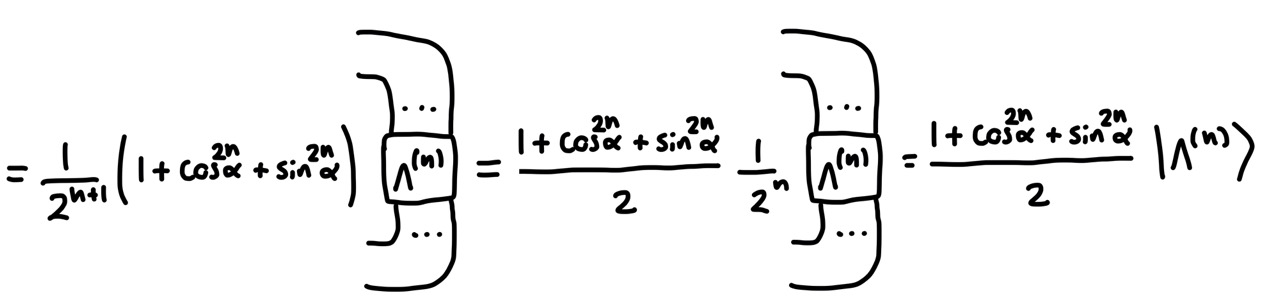},
\end{align}
Plugging the value of the scalar diagram and collecting the $
\frac{1}{2^n}$ factor with the bent $
\Lambda^{(n)}$ to get $|\Lambda^{(n)}\rangle$, finishes the proof.
\end{proof}

\newpage

\section{Long-range SRE for bipartition $B_0$}
\label{sec:app_longSRE}

In this Appendix we repeat the full exact solution of long range magic for bipartition $B_0$ from Result \ref{res:Lresult} and provide its proof. We mentioned it in Sec. \ref{sec:LSRE} and used it for Figures \ref{fig:longsreB0}, \ref{fig:Lequil}, \ref{fig:MnABB0}, \ref{fig:LT3D1234} and \ref{fig:q2Conv}. The following result is for even $T$, but a similar expression can be obtained for odd $T$. In the figures we only plot the section $J_o=J_e$, where even and odd expressions match.
\begin{result}
For partition $B_0: N_A=N_B=2T,d=2(T-1),T=2t$, the long-range SRE is: 
\begin{align}
    &L_{B_0}(J_o,J_e,T)= \tilde{M}_2(\rho_{AB}) -2\tilde{M}_2(\rho_A)= -\log\bigg(\frac{\zeta_2(\rho_{AB})}{\zeta_1(\rho_{AB})}\bigg), 
\end{align}
where the moments $\zeta_n(\rho_{AB})$ are exactly: 
\begin{align}
    &\zeta_n(\rho_{AB})=\frac{1}{2^{4T-2}}\big(1+2(f_n(J_o)f_n(J_e))^T+g_n(J_o,J_e)(f_n(J_o)f_n(J_e))^{2T-4}\big), 
\end{align}
and we have defined
\begin{alignat}{2}
    &f_n(J) := \cos^{2n}(2J)+\sin^{2n}(2J),\notag\\
    &g_n(J_o,J_e) := \frac{1}{2^{2n}}\underset{k=0}{\overset{4}{\sum}}{4\choose k}\underset{m=0}{\overset{1}{\sum}}&&\big(\cos(2J_o+2J_e)^{4-k}\sin(2J_o+2J_e)^k+\notag\\
    & &&+(-1)^{m}\cos(2J_o-2J_e)^{4-k}\sin(2J_o-2J_e)^k\big)^{2n}.
\end{alignat}
\end{result}
These satisfy $f_n(0)=1,g_n(0,0)=1$, so we can easily check that $L_{B_0}(0,0,T)=0$.

\begin{proof}
We start with Eq.~\eqref{eq:mixedLambda}:
\begin{align} &\zeta_n(\rho_{AB})=\tr((\rho_{AB}\otimes \rho_{AB}^*)^{\otimes n}(\Lambda^{(n)})^{\otimes N})=\\
    &=\frac{(\sqrt{2})^{2\frac{N}{2}T\cdot 2n}}{(\sqrt{2})^{2\frac{N}{2}\cdot 2n}}\raisebox{-4.0\baselineskip}{%
        \includegraphics[
        width=0.8\linewidth,
        keepaspectratio,
        ]{figures_compressed/LB0_proof1.jpeg}%
    }.
    \label{fig:proof_1}
\end{align}
We depicted the case of $T=6$. The denominator comes from the scalar $\frac{1}{\sqrt{2}}$ associated to each initial Bell pair, which is unnormalized in the ZX-calculus. There are $\frac{N}{2}$ Bell pairs in the initial state $|\phi^+\rangle^{\otimes N/2}$, and counting the conjugate and the $2n$ Rényi copies we end up with $2\frac{N}{2}\cdot 2n$ normalization factors. The numerator comes from the scalar $\sqrt{2}$ associated to each gate, from Eq.~\eqref{eq:Uo}. There are as many gates per layer as Bell pairs, and there are $T$ layers, so we obtain $2\frac{N}{2}T\cdot 2n$. We can simplify the overall factor to $2^{N(T-1)n}=2^{(8T-4)(T-1)n}=2^{n(8T^2-12T+4)}$. In what follows, we will keep track of the global scalar with a variable $c_i$ associated to the diagram $i$, and display diagrams that are equivalent up to a scalar. Thus, for Eq.~\eqref{fig:proof_1} the scalar is $c_\eqref{fig:proof_1} = 2^{n(8T^2-12T+4)}$.
\begin{align}
    &\raisebox{0.0\baselineskip}{%
        \includegraphics[
        width=\linewidth,
        keepaspectratio,
        ]{figures_compressed/LB0_proof2.jpeg}%
    }.
    \label{fig:proof_2}
\end{align}
Next we apply the ZX-calculus simplification for unitarity, as in Eq.~\eqref{eq:gadget_unitary}, to the qubits that are connected to their conjugate with an identity, in between regions $A,B$. Each backward light cone that we simplify contains $(T-1)\frac{T}{2}$ gates. Simplifying an unnormalized gate with its conjugate  yields a factor of $\frac{1}{2}$, as in Eq.~\eqref{eq:gadget_unitary}, and there are $2$ lightcone pairs (between $A-B$ and $B-A$) and $2n$ Rényi copies, so we get a factor of $2^{-2nT(T-1)}$. The global scalar is $c_{\eqref{fig:proof_2}} =c_{\eqref{fig:proof_1}}\cdot 2^{-2nT(T-1)}=2^{n(8T^2-12T+4)}\cdot 2^{-2nT(T-1)} = 2^{n(6T^2-10T+4)}$.

We highlighted in blue some of the loops where there is a single $\Lambda^{(n)}$. These can be simplified to $\Lambda^{(n)}_z$ using apply Lemma \ref{lem:LzI}. In orange we highlight some $\Lambda^{(n)}$ that cannot be simplified in this way, since there are two $\Lambda^{(n)}$ in the loop. The global scalar remains $c_{\eqref{fig:proof_3}}=c_{\eqref{fig:proof_2}}$.

\begin{align}
    &\raisebox{-6.0\baselineskip}{%
        \includegraphics[
        width=\linewidth,
        keepaspectratio,
        ]{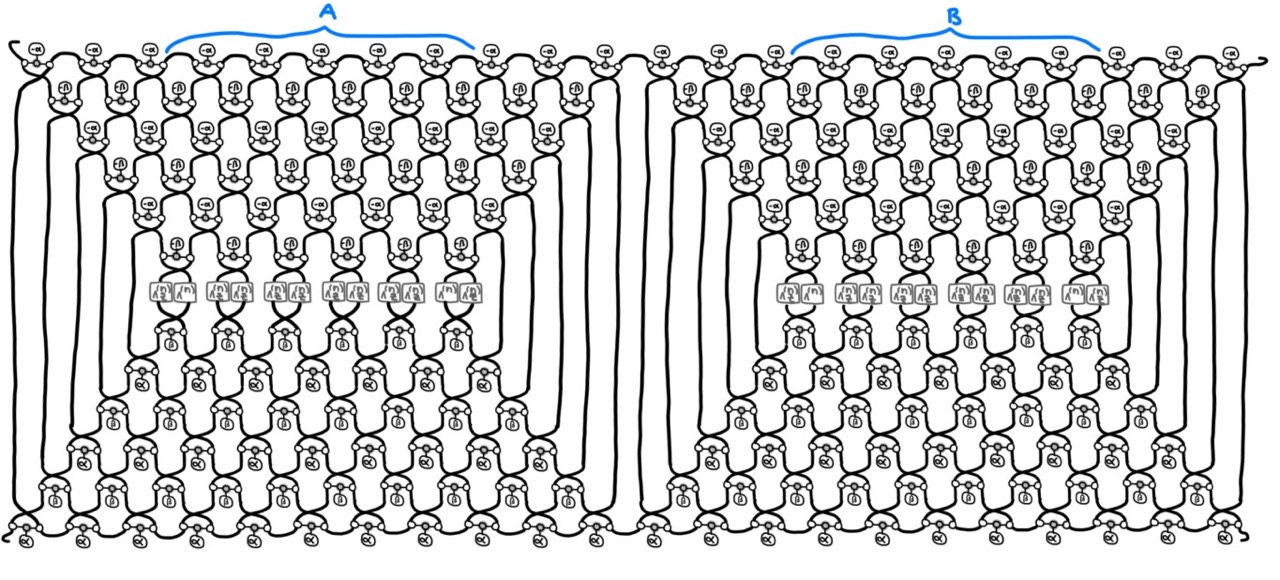}%
    }.
    \label{fig:proof_3}
\end{align}

In the next diagram we will write $\Lambda^{(n)}_z,\Lambda^{(n)}$ using the ZX-calculus. By Eqs.~\eqref{fig:LxLz_proof},~\eqref{fig:L_ZX},  each $\Lambda^{(n)}_z$ contributes $2^{n-1}$ and each $\Lambda^{(n)}$ contributes $2^{2n-1}$. Since there are $2(N_A-2)=2(2T-2)=4(T-1)$ $\Lambda^{(n)}_z$ tensors and $4$ $\Lambda^{(n)}$ tensors in diagram ~\eqref{fig:proof_3}, the global scalar is $c_{\eqref{fig:proof_4}}=c_{\eqref{fig:proof_3}}\cdot 2^{4(T-1)(n-1)+4(2n-1)}=2^{n(6T^2-10T+4)}2^{4(T-1)(n-1)+4(2n-1)} = 2^{n(6T^2-6T+8)-4T}$.

\begin{align}
    &\raisebox{-6.0\baselineskip}{%
        \includegraphics[
        width=\linewidth,
        keepaspectratio,
        ]{figures_compressed/LB0_proof4.jpeg}%
    }.
    \label{fig:proof_4}
\end{align}

We now slide the $\Lambda^{(n)}_z$ outside the regions $A,B$ as indicated by an orange arrow. The fact that the dual-unitary XXZ gates are a product of a term that is diagonal in the computational basis and a SWAP is the key to doing this: we can slide $\Lambda_z^{(n)}$ along the SWAP and commute it with the phase gadget. The global scalar is the same $c_{\eqref{fig:proof_5}}=c_{\eqref{fig:proof_4}}$.
\begin{align}
    &\raisebox{-6.0\baselineskip}{%
        \includegraphics[
        width=\linewidth,
        keepaspectratio,
        ]{figures_compressed/LB0_proof5.jpeg}%
    }.
    \label{fig:proof_5}
\end{align}

Since most of the qubits in regions $A,B$ are now connected to their conjugates with an identity, we can use Eq~\eqref{eq:gadget_unitary} to simplify a lightcone of gates in regions $A,B$. There are $(T-2)\frac{T-1}{2}$ pairs of gates that get simplified in region $A$, and as many in $B$. Thus, the new global scalar is $c_{\eqref{fig:proof_6}}=c_{\eqref{fig:proof_5}}\cdot 2^{-2n(2(T-2)\frac{T-1}{2})}=2^{n(6T^2-6T+8)-4T}2^{-2n(2(T-2)\frac{T-1}{2})}=2^{4n(T^2+1)-4T}$.\\
\begin{align}
    &\raisebox{-6.0\baselineskip}{%
        \includegraphics[
        width=\linewidth,
        keepaspectratio,
        ]{figures_compressed/LB0_proof6.jpeg}%
    }.
    \label{fig:proof_6}
\end{align}
We will use Eq.~\eqref{eq:gadget_unitary} to simplify gates along all diagonals that have only identity or gates that are diagonal in the computational basis ($Z$-spiders) connected to their conjugate, as an example see the blue and orange highlights. To make this step clearer, we provide a close-up of one of the simplifications:

\begin{align}
    &\raisebox{-6.0\baselineskip}{%
        \includegraphics[
        width=\linewidth,
        keepaspectratio,
        ]{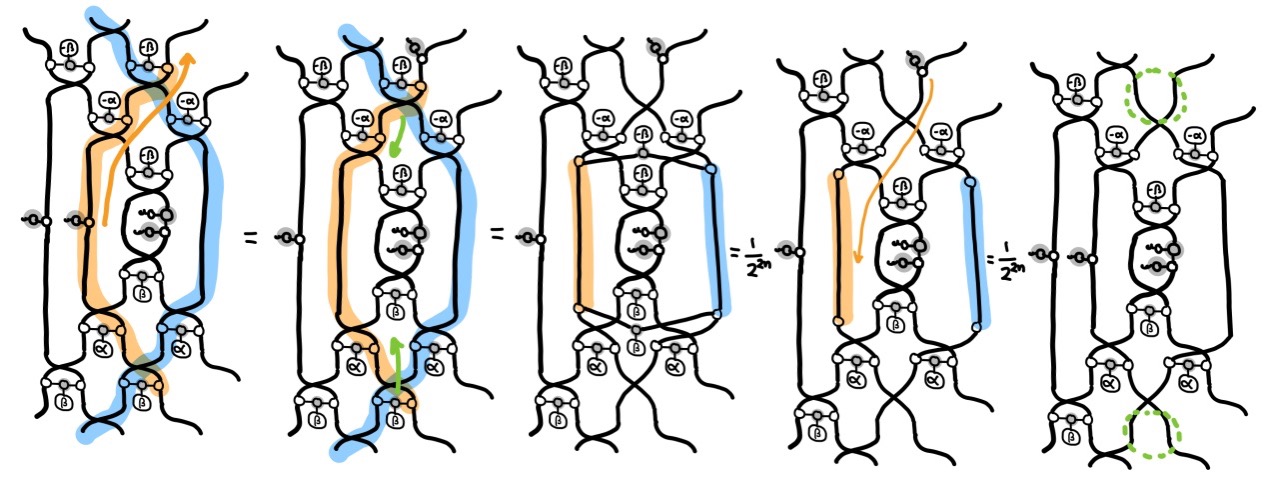}%
    }.
    \label{fig:proof_7}
\end{align}
We first slide the tensor proportional to $\Lambda^{(n)}_z$ along the orange arrow. In the second step, we move the middle $\pm\beta$ Ising interaction terms through the SWAP and the next layer of gates. It is now clear that it is connected to its conjugate only by identity, so we can apply Eq.~\eqref{eq:gadget_unitary}, obtaining a factor of $\frac{1}{2}$ from each Rényi copy. Next we slide back the tensor proportional to $\Lambda^{(n)}_z$ and highlight with a dashed circle that the desired Ising interactions are gone, leaving only the SWAPs behind. Applying this simplification to diagram ~\eqref{fig:proof_6} we can remove all gates that are not connected to their conjugate with a tensor proportional to $\Lambda_x^{(n)}$:
\begin{align}
    &\raisebox{-6.0\baselineskip}{%
        \includegraphics[
        width=\linewidth,
        keepaspectratio,
        ]{figures_compressed/LB0_proof8.jpeg}%
    }.
    \label{fig:proof_8}
\end{align}
The number of pairs of gates that have been simplified is $2(2(T-1)\frac{T}{2}-1)=2(T^2-T-1)$ and each contributes $\frac{1}{2}$, for each Rényi copy. The global scalar becomes $c_{\eqref{fig:proof_8}}=c_{\eqref{fig:proof_6}}\cdot 2^{-2n(2(T^2-T-1))}=2^{4n(T^2+1)-4T}2^{-2n(2(T^2-T-1))}=2^{4nT+8n-4T}$. We highlight in blue a loop what contains only $Z$-spiders. The right half of the loop can be contracted and fused to one of the $Z$-spiders. Here is an example of this simplification, which uses identity removal $(id)$:
\begin{align}
    &\raisebox{-6.0\baselineskip}{%
        \includegraphics[
        width=0.6\linewidth,
        keepaspectratio,
        ]{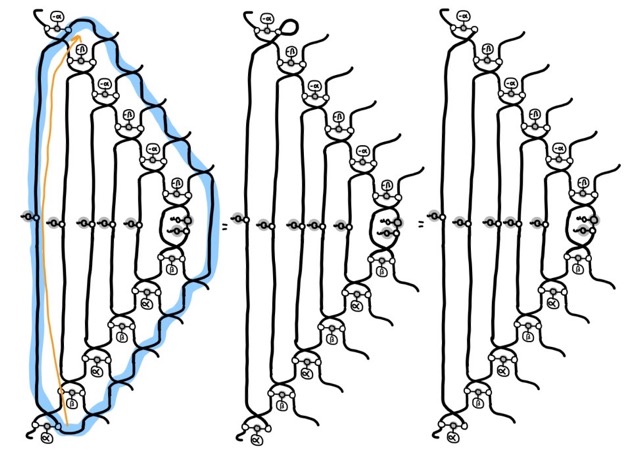}%
    }.
    \label{fig:proof_10}
\end{align}
Applying this simplification to all loops without a $\Lambda_x^{(n)}$, diagram ~\eqref{fig:proof_8} becomes:
\begin{align}
    &\raisebox{-6.0\baselineskip}{%
        \includegraphics[
        width=\linewidth,
        keepaspectratio,
        ]{figures_compressed/LB0_proof9.jpeg}%
    }.
    \label{fig:proof_9}
\end{align}
The global scalar remains the same: $c_{\eqref{fig:proof_9}}=c_{\eqref{fig:proof_8}}=2^{4nT+8n-4T}$. We highlight in blue that we can use Lemmas \ref{lem:LzI} and \ref{lem:Lacross} to remove two pairs of $\Lambda_x^{(n)},\Lambda^{(n)}_z$, obtaining a scalar $\frac{1}{(2^{2n-1})^2}$. The global scalar is then $c_{\eqref{fig:proof_11}}=c_{\eqref{fig:proof_9}}\cdot 2^{-4n+2}=2^{4nT+8n-4T}\cdot 2^{-4n+2}=2^{4nT+4n-4T+2}$. We now reshape the remaining diagram in a more compact way:
\begin{align}
    &\raisebox{-6.0\baselineskip}{%
        \includegraphics[
        width=0.8\linewidth,
        keepaspectratio,
        ]{figures_compressed/LB0n2.jpeg}%
    }.
    \label{fig:proof_11}
\end{align}
For the rest of the proof, it will be more convenient to move both endcaps to the same side. Since they are connected to the middle section via $Z$-spiders, they commute, so we can slide the right endcap through to the left:
\begin{align}
    &\raisebox{-2.0\baselineskip}{%
        \includegraphics[
        width=0.8\linewidth,
        keepaspectratio,
        ]{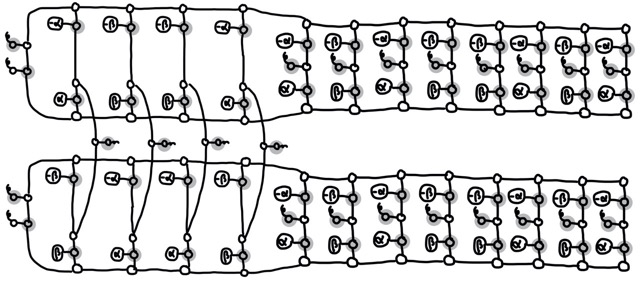}%
    }.
    \label{fig:proof_12}
\end{align}
We will evaluate this diagram by expanding the terms proportional to $\Lambda_x^{(n)}$ into sums, computing each diagram, and summing at the end:
\begin{align}
    &\raisebox{-0.2\baselineskip}{%
        \includegraphics[
        width=0.6\linewidth,
        keepaspectratio,
        ]{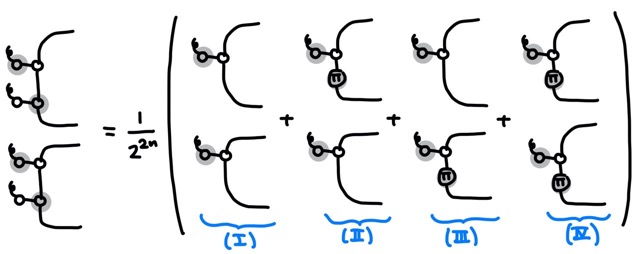}%
    }.
    \label{fig:proof_13}
\end{align}
The first term (I) is the simplest. We highlight below  in blue that the Ising interactions with phase $\pm \alpha$ are connected only by $Z$-spiders. Therefore, we can use Eq.~\eqref{eq:gadget_unitary} on the top and bottom sections to simplify them. Since all $\pm \alpha,\pm \beta$ cancel, the resulting scalar is independent of $\alpha,\beta$.
\begin{align}
    &\raisebox{-2.0\baselineskip}{%
        \includegraphics[
        width=\linewidth,
        keepaspectratio,
        ]{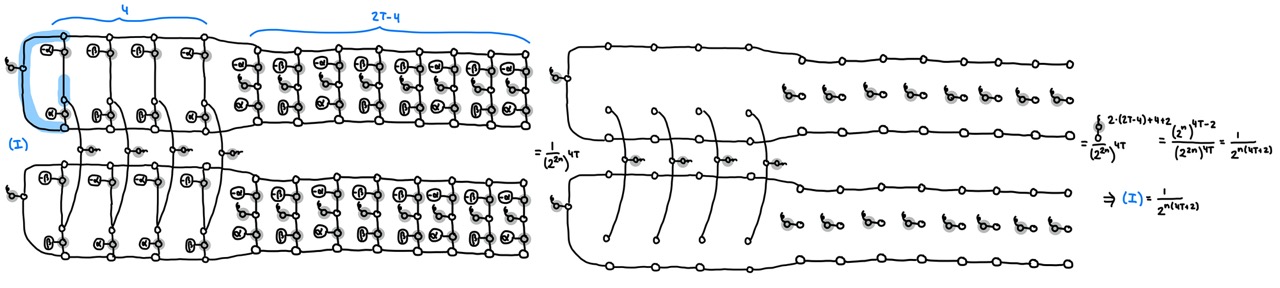}%
    }.
    \label{fig:proof_14}
\end{align}

The second term (II) has a phase-$\pi$ $X$-spider on the top half. As we show below, we slide it through the top as indicated by the orange arrow, changing all the $-$ phases into $+$ by the $\pi$-copy rule ($\pi$). This happens as well on the rest of the Rényi copies, so the $e^{-i\alpha}$ from the odd copies cancels with the $e^{i\alpha}$ from the even copies, and similarly for $\beta$. The bottom half does not have a $\pi$ $X$-spider, so we can use Eq.~\eqref{eq:gadget_unitary} as in case $(I)$ above to simplify to:
\begin{align}
    &\raisebox{-2.0\baselineskip}{%
        \includegraphics[
        width=\linewidth,
        keepaspectratio,
        ]{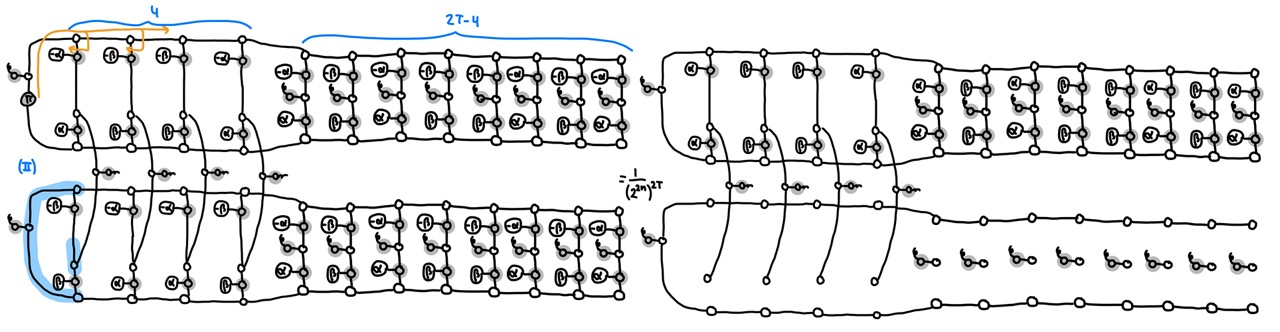}%
    }.
    \label{fig:proof_15}
\end{align}
The bottom half of the diagram above is a scalar independent of $\alpha,\beta$. We fuse the tensors proportional to $\Lambda_z^{(n)}$ to the top diagram. In the first equality below we highlight in blue that we can use Eq.~\eqref{eq:gadget_unitary}. Since the phases add up to $2\alpha,2\beta$ instead of $0$, the phase gadget with the updated phases remain in the second equality:
\begin{align}
    &\raisebox{-2.0\baselineskip}{%
        \includegraphics[
        width=0.8\linewidth,
        keepaspectratio,
        ]{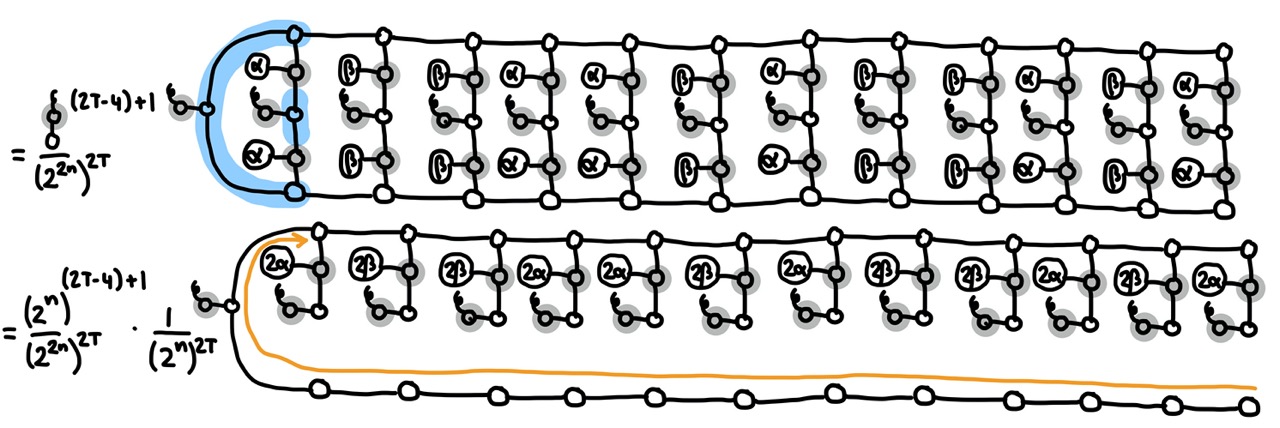}%
    }.
    \label{fig:proof_16}
\end{align}
In the next diagram, we fuse the phaseless $Z$-spiders along the orange arrow.
\begin{align}
    &\raisebox{0\baselineskip}{%
        \includegraphics[
        width=0.8\linewidth,
        keepaspectratio,
        ]{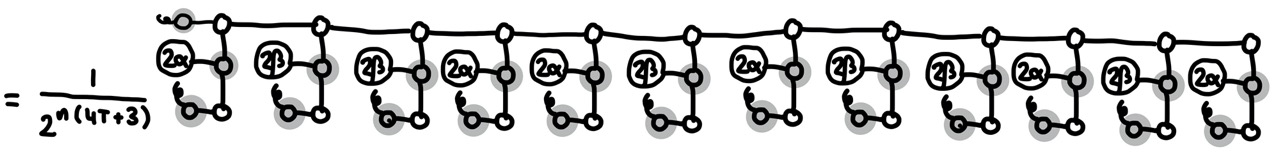}%
    }.
    \label{fig:proof_17}
\end{align}
We will apply the following simplification to each of thee phase spiders in diagram above. First we unfuse a $Z$-spider $(f)$. We indicate with an arrow that we will slide the $\Lambda_z^{(n)}$ on the branch with the phase $2\alpha$ $Z$-spider using Lemma~\ref{lem:LzSlide}. We then use the copy rule $(c)$ to split away a scalar from the rest of the diagram:
\begin{align}
    &\raisebox{-2\baselineskip}{%
        \includegraphics[
        width=0.6\linewidth,
        keepaspectratio,
        ]{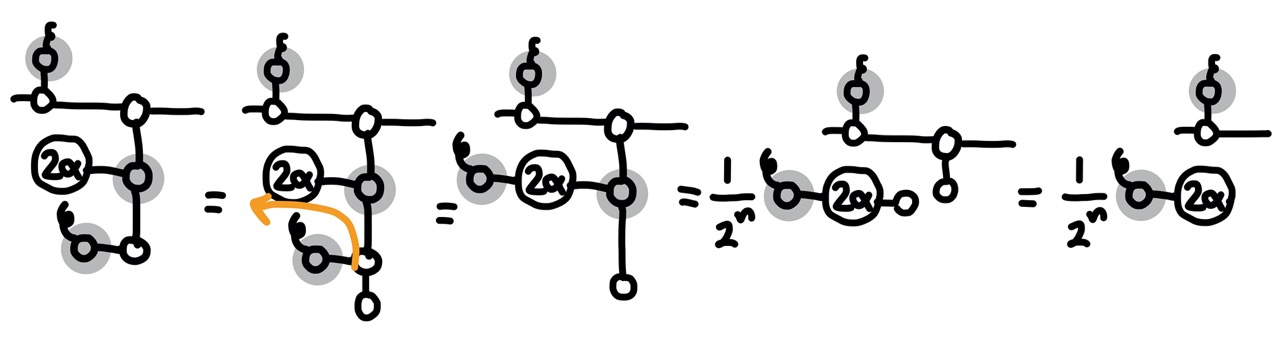}%
    }.
    \label{fig:proof_16.3}
\end{align}
Applying this to diagram \eqref{fig:proof_17} and simplifying the resulting number using diagram~\eqref{fig:tadpole} below yields the value of the second term (II).
\begin{align}
    &\raisebox{0\baselineskip}{%
        \includegraphics[
        width=0.9\linewidth,
        keepaspectratio,
        ]{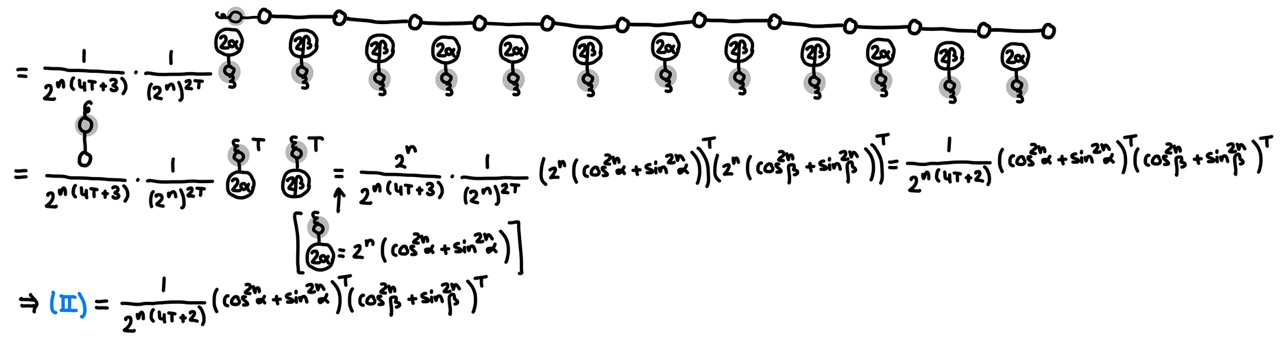}%
    }
    \label{fig:proof_18}
\end{align}

Here we compute the value of the diagram used above. We only use the fusion rule $(f)$ and write the numerical value for each expression
\begin{align}
    &\raisebox{0\baselineskip}{%
        \includegraphics[
        width=0.8\linewidth,
        keepaspectratio,
        ]{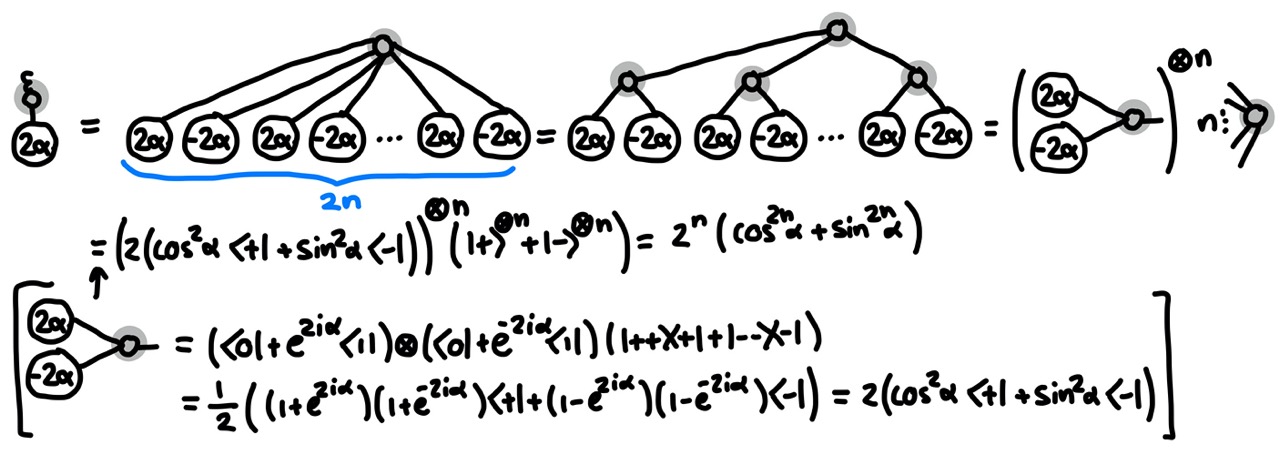}%
    }.
    \label{fig:tadpole}
\end{align}

The third term (III) can be obtained from the second term (II). We first exchange the indicated parts of the diagram in orange arrows, since they are connected by $Z$-spiders, so they commute. Then we move the top half of the diagram underneath the bottom half, arriving at (II).
\begin{align}
    &\raisebox{0\baselineskip}{%
        \includegraphics[
        width=\linewidth,
        keepaspectratio,
        ]{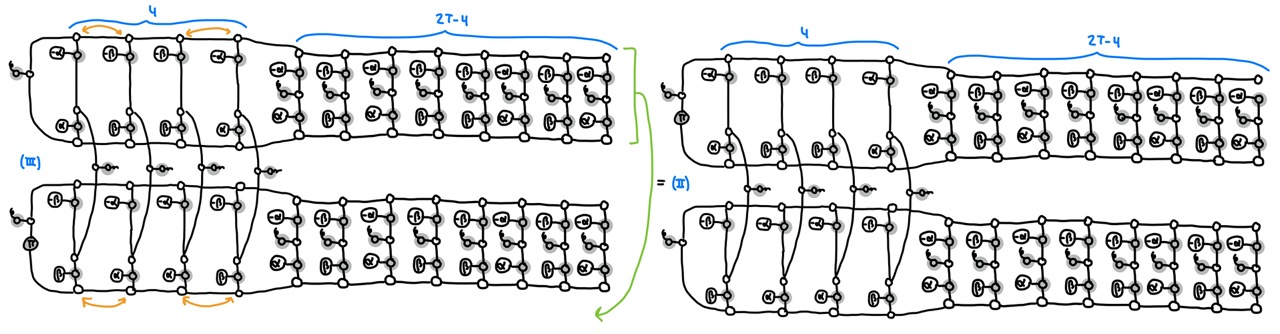}%
    }.
    \label{fig:proof_19}
\end{align}

The fourth term (IV) has a phase-$\pi$ $X$-spider on top and bottom halves. We slide them as indicated by orange arrows, making all the $-\alpha,-\beta$ into $\alpha,\beta$ by the $\pi$-copy rule ($\pi$). Again, the global phase contributions from the odd and even Rényi copies cancel. 
\begin{align}
    &\raisebox{0\baselineskip}{%
        \includegraphics[
        width=\linewidth,
        keepaspectratio,
        ]{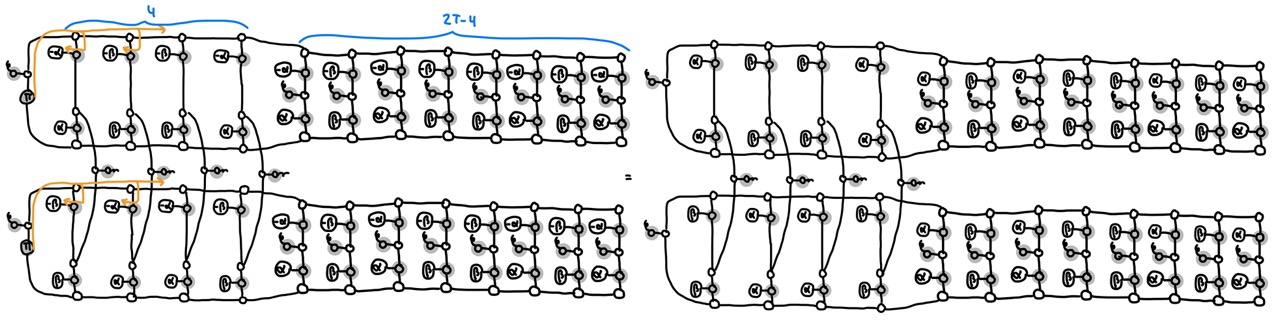}%
    }
    \label{fig:proof_20}
\end{align}

We follow the same procedure as in part (II) to slide the $\Lambda_z^{(n)}$ onto the phase spiders using Lemma~\ref{lem:LzSlide}.
\begin{align}
    &\raisebox{0\baselineskip}{%
        \includegraphics[
        width=\linewidth,
        keepaspectratio,
        ]{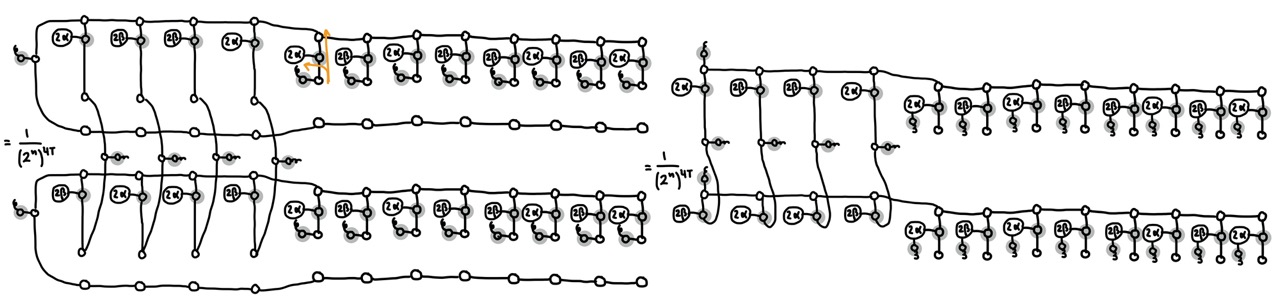}%
    }.
    \label{fig:proof_21}
\end{align}
We apply the copy rule $(c)$ to split away the subdiagrams~\eqref{fig:tadpole},  as in part (II). In the first equality, we introduce new projectors $\Lambda^{(n)}_z$.  In the second step, we use Lemma~\ref{lem:LxSquared}. 
The resulting ZX-diagram is simplified below. 
\begin{align}
    &\raisebox{0\baselineskip}{%
        \includegraphics[
        width=\linewidth,
        keepaspectratio,
        ]{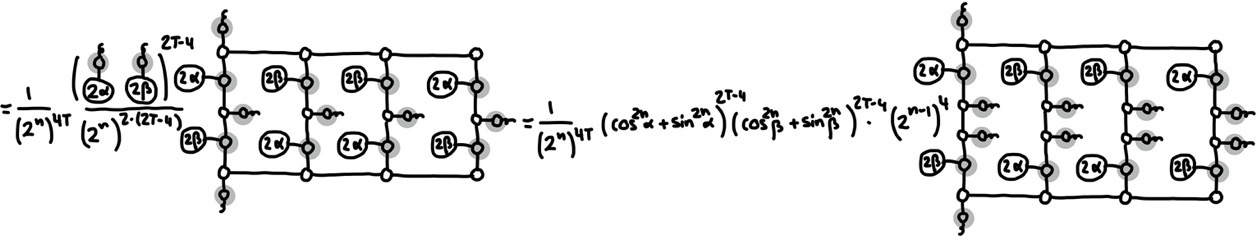}%
    }.
    \label{fig:proof_22}
\end{align}

We now compute the value of the last diagram in Eq.~\eqref{fig:proof_22}. In the first equality, we slide the $\Lambda_z^{(n)}$ in the middle section onto the phase spiders using Lemma~\ref{lem:LzSlide}. Then we use the fusion rule $(f)$ to reshape the diagram
\begin{align}
    &\raisebox{0\baselineskip}{%
        \includegraphics[
        width=0.9\linewidth,
        keepaspectratio,
        ]{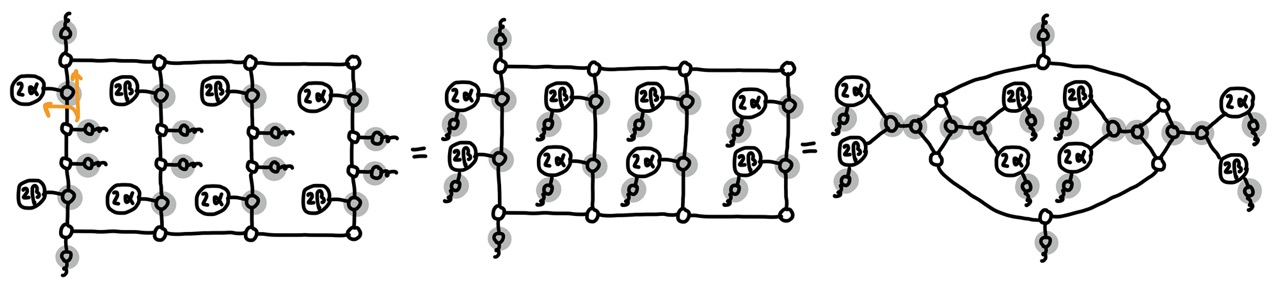}%
    }.
    \label{fig:last_1}
\end{align}
We can now use the bialgebra rule $(b)$ in the next two steps to simplify the diagram. In the last diagram we indicate that we will use Lemma~\ref{lem:LzSlide} to slide one of the $\Lambda_z^{(n)}$ along the orange arrows.
\begin{align}
    &\raisebox{0\baselineskip}{%
        \includegraphics[
        width=0.9\linewidth,
        keepaspectratio,
        ]{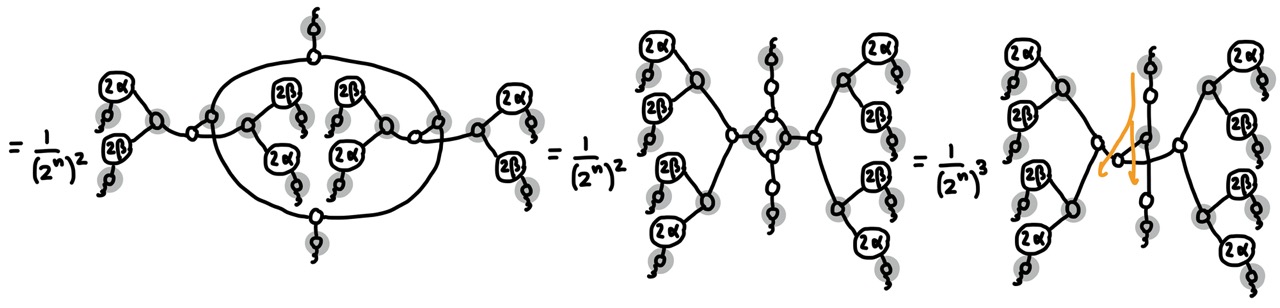}%
    }
    \label{fig:last_2}
\end{align}
In the first equality we use the copy rule $(c)$ to split away one of the $\Lambda_z^{(n)}$. In the last equality we signal that we will use Lemma~\ref{lem:LzSlide} along the orange arrows.
\begin{align}
    &\raisebox{0\baselineskip}{%
        \includegraphics[
        width=0.9\linewidth,
        keepaspectratio,
        ]{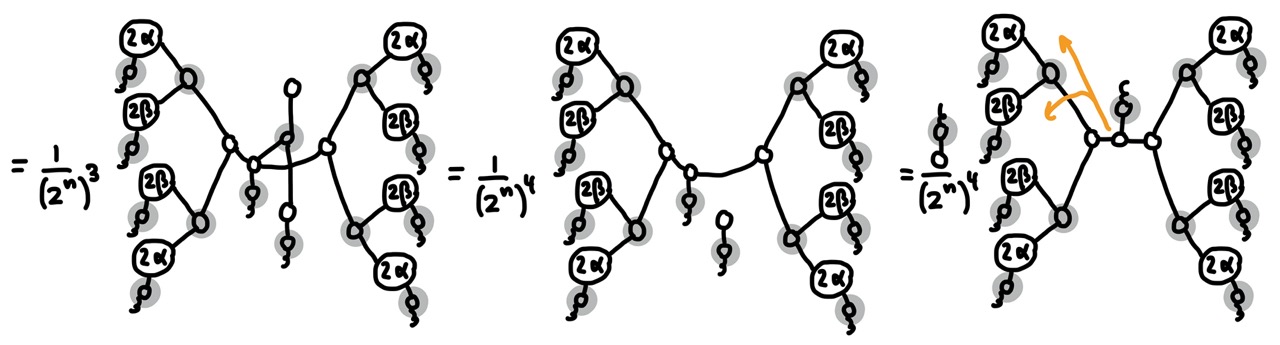}%
    }
    \label{fig:last_3}
\end{align}
Below, we first use Lemma~\ref{lem:LxSquared} to remove one of the $\Lambda_z^{(n)}$. Then we reshape the diagram using the fusion rule $(f)$. 
\begin{align}
    &\raisebox{0\baselineskip}{%
        \includegraphics[
        width=0.9\linewidth,
        keepaspectratio,
        ]{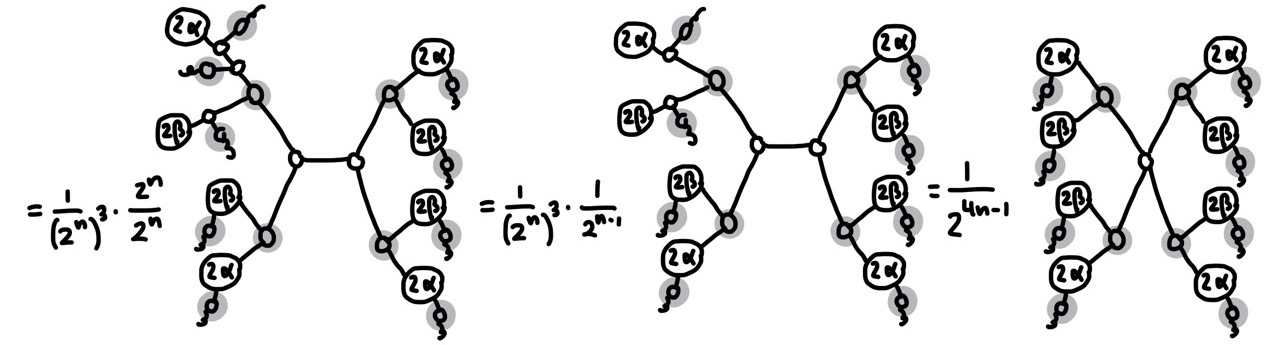}%
    }.
    \label{fig:last_4}
\end{align}

Next, we provide an expression for a useful tensor. It can be obtained by writing out the equation for each tensor and multiplying.
\begin{align}
    &\raisebox{0\baselineskip}{%
        \includegraphics[
        width=0.9\linewidth,
        keepaspectratio,
        ]{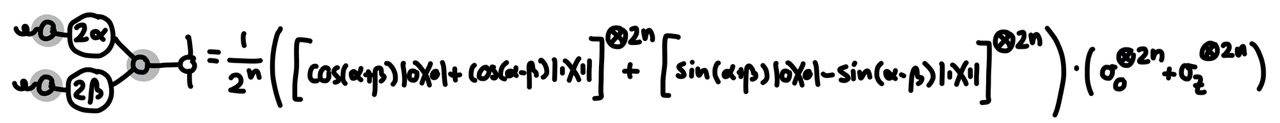}%
    }.
    \label{fig:last_5}
\end{align}
We use this to compute the value of the last diagram in Eq.~\eqref{fig:last_4}, using the notation $r=\cos(\alpha+\beta),s=\cos(\alpha-\beta),u=\sin(\alpha+\beta),v=-\sin(\alpha-\beta)$:
\begin{align}
    &\raisebox{0\baselineskip}{%
        \includegraphics[
        width=0.5\linewidth,
        keepaspectratio,
        ]{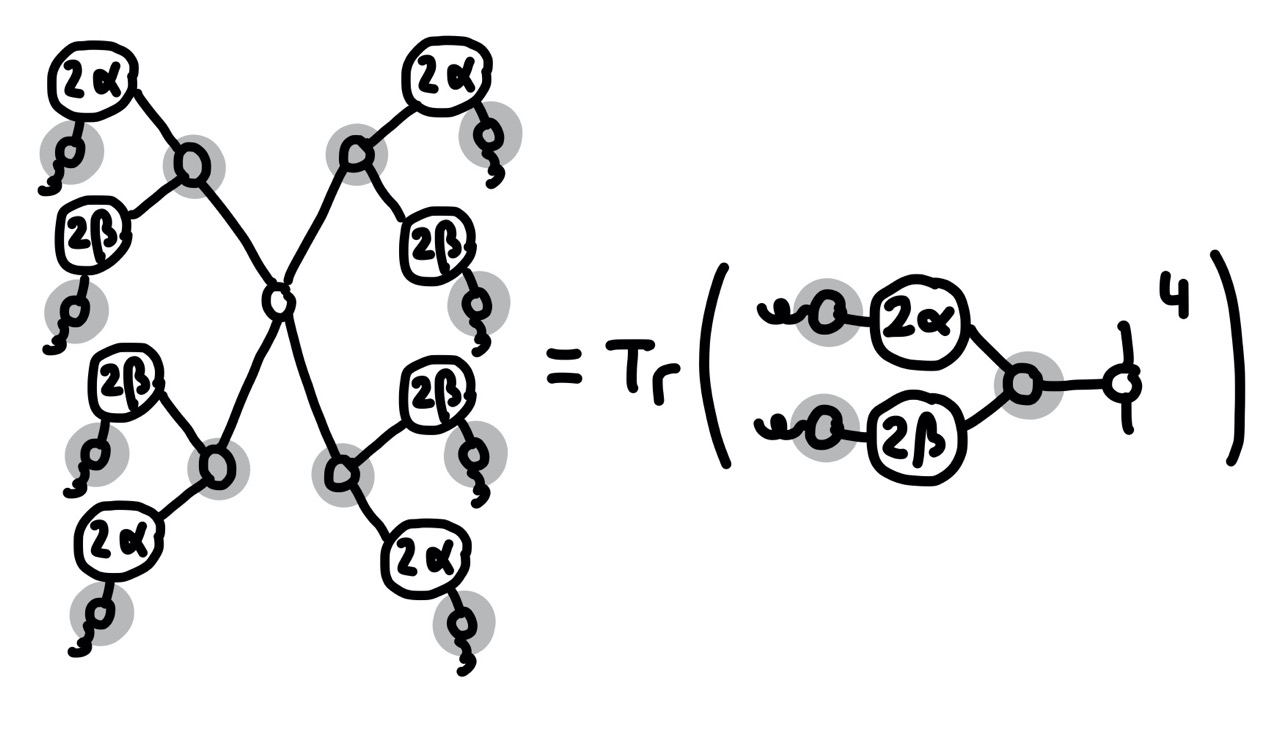}%
    }\label{fig:last_6}\\
    &=\frac{1}{2^{4n}}\tr\bigg(\Big[\big([r|0\rangle\langle 0|+s|1\rangle\langle 1|]^{\otimes 2n}+[u|0\rangle\langle 0|+v|1\rangle\langle 1|]^{\otimes 2n}\big)\big(\sigma_0^{\otimes 2n}+\sigma_z^{\otimes 2n}\big)\Big]^4\bigg)=\\
    &=\frac{1}{2^{4n}}\tr\bigg(\Big[\big([r|0\rangle\langle 0|+s|1\rangle\langle 1|]^{\otimes 2n}+[u|0\rangle\langle 0|+v|1\rangle\langle 1|]^{\otimes 2n}\big)^4\big(\sigma_0^{\otimes 2n}+\sigma_z^{\otimes 2n}\big)^4\bigg).
\end{align}
We used that $\sigma_0,\sigma_z$ commute with $|0\rangle\langle 0|,|1\rangle\langle 1|$. Next we expand the powers of $4$ and multiply them out.
\begin{align}
    &=\frac{1}{2^{4n}}\tr\bigg({4\choose 0}\Big[\big([r^4|0\rangle\langle 0|+s^4|1\rangle\langle 1|]^{\otimes 2n}+{4\choose 1}[r^3u|0\rangle\langle 0|+s^3v|1\rangle\langle 1|]^{\otimes 2n}+\\
    &+{4\choose 2}[r^2u^2|0\rangle\langle 0|+s^2v^2|1\rangle\langle 1|]^{\otimes 2n}+
    {4\choose 3}[ru^3|0\rangle\langle 0|+sv^3|1\rangle\langle 1|]^{\otimes 2n}+\\
    &+{4\choose 4}[u^4|0\rangle\langle 0|+v^4|1\rangle\langle 1|]^{\otimes 2n}\big)\big(\sigma_0^{\otimes 2n}+\sigma_z^{\otimes 2n}\big)2^3\bigg)=\\
    &=\frac{1}{2^{4n}}\tr\bigg({4\choose 0}((r^4+s^4)^{2n}+(r^4-s^4)^{2n})+{4\choose 1}((r^3u+s^3v)^{2n}+(r^4-s^4)^{2n})+\\
    &+{4\choose 2}((r^2u^2+s^2v^2)^{2n}+(r^2u^2-s^2v^2)^{2n})+
    {4\choose 3}((ru^3+sv^3)^{2n}+(ru^3-sv^3)^{2n})+\\
    &+{4\choose 4}((u^4+v^4)^{2n}+(u^4-v^4)^{2n})\bigg)=\\
    &=\frac{1}{2^{4n-3}}\underset{k=0}{\overset{4}{\sum}}{4\choose k}\underset{m=0}{\overset{1}{\sum}}(r^{4-k}u^k+(-1)^ms^{4-k}v^k)^{2n}=\\
    &=\frac{1}{2^{4n-3}}\underset{k=0}{\overset{4}{\sum}}{4\choose k}\underset{m=0}{\overset{1}{\sum}}(\cos(\alpha+\beta)^{4-k}\sin(\alpha+\beta)^k+(-1)^m\cos(\alpha-\beta)^{4-k}\sin(\alpha-\beta)^k)^{2n}.
\end{align}
We absorbed the $(-1)^k$ from $v$ into the sum over $m$. Collecting the scalars from Eq.~\eqref{fig:proof_22} we obtain the fourth term (IV):
\begin{align}
    &\raisebox{0\baselineskip}{%
        \includegraphics[
        width=0.9\linewidth,
        keepaspectratio,
        ]{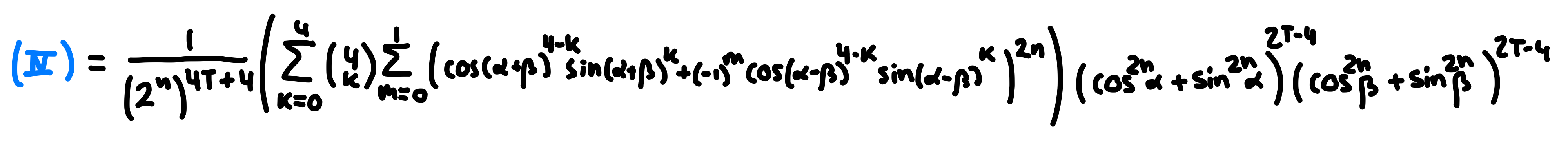}%
    }.
    \label{fig:term4}
\end{align}
Gathering the terms (I)-(IV) we obtain:
\begin{align}
    &\raisebox{0\baselineskip}{%
        \includegraphics[
        width=0.6\linewidth,
        keepaspectratio,
        ]{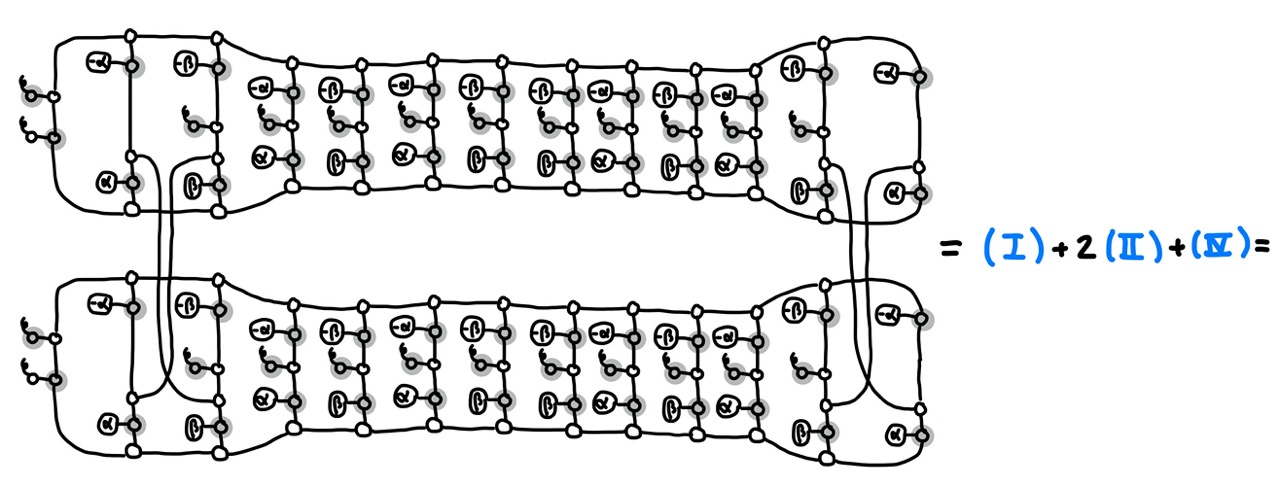}%
    }
    \label{fig:proof_23}\\
    &=\frac{1}{2^{n(4T+4)}}\big(1+2\big((\cos^{2n}(\alpha)+\sin^{2n}(\alpha))(\cos^{2n}(\beta)+\sin^{2n}(\beta))\big)^T+ \\
    &+\frac{1}{2^{2n}}\bigg(\underset{k=0}{\overset{4}{\sum}}{4\choose k}\underset{m=0}{\overset{1}{\sum}}\big(\cos(\alpha+\beta)^{4-k}\sin(\alpha+\beta)^k+(-1)^{m}\cos(\alpha-\beta)^{4-k}\sin(\alpha-\beta)^k\big)^{2n}\bigg)\cdot \notag \\
    &\cdot\big((\cos^{2n}(\alpha)+\sin^{2n}(\alpha))(\cos^{2n}(\beta)+\sin^{2n}(\beta))\big)^{2T-4}.\notag
\end{align}
Thus, together with the global scalar $c_{\eqref{fig:proof_11}}=2^{4nT+4n-4T+2}$, we obtain $\zeta_n(\rho_{AB})$:
\begin{alignat}{2}
    &\zeta_n(\rho_{AB})=\frac{1}{2^{4T-2}}\big(1+2\big((\cos^{2n}(\alpha)+\sin^{2n}(\alpha))(\cos^{2n}(\beta)+\sin^{2n}(\beta))\big)^T+ \\
    &+\frac{1}{2^{2n}}\bigg(\underset{k=0}{\overset{4}{\sum}}{4\choose k}\underset{m=0}{\overset{1}{\sum}}\big(\cos(\alpha+\beta)^{4-k}\sin(\alpha+\beta)^k+(-1)^{m}\cos(\alpha-\beta)^{4-k}\sin(\alpha-\beta)^k\big)^{2n}\bigg)\cdot \notag \\
    &\cdot\big((\cos^{2n}(\alpha)+\sin^{2n}(\alpha))(\cos^{2n}(\beta)+\sin^{2n}(\beta))\big)^{2T-4}.\notag
\end{alignat}
\end{proof}

\section{Diagonal+SWAP gates}
\label{app:duxxz}
In this Appendix we show that all two-qubit gates that can be written as a product of a SWAP and a term $D$ diagonal in the computational basis are dual-unitary XXZ gates with single-qubit $Z$-rotations.
Any unitary two-qubit gate $D$ that is diagonal in the computational basis is of the form:
\begin{align}
    &D = (e^{i a}|00\rangle\langle 00|+e^{i b}|01\rangle\langle 01|+e^{i c}|10\rangle\langle 10|+|11\rangle\langle 11|)e^{i d},
\end{align}
where $a,b,c,d\in[0,2\pi]$. We claim that $D$ can be written as:
\begin{align}
    &D=\exp(-i\frac{\alpha}{2}\sigma_z\otimes \sigma_z)(|0\rangle\langle 0|+e^{i\beta}|1\rangle\langle 1|)\otimes(|0\rangle\langle 0|+e^{i\gamma}|1\rangle\langle 1|)e^{i\nu}=\\
    &=e^{i\nu}(e^{-i\frac{\alpha}{2}}|00\rangle\langle 00|+e^{i(\frac{\alpha}{2}+\gamma)}|01\rangle\langle 01|+e^{i(\frac{\alpha}{2}+\beta)}|10\rangle\langle 10|+e^{i(-\frac{\alpha}{2}+\beta+\gamma)}|11\rangle\langle 11|)=\\
    &=e^{i(\nu-\frac{\alpha}{2}+\beta+\gamma)}(e^{-i(\beta+\gamma)}|00\rangle\langle 00|+e^{i(\alpha-\beta)}|01\rangle\langle 01|+e^{i(\alpha-\gamma)}|10\rangle\langle 10|+|11\rangle\langle 11|),
\end{align}

where $\alpha,\beta,\gamma,\nu\in[0,2\pi]$. Choosing $\alpha=\frac{b+c-a}{2},\beta =\frac{c-a-b}{2},\gamma=\frac{b-a-c}{2},\nu = d-\beta-\gamma+\frac{\alpha}{2}$ we see that any gate that is a product of a SWAP and a diagonal gate in the computational basis, $D$, is a dual-unitary XXZ gate $U(J)$, as in Eq.~\eqref{eq:V(J)DUXXZ}, with single-qubit $Z$-rotations:
\begin{align}
    &D\cdot \text{SWAP}=U(J)R_Z(\beta)\otimes R_Z(\gamma) e^{i\nu}.
\end{align}

\end{appendices}

\end{document}